\documentclass[runningheads]{llncs}

\usepackage[T1]{fontenc}

\usepackage{graphicx}

\usepackage{amsmath}

\usepackage{amssymb}

\makeatletter
\let\langle\undefined \let\rangle\undefined
\DeclareSymbolFont{yhlargesymbols}{OMX}{yhex}{m}{n}
\DeclareMathDelimiter{\rangle}
  {\mathclose}{symbols}{"69}{yhlargesymbols}{"0B}
\DeclareMathDelimiter{\langle}
  {\mathopen}{symbols}{"68}{yhlargesymbols}{"0A}

\makeatother

\usepackage{mathtools}
\usepackage{mathrsfs}
\usepackage{mathpartir}
\usepackage{tabularx}
\usepackage{scrextend}
\usepackage{tikz}
\usepackage{tikz-cd}
\tikzcdset{every label/.append style = {font = \Large}}
\usepackage{dashbox}
\usepackage{subfiles}

\usepackage{mleftright}
\mleftright

\mathtoolsset{showonlyrefs=true}

\makeatletter
\renewcommand{\@Opargbegintheorem}[4]{%
  #4\trivlist\item[\hskip\labelsep{#3#2\@thmcounterend}]}
\makeatother

\usepackage{xcolor}

\newcommand{\comment}[1]{\relax}

\newenvironment{caseindent}{\begin{addmargin}[\parindent]{0pt}}{\end{addmargin}}

\newcommand{\mam}{\mathbf{MAM}}
\newcommand{\eff}{\mathbf{EFF}_{\mathrm{one}}}
\newcommand{\del}{\mathbf{DEL}_{\mathrm{one}}}
\newcommand{\ac}{\mathbf{AC}}
\newcommand{\reflang}{\mathbf{REF}}

\newcommand{\syntacticset}[1]{\mathscr{#1}}
\newcommand{\dom}[1]{\mathrm{Dom}(#1)}
\newcommand{\im}[1]{\mathrm{Im}(#1)}

\newcommand{\getnum}[1]{\mathrm{get}\left(#1\right)}

\newcommand{\var}[1]{\mathit{#1}}
\newcommand{\unit}{\mathtt{()}}
\newcommand{\vpair}[2]{(#1,#2)}
\newcommand{\inj}[2]{\mathbf{inj}_{\mathrm{#1}}\,#2}
\newcommand{\thunk}[1]{\left\{#1\right\}}
\newcommand{\force}[1]{#1!}
\newcommand{\return}[1]{\mathbf{return}\;#1}
\newcommand{\seq}[3]{\mathbf{let}\;\var{#1} = #2\;\mathbf{in}\;#3}
\newcommand{\letin}[3]{\mathbf{let}\;\var{#1} = #2\;\mathbf{in}\;#3} %
\newcommand{\abs}[2]{\lambda \var{#1}.\;#2}

\newcommand{\app}[2]{#1\;#2}

\newcommand{\cpair}[2]{\left\langle #1,#2\right\rangle}
\newcommand{\prj}[2]{\mathbf{prj}_{#1}\,#2}
\newcommand{\pcase}[4]{\mathbf{case}\;#1\;\mathbf{of}\;(#2,#3) \mapsto #4}
\newcommand{\scase}[4]{\mathbf{case}\;#1\;\mathbf{of}\;\{(\inj{#2}{#3} \mapsto #4)_i\}}
\newcommand{\underscore}{\texttt{\_}}
\newcommand{\paren}[1]{\left(#1\right)}

\newcommand{\nil}{\mathbf{nil}}

\newcommand{\shift}[2]{\mathbf{S_0} \var{#1}.\;#2}
\newcommand{\dollar}[3]{\left\langle #1 \middle\vert \var{#2}. #3 \right\rangle}
\newcommand{\dollart}[2]{\left\langle #1 \middle\vert #2 \right\rangle}
\newcommand{\throw}[2]{\mathbf{throw}\;#1\;#2}
\newcommand{\plug}[2]{\mathcal{#1}\left[#2\right]}

\newcommand{\op}[1]{\mathtt{#1}}
\newcommand{\opcall}[2]{\app{\op{#1}}{#2}}
\newcommand{\handle}[2]{\mathbf{with}\;#1\;\mathbf{handle}\;#2}
\newcommand{\handler}[3]{\{\mathbf{return}\;\var{#1} \mapsto #2,\;#3\}}

\newcommand{\labeledc}[2]{\var{#1} : #2}
\newcommand{\labeledcp}[2]{\var{#1} : \paren{#2}}
\newcommand{\create}[1]{\mathbf{create}\;#1}
\newcommand{\resume}[2]{\mathbf{resume}\;#1\;#2}
\newcommand{\yield}[1]{\mathbf{yield}\;#1}
\newcommand{\refcell}[1]{\var{RefCell}(#1)}
\newcommand{\activelabels}[1]{\mathit{AL}(#1)}
\newcommand{\wellformed}[1]{\mathit{WellFormed}(#1)}

\newcommand{\set}[2]{\mathbf{set}\;#1\;#2}
\newcommand{\get}[1]{\mathbf{get}\;#1}

\newcommand{\config}[2]{\left\langle #1; #2\right\rangle}

\newcommand{\hole}{[\ ]}
\newcommand{\context}[1]{\mathcal{#1}}
\newcommand{\Eval}[1]{\mathrm{Eval}_{#1}}
\newcommand{\eval}[2]{\mathrm{Eval}_{#1}\left(#2\right)}

\newcommand{\redbetaM}[2]{#1 \rightarrow_{\mathbf{M}}^{\beta} #2}

\newcommand{\redbetaE}[2]{{#1} \rightarrow_{\mathbf{E}}^{\beta} {#2}}
\newcommand{\redbetaEp}[4]{\config{#1}{#2} \rightarrow_{\mathbf{E}}^{\beta} \config{#3}{#4}}

\newcommand{\redbetaD}[2]{{#1} \rightarrow_{\mathbf{D}}^{\beta} {#2}}
\newcommand{\redbetaDp}[4]{\config{#1}{#2} \rightarrow_{\mathbf{D}}^{\beta} \config{#3}{#4}}

\newcommand{\redbetaAC}[2]{{#1} \rightarrow_{\mathbf{AC}}^{\beta} {#2}}
\newcommand{\redbetaACp}[4]{\config{#1}{#2} \rightarrow_{\mathbf{AC}}^{\beta} \config{#3}{#4}}

\newcommand{\redbetaRp}[4]{\config{#1}{#2} \rightarrow_{\mathbf{R}}^{\beta} \config{#3}{#4}}

\newcommand{\redM}[2]{#1 \rightarrow_{\mathbf{M}} #2}
\newcommand{\redMclos}[2]{#1 \rightarrow^{*}_{\mathbf{M}} #2}

\newcommand{\redE}[2]{#1 \rightarrow_{\mathbf{E}} #2}
\newcommand{\arrE}{\rightarrow_{\mathbf{E}}}
\newcommand{\redEp}[4]{\config{#1}{#2} \rightarrow_{\mathbf{E}} \config{#3}{#4}}
\newcommand{\redEclos}[2]{#1 \rightarrow^{*}_{\mathbf{E}} #2}
\newcommand{\redEplus}[2]{#1 \rightarrow^{+}_{\mathbf{E}} #2}

\newcommand{\redD}[2]{#1 \rightarrow_{\mathbf{D}} #2}
\newcommand{\arrD}{\rightarrow_{\mathbf{D}}}
\newcommand{\redDp}[4]{\config{#1}{#2} \rightarrow_{\mathbf{D}} \config{#3}{#4}}
\newcommand{\redDclos}[2]{#1 \rightarrow^{*}_{\mathbf{D}} #2}
\newcommand{\redDplus}[2]{#1 \rightarrow^{+}_{\mathbf{D}} #2}

\newcommand{\redAC}[2]{#1 \rightarrow_{\mathbf{AC}} #2}
\newcommand{\arrAC}{\rightarrow_{\mathbf{AC}}}
\newcommand{\redACp}[4]{\config{#1}{#2} \rightarrow_{\mathbf{AC}} \config{#3}{#4}}
\newcommand{\redACclos}[2]{#1 \rightarrow^{*}_{\mathbf{AC}} #2}
\newcommand{\redACplus}[2]{#1 \rightarrow^{+}_{\mathbf{AC}} #2}

\newcommand{\redR}[2]{#1 \rightarrow_{\mathbf{R}} #2}
\newcommand{\redRp}[4]{\config{#1}{#2} \rightarrow_{\mathbf{R}} \config{#3}{#4}}
\newcommand{\redRclos}[2]{#1 \rightarrow^{*}_{\mathbf{R}} #2}
\newcommand{\redRplus}[2]{#1 \rightarrow^{+}_{\mathbf{R}} #2}

\newcommand{\mt}[1]{\underline{#1}}
\newcommand{\mte}[2]{\underline{#1}_{#2}}
\newcommand{\defines}{\coloneq}

\newcommand{\mtempty}{\mt{\makebox[2mm][c]{$\cdot$}}}

\newcommand{\Succ}[1]{\app{\var{Succ}}{#1}}

\newcommand{\pri}[1]{\mathrm{pr}_1(#1)}
\newcommand{\prii}[1]{\mathrm{pr}_2(#1)}
\newcommand{\priii}[1]{\mathrm{pr}_3(#1)}
\newcommand{\simu}[2]{#1 \sim #2}

\newcommand{\simuek}[4]{#1\;\overunderset{{}^{#3}}{#4}{\sim}\;#2}
\newcommand{\simuekc}[4]{#1\;\overunderset{{}^{#3}}{#4}{\sim_{\mathsf{c}}}\;#2}

\newcommand{\rsimuek}[2]{\overunderset{{}^{#1}}{#2}{\sim}}
\newcommand{\rsimuekc}[2]{\overunderset{{}^{#1}}{#2}{\sim_{\mathsf{c}}}}

\newcommand{\trconst}[1]{\underline{\mathbf{#1}}}

\usepackage[%
  setpagesize=false,%
  bookmarks=true,%
  bookmarksdepth=2,%
  bookmarksnumbered=true,%
  colorlinks=false,%
  hidelinks,%
  pdftitle={},%
  pdfsubject={},%
  pdfauthor={},%
  pdfkeywords={},%
  pdfpagelabels=true,%
  plainpages=false%
  ]{hyperref}
\usepackage{hyperref}
\usepackage{color}

\AddToHook{cmd/appendix/before}{}

\begin{document}

\title{Expressive Power of One-Shot Control Operators and Coroutines} 

\author{Kentaro Kobayashi\inst{1}\orcidID{0009-0004-3505-7151} \and Yukiyoshi Kameyama\inst{1}\orcidID{0000-0002-2693-5133}}
\authorrunning{K. Kobayashi and Y. Kameyama}

\institute{University of Tsukuba, Ibaraki, Japan\\
\email{\{kentaro.kobayashi,kameyama\}@acm.org}}

\maketitle

\begin{abstract}
Control operators, such as exceptions and effect handlers, provide a means of representing computational effects in programs abstractly and modularly.
While most theoretical studies have focused on multi-shot control operators, one-shot control operators -- which restrict the use of captured continuations to at most once -- are gaining attention for their balance between expressiveness and efficiency.
This study aims to fill the gap. We present a mathematically rigorous comparison of the expressive power among one-shot control operators, including effect handlers, delimited continuations, and even asymmetric coroutines. 
Following previous studies on multi-shot control operators, we adopt Felleisen's macro-expressiveness as our measure of expressiveness.
We verify the folklore that one-shot effect handlers and one-shot delimited-control operators can be macro-expressed by asymmetric coroutines, but not vice versa.
We explain why a previous informal argument fails, and how to revise it to make a valid macro-translation.
\end{abstract}

\section{Introduction}

Control operators are powerful tools for representing computational effects.
Exceptions and coroutines are classic control operators,
implemented in many languages.
Delimited-control operators (e.g., shift/reset)
have been extensively studied in the literature.
The last decade has seen growing interest in effect handlers, which support modular abstraction of computational effects~\cite{plotkin2003algebraic,plotkin2009handlers}.

This paper studies the theoretical foundation of \emph{one-shot} variants of control operators, where captured continuations are restricted to at most one use.
While most studies\footnote{A notable exception is Berdine et al., who proposed linearly-used continuations~\cite{berdine2002linear}.} on control operators focused on unrestricted (i.e., \emph{multi-shot}) control operators, one-shot variants have been recently gaining attention.
There are several reasons to consider one-shot variants:
First, they can be implemented more efficiently than
multi-shot, as they avoid stack copying~\cite{bruggeman1996oneshot}.
Second, they may alleviate the verification burden by reducing the complexity of reasoning about sensitive resources~\cite{vilhena2021separation}.
Third, one-shotness is key to relating control operators based on continuations to classic ones found in many dynamic languages.
For instance, a recent example in the former category is one-shot effect handlers in OCaml Version 5.x,\footnote{\url{https://ocaml.org}} while the latter category includes coroutines and the yield operator, both of which are intrinsically one-shot.\footnote{James and Sabry argued that a multi-shot variant of the yield operator is as expressive as delimited-control operators~\cite{yield2011}.}

Despite these advantages, few authors have studied the theoretical foundation of one-shot control operators.
Indeed, it is folklore that results for multi-shot control operators carry over easily to their one-shot counterparts; however, this is not the case.
Since one-shotness is a dynamic property, a formal calculus for one-shot control operators should track the validity of each continuation, complicating both the semantics and precise reasoning about them.
Among the few authors, de Moura and Ierusalimschy demonstrated a connection between one-shot delimited-control operators and coroutines~\cite{moura2009revisiting}.
However, this correspondence relies heavily on mutable states that can store higher-order functions, which, in our view, obscures the raw expressive power of these control operators.

We note that comparing the expressiveness of control operators is surprisingly difficult.
For the case of one-shot control operators, the expressiveness results in the literature often lacked correctness proofs (e.g.,\cite{kawahara2020one}), or were incorrect.
In this paper, we explain why a simple and seemingly correct translation from delimited-control operators to coroutines fails to preserve semantics.

This paper studies the relative expressiveness among three one-shot control operators: one-shot effect handlers, one-shot delimited-control operators, and asymmetric coroutines~\cite{moura2009revisiting}.
We adopt macro-expressibility~\cite{felleisen1991expressive} as the basis for our comparison,
since it is well-established in the literature, particularly in the study of control operators, such as
Forster, Kammar, Lindley and Pretnar~\cite{forster2019expressive}.
To our knowledge, this is the first systematic study to rigorously analyze the expressiveness of one-shot control operators.

Our contributions are threefold:
\begin{enumerate}
\item We prove that
one-shot delimited-control operators
can be macro-expressed by asymmetric coroutines.
\item We also prove that
one-shot effect handlers
can be macro-expressed by asymmetric coroutines.
\item We show that the converse direction does not hold:
asymmetric coroutines cannot be macro-expressed
by either one-shot delimited-control operators or
one-shot effect handlers.
\end{enumerate}
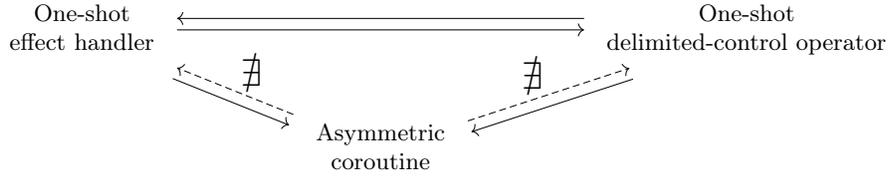
\begin{figure}[h]
  \centering
  \begin{tikzcd}
    \text{
      \begin{tabular}{c}
        One-shot \\
        effect handler
      \end{tabular}
    } \arrow[rrrr] \arrow[rrd, shift right=2] &  &                                                                                                &  & \text{
      \begin{tabular}{c}
        One-shot \\
        delimited-control operator
      \end{tabular}
    } \arrow[llll, shift right=2] \arrow[lld, shift left=2] \\
    &  & \text{
      \begin{tabular}{c}
        Asymmetric\\
        coroutine
      \end{tabular}
    } \arrow[rru, "\nexists", dashed] \arrow[llu, "\nexists"', dashed] &  &
  \end{tikzcd}
  \caption{Macro-expressibility among control operators. Solid lines indicate the existence of a macro-translation; dashed lines indicate its non-existence.}
  \label{fig:relations}
\end{figure}
Figure~\ref{fig:relations} illustrates the macro-expressibility results established in this paper.

The remainder of this paper is organized as follows.
Section~\ref{chap:framework} introduces macro-translations and the core calculus.
Sections~\ref{chap:DELoneToAC} and \ref{chap:EFFoneToDELone} present three extensions to the core calculus and establish the macro-expressibility results, while Section~\ref{chap:nonexistent} disproves the macro-expressibility of the converse direction.
Section~\ref{chap:conclusion} concludes the paper.

\section{Macro-expressibility and Core Calculus}\label{chap:framework}

\subsection{Macro-expressibility}
Felleisen~\cite{felleisen1991expressive} formalized the notion of
\textit{macro-expressibility} to compare the expressive power of two
programming languages when one is an extension of the other.
The notion has later been adjusted to compare two languages when both
are extensions of a common language~\cite{forster2019expressive}.
Below, we show a definition based on operational semantics, in the spirit of Felleisen's formalization.

We assume that a programming language $\mathscr{L}$
is equipped with a set of $\mathscr{L}$-phrases,
a set of $\mathscr{L}$-programs, which is a
non-empty subset of $\mathscr{L}$-phrases,
and an evaluation function $\eval{\mathscr{L}}{\cdot}$
that maps $\mathscr{L}$-programs to (suitably defined)
values.
Phrases and programs may be distinct in some languages.
For instance,
phrases can contain runtime values such as
references.
An $n$-hole syntactic abstraction in $\mathscr{L}$ is
an $\mathscr{L}$-phrase that has $n$ holes.

\begin{definition}\label{def:macro-expressibility}
  Let $\mathscr{L}_1$ and $\mathscr{L}_2$ be conservative extensions\footnote{The formal definition of a conservative extension is provided in Appendix~\ref{app:conservative_extensions}.} of $\mathscr{L}$.
  A partial map $\phi$ from $\mathscr{L}_1$-phrases to $\mathscr{L}_2$-phrases is a macro-translation if
  and only if all the following conditions are satisfied:
  \begin{enumerate}
  \item If $M$ is an $\mathscr{L}_1$-program, $\phi(M)$ is defined and an $\mathscr{L}_2$-program.
  \item If $F$ is an $n$-ary function symbol of $\mathscr{L}$, for any $\mathscr{L}_1$-phrases $M_1, \ldots, M_n$,
    $\phi(F(M_1, \ldots, M_n)) = F(\phi(M_1), \ldots, \phi(M_n))$.
  \item For each $n$-ary function symbol $F \in \mathscr{L}_1\setminus\mathscr{L}$, there is an $n$-hole syntactic abstraction $A$ in $\mathscr{L}_2$ such that $\phi(F(M_1, \ldots, M_n)) = A[\phi(M_1), \ldots, \phi(M_n)]$ for any $\mathscr{L}_1$-phrases $M_1, \ldots, M_n$.
  \item $\eval{\mathscr{L}_1}{M}$ terminates
if and only if $\eval{\mathscr{L}_2}{\phi(M)}$ terminates.
  \end{enumerate}
We say $\mathscr{L}_1$ is (strongly)
macro-expressible in $\mathscr{L}_2$
if a macro-translation from $\mathscr{L}_1$ to $\mathscr{L}_2$ exists.
We can define a \text{weak} macro-translation
by replacing the ``if and only if''-clause in
the last condition by ``only if''.
\end{definition}

An example of macro-expressibility is
the let-construct in call-by-value lambda calculus.
We can define a macro-translation $\phi$ in
such a way that
$\phi(\letin{x}{M}{N})
= \app{(\abs{x}{\phi(N)})}{\phi(M)}$ holds.
Macro-expressibility is transitive: the composition of macro-translations is also a macro-translation.

\subsection{Core Calculus \texorpdfstring{$\mam$}{MAM}}

We use the language $\mam$ (multi-adjunctive language)
by Forster et al.\ \cite{forster2019expressive},
which was designed after Levy's Call-By-Push-Value calculus~\cite{levy2004cbpv}.
It serves as the common core calculus for our extensions.

\begin{figure}[tb]
  \setlength{\parindent}{0cm}
  \setlength{\arraycolsep}{4pt}%
  \renewcommand{\arraystretch}{0.9}%
  \noindent
  \begin{minipage}[t]{0.645\columnwidth}
    \begin{tabular}[t]{@{}l@{\hspace{0.1em}}l@{\hspace{0.2em}}l@{}}
      $V,W$ & $::=$ & \textsf{value} \\
            & \textbar\ $x \in \syntacticset{V}$ & variable \\
            & \textbar\ $\unit$ & unit \\[1em]
      $M,N$ & $::=$ & \textsf{computation} \\
            & \textbar\ $\pcase{V}{x_1}{x_2}{M}$ & product matching \\      
            & \textbar\ $\scase{V}{L_{\mathnormal{i}}}{x_i}{M_i}$ & variant matching \\
            & \textbar\ $\force{V}$ & force  \\
            & \textbar\ $\return{V}$ & returner
    \end{tabular}
  \end{minipage}%
  \begin{minipage}[t]{0.35\columnwidth}
    \begin{tabular}[t]{@{}l@{\hspace{0.2em}}l@{}}  
      \textbar\ $\vpair{V}{W}$ & pairing \\
      \textbar\ $\inj{L}{V}\quad(\mathrm{L} \in \syntacticset{C})$ & variant \\
      \textbar\ $\thunk{M}$ & thunk \\[1em]
      \textbar\ $\seq{x}{M}{N}$ & sequencing \\
      \textbar\ $\abs{x}{M}$ & abstraction \\
      \textbar\ $\app{M}{V}$ & application \\
      \textbar\ $\cpair{M}{N}$ & pairing \\
      \textbar\ $\prj{i}{M}$ & projection
    \end{tabular}
  \end{minipage}%
  \caption{Syntax of $\mam$}
  \label{fig:mam_syntax}
\end{figure}

Figure~\ref{fig:mam_syntax} gives the syntax of $\mam$.
$\mam$ differs from
the untyped lambda calculus
in that
values and computations are clearly separated,
and its own constructs are as follows:
The pairing of computations $\cpair{M}{N}$
is \textit{lazy}, 
a thunk freezes a computation,
and a force thaws the thunk out:
$\redbetaM{\force{\thunk{M}}}{M}$.

We consider terms modulo renaming of bound variables as usual.
We define $\mam$-Phrases as the union of the set of values and that of computations.
$\mam$-Programs are the set of computations.

To present the operational semantics of $\mam$, we define frames and contexts in Figure~\ref{fig:mam_frame}, and the
beta reduction rules $\redbetaM{}{}$ on computations in Figure~\ref{fig:mam_beta}.
\begin{figure}[tb]
  \setlength{\arraycolsep}{4pt}%
  \renewcommand{\arraystretch}{0.9}%
  \begin{tabular}{rrcl}
    \textsf{pure frame} & $\mathcal{P}$ & ::= & $\seq{x}{\hole}{N}$ \textbar\ $\app{\hole}{V}$ \textbar\ $\prj{i}{\hole}$ \\
    \textsf{computational frame} & $\mathcal{F}$ & ::= & $\mathcal{P}$ \\
    \textsf{pure context} & $\mathcal{H}$ & ::= & $\hole$ \textbar\ $\mathcal{P}[\mathcal{H}\hole]$ \\
    \textsf{evaluation context} & $\mathcal{C}$ & ::= & $\hole$ \textbar\ $\mathcal{F}[\mathcal{C}\hole]$
  \end{tabular}%
  \caption{Frames and Contexts of $\mam$}
  \label{fig:mam_frame}
\end{figure}
\begin{figure}[tb]
  \centering
  \begin{minipage}[t]{0.49\linewidth}
    \begin{tabular}{l}
    $(\times)\quad\begin{array}{l}
      \redbetaM{\pcase{(V_1,V_2)}{x_1}{x_2}{M}\\}{M[V_1/x_1,V_2/x_2]}
                 \end{array}$ \\
    $(+)\quad\begin{array}{l}
              \redbetaM{\mathbf{case}\;\mathbf{inj}_{\mathrm{L}_k}\,{V}\;\mathbf{of}\;\{(\mathbf{inj}_{\mathrm{L}_i}\,{x_i} \mapsto M_i)_i\}\\}{M_k[V/x_k]}
            \end{array}$
  \end{tabular}
  \end{minipage}%
  \begin{minipage}[t]{0.51\linewidth}
    \begin{tabular}{l}
    $(F)\quad\redbetaM{\seq{x}{\return{V}}{M}}{M[V/x]}$ \\      
    $(U)\quad\redbetaM{\force{\thunk{M}}}{M}$ \\
    $(\to)\quad\redbetaM{\app{(\abs{x}{M})}{V}}{M[V/x]}$ \\
    $(\&)\quad\redbetaM{\prj{i}{\cpair{M_1}{M_2}}}{M_i}$
  \end{tabular}
  \end{minipage}%
  \caption{Beta Reduction Rules of $\mam$}
  \label{fig:mam_beta}
\end{figure}
We define the transition relation $\rightarrow_{\mathbf{M}}$ on computations by using evaluation contexts~\cite{felleisen1987reduction} as follows:
\[
  \inferrule
    {\redbetaM{M}{M'}}
    {\redM{\mathcal{C}[M]}{\mathcal{C}[M']}}
\]

Evaluation of a program is defined by:
$\eval{\mam}{M} \defines V \quad\text{if}\quad \redMclos{M}{\return{V}}$.
$\eval{\mam}{M}$ is a well-defined partial function, since $\rightarrow^{\mathbf{M}}$ is deterministic.

\section{One-Shot Delimited Continuations as Asymmetric Coroutines}\label{chap:DELoneToAC}

This section investigates the macro-expressibility of one-shot delimited continuations in terms of asymmetric coroutines, which has been considered folklore, but this proves unexpectedly complicated.
We first introduce two calculi: $\del$ for one-shot delimited continuations and $\ac$ for asymmetric coroutines, then
present a macro-translation from $\del$ to $\ac$.

\subsection{The Calculus for One-Shot Delimited Continuations}
\label{subsec:del}

We present the calculus $\del$, which incorporates a one-shot version of the control operator $\mathit{shift}_0/\mathit{dollar}$.
The $\mathit{shift}_0/\mathit{dollar}$ operator, proposed by Materzok and Biernacki~\cite{materzok2012ADI}, is a variant of $\mathit{shift}_0/\mathit{reset_0}$~\cite{danvy1989functional} and has the same macro-expressive power.
We adopt the calculus $\mathbf{DEL}$ defined by Forster et al.~\cite{forster2019expressive}, restricting each captured continuation to one-shot use.

Figure~\ref{fig:del_syntax} shows the syntax of $\del$ as an extension of $\mam$, where the ellipses (\ldots) indicate the syntax from $\mam$.
\begin{figure}[tb]
  \begin{minipage}[t]{0.42\columnwidth}
    \begin{tabular}[t]{@{}l@{\hspace{0.1em}}l@{\hspace{0.2em}}l@{}}
      $V,W$ & $::= \ldots$ & \textsf{value} \\
      & \textbar\ $l\in\syntacticset{L}_{\mathbf{D}}$ & continuation label
    \end{tabular}
  \end{minipage}%
    \begin{minipage}[t]{0.33\columnwidth}
    \begin{tabular}[t]{@{}l@{\hspace{0.1em}}l@{\hspace{0.2em}}l@{}}
      $M,N$ & $::= \ldots$ & \textsf{computation} \\
      & \textbar\ $\shift{k}{M}$ & shift0
    \end{tabular}%
  \end{minipage}%
    \begin{minipage}[t]{0.2\columnwidth}
    \begin{tabular}[t]{@{}l@{\hspace{0.2em}}l@{}}
      \textbar\ $\dollar{M}{x}{N}$ & dollar \\
      \textbar\ $\throw{V}{W}$ & throw
    \end{tabular}%
  \end{minipage}\\[1em]
  \begin{minipage}[t]{0.5\columnwidth}
    \begin{tabular}{rrcl}
      \textsf{pure frame} & $\mathcal{P}$ & ::= & \ldots \\
      \textsf{computational frame} & $\mathcal{F}$ & ::= & \ldots \textbar\  $\dollar{\hole}{x}{N}$ \\

    \end{tabular}
  \end{minipage}
  \begin{minipage}[t]{0.5\columnwidth}
    \begin{tabular}{rrcl}
      \textsf{pure context} & $\mathcal{H}$ & ::= & \ldots \\
      \textsf{evaluation context} & $\mathcal{C}$ & ::= & \ldots \\
    \end{tabular}
  \end{minipage}
\caption{Syntax of $\del$}
\label{fig:del_syntax}
\end{figure}
$\syntacticset{L}_{\mathbf{D}}$ is a countable set of continuation labels, where $l$ is a dynamically generated label for a continuation.
The dollar term $\dollar{M}{x}{N}$ is similar to reset0: when evaluated,
it installs a delimiter for continuations captured in $M$.
Unlike reset0, it has an additional part $x. N$ where $x$ is bound in $N$.
When $M$ evaluates to a value, its result is bound to $x$, and $N$ is evaluated.
When the term $\shift{k}{M}$ is evaluated, a continuation delimited by the nearest dollar term is captured, $k$ is bound to the continuation, and the body $M$ is evaluated.
If there is no surrounding dollar term, the evaluation gets stuck.
$\throw{l}{V}$ invokes the continuation represented by $l$, passing $V$ as an argument.
$\del$-Phrases is the set of all $\del$-values and $\del$-computations, and $\del$-Programs is the set of computations without continuation labels.

In $\del$, continuations can be invoked at most once.
Consider the example:
\[
  \dollar{
    \begin{aligned}
      \shift{k}{
      &\letin{a}{\throw{k}{1}}{\\&\letin{b}{\throw{k}{2}}{\return{\vpair{a}{b}}}}
      }
    \end{aligned}       
  }{x}{\return{x}}
\]
The term attempts to use the continuation $k$ twice, first to compute $a$ and then to compute $b$.
Such usage violates the one-shotness constraint, and
to detect it, we use a \textit{store} to record the content and the status of continuations.

A store is a partial function
$\theta : \syntacticset{L}_{\mathbf{D}}\;\rightharpoonup\;\mathsf{computation}\sqcup\{\nil\}$.
$\dom{\theta}$ is the set of continuation labels $l$ such that $\theta(l)$ is defined.
Note that, $\nil$ is not an undefined element, and the complement of $\dom{\theta}$ is \textit{not} equal to $\theta^{-1}(\nil)$.
If $\theta(l)=\nil$, which means that the continuation labeled $l$ has already been invoked.
The empty set $\emptyset$ represents the store with no bindings.

We introduce \textit{configurations} to describe the runtime states of programs.
A configuration $C$ is a pair of a $\del$-computation $M$ and a store $\theta$, or the error state $\bot$.
The beta reduction rules $\rightarrow^{\mathbf{D}}_{\beta}$ on configurations are defined in Figure~\ref{fig:del_beta}.\footnote{In $(\mathrm{shift})$, we assume that $y$ does not freely occur in the context $\context{H}$. Similarly, throughout the rest of this paper, we assume that the same condition holds for variables that are freshly generated, whether in reduction rules or in translations.}

\begin{figure}[tb]
  \centering
  \begin{tabular}{cl}
    $(\mam)$ & $\inferrule{\redbetaM{M}{M'}}{\redbetaDp{M}{\theta}{M'}{\theta}}$ \\
    $(\mathrm{ret})$ & $\redbetaDp{\dollar{\return{V}}{x}{M}}{\theta}{M[V/x]}{\theta}$ \\
    $(\mathrm{shift})$ &
    $\inferrule
    {
      \text{$l$ is a fresh label}
    }
    { 
      \redbetaDp{\dollar{\plug{H}{\shift{k}{M}}}{x}{N}}{\theta}{M[l/k]}{\theta[l \coloneq \abs{y}{\dollar{\plug{H}{\return{y}}}{x}{N}}]}
    }$ \\
    $(\mathrm{throw})$ &
    $\inferrule
    {
      \theta(l) = \abs{y}{\dollar{\plug{H}{\return{y}}}{x}{N}}
    }
    {
      \redbetaDp{\throw{l}{V}}{\theta}{\dollar{\plug{H}{\return{V}}}{x}{N}}{\theta[l \coloneq \nil]}
    }$ \\
    $(\mathrm{fail})$ &
    $\inferrule
    {
      \theta(l) = \nil
    }
    {
      \config{\throw{l}{V}}{\theta} \rightarrow_{\mathbf{D}}^{\beta} \bot
    }$
  \end{tabular}
  \caption{Beta reduction rules of $\del$}
  \label{fig:del_beta}
\end{figure}

The reduction rules of $\del$ are defined as follows:
\[
  \inferrule{\redbetaDp{M}{\theta}{M'}{\theta'}}{\redDp{\plug{C}{M}}{\theta}{\plug{C}{M'}}{\theta'}} \hspace {4em} \inferrule{\redbetaD{\config{M}{\theta}}{\bot}}{\redD{\config{\plug{C}{M}}{\theta}}{\bot}}
\]
Evaluation of a $\del$-program is defined by: $\eval{\del}{M} \defines V$ if there exists a store $\theta$ such that $\redDclos{\config{M}{\emptyset}}{\config{\return{V}}{\theta}}$.
$\eval{\del}{M}$ is a well-defined partial function, since $\arrD{}$ is deterministic.

\subsection{The Calculus for Asymmetric Coroutines}

De Moura and Ierusalimschy~\cite{moura2009revisiting} studied several variations of coroutines and introduced two calculi, symmetric coroutines and asymmetric coroutines.
In this paper, we present the calculus $\ac$ for asymmetric coroutines, which are prevalent in modern programming languages.

Figure~\ref{fig:ac_syntax} presents the syntax of $\ac$.
\begin{figure}[tb]
  \begin{minipage}[t]{0.37\columnwidth}
    \begin{tabular}[t]{@{}l@{\hspace{0.1em}}l@{\hspace{0.3em}}l@{}}
      $V,W$ & $::= \ldots$ & \textsf{value} \\
       & \textbar\ $l\in\syntacticset{L}_{\ac}$ & coroutine label
    \end{tabular}
  \end{minipage}%
    \begin{minipage}[t]{0.6\columnwidth}
    \begin{tabular}[t]{@{}l@{\hspace{0.15em}}l@{\hspace{0.2em}}l@{\hspace{0.2em}}l@{\hspace{0.2em}}l@{}}
      $M,N$ & $::= \ldots$ & \textsf{computation} & \textbar\ $\yield{V}$ & yield \\
            & \textbar\ $\create{V}$ & create & \textbar\ $\resume{V}{W}$ & resume \\
            & \textbar\ $\labeledc{l}{M}$ & \makebox[0em][l]{labeled computation} & & 
    \end{tabular}%
  \end{minipage}\\
  \begin{minipage}[t]{0.5\columnwidth}
    \begin{tabular}{rrcl}
      \textsf{pure frame} & $\mathcal{P}$ & ::= & \ldots \\
      \textsf{computational frame} & $\mathcal{F}$ & ::= & \ldots \textbar\  $\labeledc{l}{\hole}$ \\

    \end{tabular}
  \end{minipage}
  \begin{minipage}[t]{0.5\columnwidth}
    \begin{tabular}{rrcl}
      \textsf{pure context} & $\mathcal{H}$ & ::= & \ldots \\
      \textsf{evaluation context} & $\mathcal{C}$ & ::= & \ldots \\
    \end{tabular}
  \end{minipage}
\caption{Syntax of $\ac$}
\label{fig:ac_syntax}
\end{figure}
Besides the constructors of $\mam$, $\ac$ has coroutine labels as values, and four constructs to manipulate coroutines as computations: labeled computation, \textbf{create}, \textbf{resume}, and \textbf{yield}.
$\syntacticset{L}_{\ac}$ is a countable set of coroutine labels.
The labeled computation $\labeledc{l}{M}$ represents the coroutine $l$ executing the computation $M$.
$\create{V}$ produces a new coroutine whose computation is $V$, and $\resume{V}{W}$ starts (or resumes) a coroutine whose label is $V$ with a parameter $W$.
$\yield{V}$ suspends the current coroutine, yielding $V$ to its caller.
$\ac$-Phrases are the values and computations of $\ac$, and $\ac$-Programs are the computations without coroutine labels.

A store $\theta$ is a partial function that maps labels to \textit{values} or \texttt{nil}.
Note the difference from the store in $\del$,
which maps labels to \textit{computations} or \texttt{nil}.
As in $\del$, a configuration in $\ac$ is a pair $\config{M}{\theta}$ of a computation $M$ and a store $\theta$, or the error state $\bot$.

\begin{figure}[tb]
  \begin{tabular}{cl}
    $(\mam)$ & $\inferrule{\redbetaM{M}{M'}}{\redbetaACp{M}{\theta}{M'}{\theta}}$ \\

    $(\mathrm{create})$ &
    $\inferrule
    {
      \text{$l$ is a fresh label}
    }
    {
      \redbetaACp{\create{V}}{\theta}{\return{l}}{\theta[l := V]}
    }$ \\

    $(\mathrm{resume})$ &
    $\inferrule
    {
      l \in \dom{\theta} \\ \theta(l) \neq \nil
    }
    {
      \redbetaACp{\resume{l}{V}}{\theta}{\labeledc{l}{(\app{\force{\theta(l)}}{V})}}{\theta[l := \nil]}
    }$ \\

    $(\mathrm{fail})$ &
    $\inferrule
    {
      \theta(l) = \nil
    }
    {
      \redbetaAC{\config{\resume{l}{V}}{\theta}}{\bot}
    }$ \\

    $(\mathrm{ret})$ & $\redbetaACp{\labeledc{l}{\return{V}}}{\theta}{\return{V}}{\theta}$ \\
    $(\mathrm{yield})$ & $\redbetaACp{\labeledc{l}{\context{H}[\yield{V}]}}{\theta}{\return{V}}{\theta[l := \thunk{\abs{y}{\context{H}[\return{y}]}}]}$
  \end{tabular}
  \[
  \inferrule{\redbetaACp{M}{\theta}{M'}{\theta'}}{\redACp{\plug{C}{M}}{\theta}{\plug{C}{M'}}{\theta'}} \hspace {4em} \inferrule{\redbetaAC{\config{M}{\theta}}{\bot}}{\redAC{\config{\plug{C}{M}}{\theta}}{\bot}}
  \]
  \caption{Semantics of $\ac$}
  \label{fig:ac_beta}
\end{figure}

Figure~\ref{fig:ac_beta} defines the operational semantics of $\ac$
which includes
the beta reduction rules $\arrAC^{\beta}$ and
the reduction rules $\arrAC$.
Note that coroutines are inherently one-shot: while a coroutine may be called multiple times by suspending and resuming it, it cannot be duplicated and reused.

Evaluation of an $\ac$-program is defined by: $\eval{\ac}{M} \defines V$ if there exists a store $\theta$ such that $\redACclos{\config{M}{\emptyset}}{\config{\return{V}}{\theta}}$.
Since $\arrAC{}$ is deterministic, $\Eval{\ac}$ is a well-defined partial function.

\subsection{Naive Translation and its Failure}

Given the similarity between delimited continuations and coroutines, one may think that it is straightforward to macro-translate $\del$ to $\ac$ with the following correspondence: a dollar term in $\del$ is translated to a create term in $\ac$, a shift0 term to a yield term, and a throw term to a resume term.
Based on this intuition, we can define the naive translation from $\del$ to $\ac$ in Figure~\ref{fig:deltoac:naive}.
Note that on the right-hand side of $\mt{\dollar{M}{x}{N}}$, $x$ corresponds to the binding occurrence of $x$ in the original term $\dollar{M}{x}{N}$.
Figure~\ref{fig:deltoac:naive} only shows non-trivial cases.
Other cases are translated homomorphically except for continuation labels, whose translations are not defined.
This is not problematic, since a macro-translation only has to translate all $\del$-programs, which do not contain continuation labels.
\begin{figure}[tb]
  \centering
    $\begin{array}{rll}
    \mt{\dollar{M}{x}{N}} & \defines &
                                       \left(\begin{array}{l}
                                         \letin{z}{\create{\thunk{\abs{\underscore}{\letin{x}{\mt{M}}{\return{\thunk{\abs{\underscore}{\mt{N}}}}}}}}}
                                         {\\\letin{res}{\resume{z}{\unit}}
                                         {\\\app{\force{\var{res}}}{z}}}
                                       \end{array}\right) \\
    \mt{\shift{k}{L}} & \defines & \yield{\thunk{\abs{k}{\mt{L}}}} \\
    \mt{\throw{V}{W}} & \defines &
                                   \left(\begin{array}{l}
                                     \letin{res}{\resume{\mt{V}}{\mt{W}}}
                                     {\\\app{\force{\var{res}}}{\mt{V}}}
                                   \end{array}\right)
  \end{array}$
  \caption{Naive translation from $\del$ to $\ac$}
  \label{fig:deltoac:naive}
\end{figure}

The naive translation works for simple expressions.
However, we found a counterexaple to it.
Consider the following $\del$-program $M$.
\[
  M \defines
  \left\langle \begin{array}{@{}l@{}}
      \mathbf{let}\;\var{j} = (\shift{k_{\mathrm{1}}}{\mathbf{let}\;\var{r_{\mathrm{1}}} = \throw{k_1}{10}\;\mathbf{in}} \\
      \phantom{\mathbf{let}\;\var{j} = (\mathbf{S_0} \var{k_{\mathrm{1}}}.\;}\mathbf{let}\;\mathrlap{\var{r_{\mathrm{2}}}}\phantom{\var{r_{\mathrm{1}}}} = \throw{k_1}{20}\;\mathbf{in}\;\return{\var{r_{\mathrm{1}}}})\;\mathbf{in} \\
      \shift{k_{\mathrm{2}}}{\return{30}}
    \end{array}\;\middle\vert\;\var{i}.\,\return{i} \right\rangle
\]
In $\del$, $M$ is evaluated as follows:
First, the shift0 term $\shift{k_1}{\cdots}$ is invoked, then the pure context surrounding it is captured and stored under a fresh label $l_1$:\footnote{For brevity, the full content of the store is omitted from the evaluation trace. We will only mention the changes relevant to the main computation.}
\begin{equation}
  M \arrD
  \begin{array}{@{}l@{}}
    \mathbf{let}\;\var{r_{\mathrm{1}}} = \throw{l_1}{10}\;\mathbf{in}\;
    \mathbf{let}\;\mathrlap{\var{r_{\mathrm{2}}}}\phantom{\var{r_{\mathrm{1}}}} = \throw{l_1}{20}\;\mathbf{in}\;\return{\var{r_{\mathrm{1}}}}      
  \end{array}\label{eq:evalM1}  
\end{equation}
Next, $\throw{l_1}{10}$ invokes the captured continuation labeled $l_1$, rendering it invalid:
\begin{align}
    \ldots &\arrD^+
    \paren{\begin{array}{@{}l@{}}
    \mathbf{let}\;\var{r_{\mathrm{1}}} = \left\langle
      \begin{array}{@{}l@{}}
      \mathbf{let}\;\var{j} = \return{10}\;\mathbf{in}\;\\
      \shift{k_{\mathrm{2}}}{\return{30}}
      \end{array}
    \;\middle\vert\;\var{i}.\,\return{i} \right\rangle\;\mathbf{in}\\
    \mathbf{let}\;\mathrlap{\var{r_{\mathrm{2}}}}\phantom{\var{r_{\mathrm{1}}}} = \throw{l_1}{20}\;\mathbf{in}\;\return{\var{r_{\mathrm{1}}}}      
    \end{array}}\label{eq:evalM2} \\
  &\arrD\paren{\begin{array}{@{}l@{}}
    \mathbf{let}\;\var{r_{\mathrm{1}}} = \left\langle
      \shift{k_{\mathrm{2}}}{\return{30}}
    \;\middle\vert\;\var{i}.\,\return{i} \right\rangle\;\mathbf{in}\\
    \mathbf{let}\;\mathrlap{\var{r_{\mathrm{2}}}}\phantom{\var{r_{\mathrm{1}}}} = \throw{l_1}{20}\;\mathbf{in}\;\return{\var{r_{\mathrm{1}}}}      
    \end{array}}\label{eq:evalM3}
\end{align}
Then, the shift0 term $\shift{k_{\mathrm{2}}}{\cdots}$ is invoked.
This captures the context and stores it under another fresh label $l_2$.
The evaluation then continues as follows (note that the continuation labeled $l_2$ is not used):
\begin{align}
    \ldots
     &\arrD^+\begin{array}{@{}l@{}}
      \mathbf{let}\;\var{r_{\mathrm{1}}} = \return{30}\;\mathbf{in}\;
      \mathbf{let}\;\mathrlap{\var{r_{\mathrm{2}}}}\phantom{\var{r_{\mathrm{1}}}} = \throw{l_1}{20}\;\mathbf{in}\;\return{\var{r_{\mathrm{1}}}}
    \end{array}\label{eq:evalM4} \\
    &\arrD\begin{array}{@{}l@{}}
      \mathbf{let}\;\mathrlap{\var{r_{\mathrm{2}}}}\phantom{\var{r_{\mathrm{1}}}} = \throw{l_1}{20}\;\mathbf{in}\;\return{30}      
    \end{array}\label{eq:evalM5}
\end{align}
Finally, $\throw{l_1}{20}$ is invoked, but since the continuation labeled $l_1$ has already been consumed, the evaluation fails.

\begin{figure}[tb]
\begin{align}
    &\mt{M} \arrAC^{+}  \paren{\begin{array}{@{}l@{}}
                            \seq{r_{\mathrm{1}}}{\app{\app{\mt{\mathbf{throw}}}{(l\textcolor{red}{, l_c, 0})}}{10}}
                            {\\\seq{r_{\mathrm{2}}}{\app{\app{\mt{\mathbf{throw}}}{(l\textcolor{red}{, l_c, 0})}}{20}}
                            {\\\return{r_{\mathrm{1}}}}}
                            \end{array}}
                            \left[\begin{array}{@{}l@{}}
                              \text{where} \\
                              l\mapsto\thunk{\begin{array}{@{}l@{}}
                                \abs{y}{\plug{P}{
                                \begin{array}{@{}l@{}}
                                  \seq{j}{\return{y}}
                                  {\\\yield{\thunk{\abs{k_{\mathrm{2}}}{\return{30}}}}}
                                \end{array}
                                }}
                              \end{array}} \\
                              \textcolor{red}{l_c\mapsto\thunk{\refcell{0}}}
                            \end{array}\right] \nonumber \\
    & \text{($\mt{\mathbf{throw}}$ \textcolor{red}{increments $l_c$ and} resumes $l$.)} \nonumber \\
    & \arrAC^{+} \paren{\begin{array}{@{}l@{}}
                            \seq{r_{\mathrm{1}}}{\paren{
                            \begin{array}{@{}l@{}}
                                \seq{res}{\labeledc{l}{\paren{
                                \plug{P}{\begin{array}{@{}l@{}}
                                    \seq{j}{\return{10}}
                                    {\\\yield{\thunk{\abs{k_{\mathrm{2}}}{\return{30}}}}}
                                \end{array}}
                                }} }
                                {\\\app{\force{\var{res}}}{\underbrace{(l\textcolor{red}{, l_c, 1})}_{\text{\scriptsize valid}}}}
                            \end{array}
                            }}
                            {\\\seq{r_{\mathrm{2}}}{\app{\app{\mt{\mathbf{throw}}}{\underbrace{(l\textcolor{red}{, l_c, 0})}_{\text{\scriptsize invalid}}}}{20}}
                            {\\\return{r_{\mathrm{1}}}}}
                            \end{array}} \label{eq:evalmtM1}\\
                            &\left[\begin{array}{@{}l@{}}
                              \text{where} \\
                              l \mapsto \nil \\
                              \textcolor{red}{l_c \mapsto \thunk{\refcell{1}}}
                            \end{array}\right] \\
    &  \text{(The coroutine labeled $l$ is suspended by $\mathbf{yield}$.)} \nonumber \\
    & \arrAC^{+} \paren{\begin{array}{@{}l@{}}
                            \seq{r_{\mathrm{1}}}{\app{\force{\thunk{\abs{k_{\mathrm{2}}}{\return{30}}}}}{(l\textcolor{red}{, l_c, 1})}}
                            {\\\seq{r_{\mathrm{2}}}{\app{\app{\mt{\mathbf{throw}}}{(l\textcolor{red}{, l_c, 0})}}{20}}
                            {\\\return{r_{\mathrm{1}}}}}
                            \end{array}}
                            \left[\begin{array}{@{}l@{}}
                              \text{where} \\
                              l \mapsto \thunk{\begin{array}{@{}l@{}}
                                \abs{y}{\plug{P}{\return{y}}}
                              \end{array}} \\
                              \textcolor{red}{l_c \mapsto \thunk{\refcell{1}}}
                            \end{array}\right] \label{eq:evalmtM2} \\
    &  \text{(The continuation $(l\textcolor{red}{, l_c, 1})$ is not used and $l$ remains active)} \nonumber \\
    &  \arrAC^{+} \paren{\begin{array}{@{}l@{}}
                            \seq{r_{\mathrm{2}}}{\app{\app{\mt{\mathbf{throw}}}{(l\textcolor{red}{, l_c, 0})}}{20}}
                            {\\\return{30}}
                            \end{array}}
                            \left[\begin{array}{@{}l@{}}
                              \text{where} \\
                              l \mapsto \thunk{\begin{array}{@{}l@{}}
                                \abs{y}{\plug{P}{\return{y}}}
                              \end{array}} \\
                              \textcolor{red}{l_c \mapsto \thunk{\refcell{1}}}
                            \end{array}\right] \label{eq:evalmtM3}
\end{align}
\caption{Evaluation of $\mt{M}$ where $\context{P} \equiv \seq{i}{\hole{}}{\return{\thunk{\abs{\underscore}{\return{i}}}}}$. The black parts show the evaluation using the naive translation; the red parts are additions by our refined translation in Figure~\ref{fig:deltoac}.}
\label{fig:deltoac:evalM}
\end{figure}

Therefore, in $\ac$, $\mt{M}$ must not successfully terminate since macro-translations preserve semantics; however, this is not the case.
To understand the reason, consider the evaluation trace shown in the black parts of Figure~\ref{fig:deltoac:evalM} (ignore the red\footnote{In the printed (monochrome) version, the red parts may appear as gray.} parts for the moment).
First, after applying the translation, $\mt{M}$ reduces to the term on the first line (corresponding to \eqref{eq:evalM1}) where the continuation captured by $\shift{k_{\mathrm{1}}}{\cdots}$ in $M$ is represented by a coroutine labeled $l$.
Next, $\app{\mt{\mathbf{throw}}}{\app{l}{10}}$ is evaluated.
This resumes the coroutine labeled $l$ and invalidates $l$ by mapping it to $\nil$.
The evaluation then reaches \eqref{eq:evalmtM1}, which corresponds to \eqref{eq:evalM2}.
Then, $\letin{j}{\return{10}}{\cdots}$ is evaluated and the term $\yield{\thunk{\abs{k_{\mathrm{2}}}{\return{30}}}} (\equiv \mt{\shift{k_{\mathrm{2}}}{\return{30}}})$ is invoked.
This suspends the coroutine labeled $l$, \emph{reactivating} it in the store, as shown in \eqref{eq:evalmtM2}.

This reacvation is the source of the failure.
In the evaluation of $M$ in $\del$, the two shift0 invocations result in two distinct continuation labels, $l_1$ and $l_2$.
Our naive translation, however, maps both of these labels to the same coroutine label $l$.
Because of this, the reactivation of $l$ makes the stale continuation (corresponding to $l_1$) available again.
This would be harmless if the thunk returned by \textbf{yield} eventually resumed $l$, as that would invalidate $l$ again.
In our counterexample, however, this thunk does not resume $l$.
Therefore, the coroutine labeled $l$ remains active and the final \textbf{throw} in \eqref{eq:evalmtM3} succeeds incorrectly.

This phenomenon had been overlooked for years.
In fact, there was folklore that a simple macro-translation from $\del$ to $\ac$ should exist.
For example, Kawahara and Kameyama proposed such a translation from
one-shot effect handlers to asymmetric coroutines, which has been
implemented in Lua, Go, and several other
languages~\cite{kawahara2020one}.
Our analysis, however, reveals that these simple translations fail to preserve semantics and therefore cannot be macro-translations.
The problem becomes apparent when we examine a dollar term that contains
more than one occurrence of shift0.\footnote{It can be shown that Kawahara and Kameyama's translation fails to preserve semantics by considering a similar counterexample.}

\subsection{Refined Translation from \texorpdfstring{$\del$}{DELone} to \texorpdfstring{$\ac$}{AC}}

Our key idea to handle this problem is to introduce a counter mechanism into the translation so that one can distinguish valid continuations from stale ones.
The red parts of Figure~\ref{fig:deltoac:evalM} illustrate how these counters are introduced: $l_c$ is a counter associated with a coroutine $l$, and holds a counter object in the form $\refcell{n}$ where $n$ is a natural number.
Each counter is incremented every time the continuation associated with the corresponding coroutine is used.
Each continuation also carries an index -- a natural number.
If this index does not match the counter's value, it indicates that the continuation has already been used, causing the computation to fail.
In the second use of $k_1$, $k_1$'s index is 0 while $l_c$'s value is 1, correctly invalidating the invocation.

\begin{figure}[tb]
  \noindent
  \begin{minipage}[t]{0.42\linewidth}
      $\begin{array}{l}
        \mt{\shift{k}{M}} \defines \yield{\thunk{\abs{k}{\mt{M}}}} \\
        \mt{\dollar{M}{x}{N}} \defines \\
        \left(\begin{array}{l}
          \letin{z}{\mathbf{create}\\\;\;\thunk{
          \begin{array}{@{}l@{}}
            \lambda \var{\underscore}.\;\mathbf{let}\;\var{x} = M\;\mathbf{in} \\
            \phantom{\lambda \var{\underscore}.\;}\return{\thunk{\abs{\underscore}{\mt{N}}}}
          \end{array}
          }}
          {\\\letin{zc}{\app{\force{\var{ref}}}{0}}
          {\\\letin{res}{\resume{z}{\unit}}
          {\\\app{\force{\var{res}}}{((\var{z}, \var{zc}), 0)}}}}
        \end{array}\right) \\
        \mt{\throw{V}{W}} \defines \\
        \left(\begin{array}{l}
          \mathbf{case}\;\mt{V}\;\mathbf{of}\;\{\\
          \;\;((\var{z}, \var{zc}), \var{i}) \mapsto\\
          \;\;\;\;\mathbf{let}\;\var{j} = \app{\force{\var{get}}}{\var{zc}}\;\mathbf{in} \\
          \;\;\;\;\mathbf{let}\;\var{b} = \app{\app{\force{\var{compare}}}{i}}{j}\;\mathbf{in} \\
          \;\;\;\;\mathbf{case}\;\var{b}\;\mathbf{of}\;\{\\
          \;\;\;\;\;\;\paren{\inj{True}{\unit}} \mapsto\\
          \;\;\;\;\;\;\;\;\mathbf{let}\;\var{i'} = \app{\force{\var{incr}}}{i}\;\mathbf{in}\\
          \;\;\;\;\;\;\;\;\mathbf{let}\;\unit = \app{\app{\force{\var{set}}}{\var{zc}}}{i'}\;\mathbf{in}\\
          \;\;\;\;\;\;\;\;\mathbf{let}\;\var{res} = \app{\app{\mathbf{resume}}{\var{z}}}{\mt{W}}\;\mathbf{in}\\
          \;\;\;\;\;\;\;\;\app{\force{\var{res}}}{((\var{z}, \var{zc}), \var{i'})} \\
          \;\;\;\;\;\;\paren{\inj{False}{\unit}} \mapsto \force{\var{fail}}\\
          \;\;\;\;\} \\
          \}
        \end{array} \right) \\
      \end{array}$
    \end{minipage}
  \begin{minipage}[t]{0.46\linewidth}
   \noindent$\begin{array}{l}
        \text{where} \\
        \var{fail} \defines \\
        \left\{ \begin{array}{l}
          \letin{z}{\create{\thunk{\abs{\underscore}{\return{\unit}}}}}
          {\\\letin{\underscore}{\resume{z}{\unit}}
          {\\\resume{z}{\unit}}}
        \end{array} \right\}\\
        0 \defines \inj{Zero}{\unit} \\
        \var{Succ} \defines \inj{Succ} \\
        \var{incr} \defines \thunk{\abs{n}{\return{\paren{\app{\var{Succ}}{n}}}}} \\
        \var{compare} \defines \\\thunk{\app{\paren{\abs{x}{\app{\force{\var{cmp}}}{\thunk{\app{\force{\var{x}}}{\var{x}}}}}}}{\thunk{\abs{x}{\app{\force{\var{cmp}}}{\thunk{\app{\force{\var{x}}}{\var{x}}}}}}}} \\
        \var{cmp} \defines \\
                               \left\{\begin{array}{l}
                                 \lambda \var{f}.\lambda \var{n}.\lambda \var{m}.\;\mathbf{case}\;\var{(\var{n},\var{m})}\;\mathbf{of}\;\{ \\
                                 \phantom{\lambda \var{f}.}(0, 0) \mapsto \inj{True}{\unit} \\
                                 \phantom{\lambda \var{f}.}(0, \Succ{\var{n'}}) \mapsto \inj{False}{\unit} \\
                                 \phantom{\lambda \var{f}.}(\Succ{\var{n'}}, 0) \mapsto \inj{False}{\unit} \\
                                 \phantom{\lambda \var{f}.}(\Succ{\var{n'}}, \Succ{\var{m'}}) \mapsto \app{\app{\force{\var{f}}}{\var{n'}}}{\var{m'}}\} \\
                               \end{array}\right\} \\
        \var{ref} \defines 
                               \begin{array}{l}
                                 \thunk{\lambda \var{v}.\;\create{\refcell{\var{v}}}}
                               \end{array}\\
        \refcell{v} \defines \\
        \thunk{
        \begin{array}{@{}l@{}}
          \abs{y}{\letin{q'}{\return{\var{y}}}
          {\\\phantom{\lambda \var{y}.\;}\app{\force{\thunk{\app{\paren{\abs{x}{\app{\force{\var{th}}}{\thunk{\app{\force{\var{x}}}{\var{x}}}}}}}{\thunk{\abs{x}{\app{\force{\var{th}}}{\thunk{\app{\force{\var{x}}}{\var{x}}}}}}}}}}{\app{\var{v}}{\var{q'}}}}}
        \end{array}} \\
        \var{th} \defines \\
                              \left\{\begin{array}{@{}l@{}}
                                \lambda \var{f}. \lambda \var{s}. \lambda \var{q}.\;\mathbf{case}\;\var{q}\;\mathbf{of}\;\{\\
                                \quad(\inj{Set}{\var{v}}) \mapsto \letin{q'}{\yield{\unit}}{\app{\app{\force{\var{f}}}{\var{v}}}{\var{q'}}} \\
                                \quad(\inj{Get}{\unit}) \mapsto \letin{q'}{\yield{\var{s}}}{\app{\app{\force{\var{f}}}{\var{s}}}{\var{q'}}}\}
                              \end{array}\right\}\\
        \var{get} \defines \thunk{\abs{c}{\resume{\var{c}}{\inj{Get}{\unit}}}} \\
        \var{set} \defines \thunk{\abs{c}{\abs{v}{\resume{\var{c}}{\inj{Set}{\var{v}}}}}}
      \end{array}$
  \end{minipage}
  \caption{Refined translation from $\del$ to $\ac$}
  \label{fig:deltoac}
\end{figure}

Realizing this idea as a macro-translation requires a bit of programming:
First, we encode natural numbers by regarding $\inj{Zero}{\unit}$ as 0 and $\mathbf{inj}_{\mathrm{Succ}}$ as the successor function, then we can write increment and comparison functions in $\ac$.
Second, we encode counters as mutable cells that are expressible by coroutines (see Section~\ref{chap:nonexistent}).
Figure~\ref{fig:deltoac} shows the complete translation.
A dollar term is translated into a term that creates a new
counter by $\app{\force{\var{ref}}}{\inj{Zero}{\unit}}$, and makes a
pair $(z, zc)$ consisting of the coroutine and the corresponding
counter.
The last element in the argument passed to $\force{\var{res}}$ is
$\inj{Zero}{\unit}$, indicating that the first continuation label to
be generated should have index 0.
A throw term is translated into a term whose first argument $V$ is the
tuple $((z, zc), i)$.
The translated term checks whether the value of the counter $zc$
matches $i$ using $\app{\app{\force{\var{compare}}}{i}}{j}$.
If it holds, the counter is incremented, and the continuation stored
in $z$ is resumed with the argument $W$.
Otherwise, an invalidated continuation is about to be invoked, and the
execution fails.

\subsection{Simulation}

To prove that the translation in Figure~\ref{fig:deltoac} is a valid macro-translation, we use \emph{simulation}.
For two transition systems $\mathscr{L}$ and $\mathscr{L}'$, a binary relation on $\mathscr{L}$-terms and $\mathscr{L}'$-terms is a simulation relation if the following condition holds:
\begin{quote}
  If $M \rightarrow_{\mathscr{L}} M'$ and $M \sim N$ hold, there exists an $N'$ such that $N \rightarrow^{*}_{\mathscr{L}'} N'$ and $M' \sim N'$ hold.
\end{quote}
If such a simulation relation exists, we say $\mathscr{L}'$ \emph{simulates} $\mathscr{L}$.
In our case, we shall construct a simulation relation $\sim$ on $\del$-configurations and $\ac$-configurations that respects the translation in Figure~\ref{fig:deltoac}.
We then use this to show that the translation preserves the semantics.

To establish the simulation, we must resolve the fundamental mismatch between $\del$ and $\ac$: the former generates a fresh label for each continuation capture, while the latter reuses the same coroutine labels.
In our translation, this mismatch is addressed by the counter mechanism.
Therefore, the simulation relation must incorporate the logic of the counter mechanism.

We introduce several auxiliary notions that are necessary to define a simulation relation.
First, since configurations may contain runtime labels, we must extend the translation $\mtempty{}$, which is only defined for label-free $\del$-programs.
For this purpose, we introduce a mapping $\eta$ as a partial function from $\syntacticset{L}_{\del}$ to $\syntacticset{L}_{\ac} \times \syntacticset{L}_{\ac} \times \mathbb{N}$.
The extended translation $\mte{\cdot}{\eta}$ is then defined by $\mte{l}{\eta} \defines \eta(l)$ for continuation labels $l$, while acting identically to $\mtempty{}$ for all other terms.
For convenience, we identify the $\ac$-representations of Peano numbers with natural numbers ($\mathbb{N}$).

Second, to track the association between a coroutine and its dedicated counter, we use a partial function $\kappa$ from $\syntacticset{L}_{\ac}$ to $\syntacticset{L}_{\ac}$.
With the partial functions $\eta$ and $\kappa$, we define a binary relation $\rsimuek{\eta}{\kappa}$ between configurations.
We only present two crucial cases, and the complete definition can be found in Appendix~\ref{app:DELoneToAC:contextual_relation}.
\begin{definition}[excerpt]\label{def:rsim-excerpt}
  Let $C$ be a $\del$-configuration and $D$ be an $\ac$-configuration.
  We inductively define a binary relation $\simuek{C}{D}{\eta}{\kappa}$, which is parameterized by $\eta$ and $\kappa$, as follows:
  
\medskip
\noindent{}(Case: return term)
\[
  \simuek{\config{\return{V}}{\theta}}{\config{\mte{\return{V}}{\eta}}{\tau}}{\eta}{\kappa}
\]
\noindent{}(Case: dollar term)
\[
  \inferrule
    {
      \simuek{\config{M_1}{\theta}}{\config{N_1}{\tau}}{\eta}{\kappa} \\
      \getnum{\kappa(m), \tau} = i \\
      \tau(m) = \nil \\
      \eta^{-1}(m, \kappa(m), \_) \subseteq \theta^{-1}(\nil)
    }
    {
      \simuek
      {\config{\dollar{M_1}{x}{M_2}}{\theta}}
      {\config
        {\begin{array}{l}
        \letin{res}{\labeledc{m}{(\letin{x}{N_1}{\app{{\return{\thunk{\abs{\underscore}{\mte{M_2}{\eta}}}}}}})}}
        {\\\app{\force{\var{res}}}{\var{(m, \kappa(m), i)}}}
        \end{array}}{\tau}}{\eta}{\kappa}}
\]
\end{definition}
$C \rsimuek{\eta}{\kappa} D$ relates each $\del$-computation to one or more corresponding $\ac$-computations.
In most of the cases, it does so structurally; that is, $\simuek{\config{M}{\theta}}{\config{\mte{M}{\eta}}{\tau}}{\eta}{\kappa}$.
However, the case for a dollar term is an exception: it involves an active coroutine label $m$, which lacks a corresponding continuation label.
This imposes a few constraints on stores and labels.
For example, $\getnum{\kappa(m), \tau} = i$ specifies that the index of the continuation label to be generated must equal the value of its counter; and $\eta^{-1}(m, \kappa(m), \_) \subseteq \theta^{-1}(\nil)$ requires that all continuation labels corresponding to $m$ be invalid.

Finally, we lift the relation $\rsimuek{\eta}{\kappa}$ to the simulation relation $\sim$.
The relation $\rsimuek{\eta}{\kappa}$ checks if the structures of two configurations correspond to each other.
However, it is not sufficient; we also need to ensure that the global properties for the counter mechanism are satisfied.

For example, a label for an invalidated continuation in $\del$ must correspond to a triple in $\ac$ whose index is strictly smaller than its counter's value, but this property is not ensured by the relation $\rsimuek{\eta}{\kappa}$.
To capture such properties, we define $\simu{C}{D}$ as a binary relation that holds if and only if there exist partial functions $\eta$ and $\kappa$ such that $\simuek{C}{D}{\eta}{\kappa}$ holds and certain \emph{invariant conditions} are satisfied.
A key invariant condition is ``$\theta(l) = \nil \implies \getnum{\mathrm{pr}_2(\eta(l)), \tau} > \mathrm{pr}_3(\eta(l))$'', which formalizes exactly the rule mentioned above.
For the complete definition, see Appendix~\ref{app:DELoneToAC:contextual_relation}.
We remark that $\simu{C}{D}$ has the following property.
\begin{lemma}\label{lem:deltoac_initial}
  \begin{enumerate}
  \item   For any $\del$ computation $M$,
  $\simuek{\config{M}{\emptyset}}{\config{\mt{M}}{\emptyset}}{\emptyset}{\emptyset}$.
  Therefore,
  $\simu{\config{M}{\emptyset}}{\config{\mt{M}}{\emptyset}}$.

  \item If $\simu{\config{\return{V}}{\theta}}{\config{N}{\tau}}$, then there exist a partial function $\eta : \syntacticset{L}_{\del} \rightharpoonup \syntacticset{L}_{\ac} \times \syntacticset{L}_{\ac} \times \mathbb{N}$ such that $N \equiv \return{\mte{V}{\eta}}$.
  \end{enumerate}
  \end{lemma}

We can prove that the relation $\sim$ is a simulation relation from $\del$ to $\ac$.
\begin{theorem}\label{thm:deltoac_simulation}
  Let $C$ be a $\del$-configuration and $D$ be an $\ac$-configuration and assume that $\simu{C}{D}$ and $\redD{C}{C'}$.  Then, there exists an $\ac$-configuration $D'$ such that $\redACplus{D}{D'}$ and
  $\simu{C'}{D'}$ hold.
\end{theorem}
\begin{proof}
  We can prove this theorm by induction on case analysis on $\redD{C}{C'}$.
  See Appendix~\ref{app:DELoneToAC:simulation} for the complete proof.
\end{proof}

\begin{theorem}\label{thm:deltoac}
  $\del$ is macro-expressible in $\ac$.
\end{theorem}
\begin{proof}
  We show that $A \mapsto \mt{A}$ is a valid macro-translation.
  Here, we only prove semantic preservation, i.e., $\eval{\del}{M}$ terminates if and only if $\eval{\ac}{\mt{M}}$ does, since the other conditions are straightforward to check.
  We first prove the ``only if'' direction.
  Suppose that there exist a value $V$ and a store $\theta$ such that $\redDplus{\config{M}{\emptyset}}{\config{\return{V}}{\theta}}$.
  By Lemma~\ref{lem:deltoac_initial}, we have $\simu{\config{M}{\emptyset}}{\config{\mt{M}}{\emptyset}}$.
  By repeatedly applying Theorem~\ref{thm:deltoac_simulation}, we obtain $\redACplus{\config{\mt{M}}{\emptyset}}{\config{N}{\tau}}$ and $\simu{\config{\return{V}}{\theta}}{\config{N}{\tau}}$ for some $N$ and $\tau$.
  By Lemma~\ref{lem:deltoac_initial}, we have $N \equiv \return{\mte{V}{\eta}}$, which completes the ``only if'' direction.
  The proof of the ``if'' direction is provided in Appendix~\ref{sec:app_deltoac:strong}.
\end{proof}

The complexity of the simulation arises from the fundamental mismatch between $\del$ and $\ac$.
The naive translation may suffice for establishing weak macro-expressibility of $\del$ in $\ac$; however, its simulation proof must still address the mismatch.
In particular, it must track the correspondence between continuation labels and coroutine labels, via $\eta$ and certain invariant conditions.
This reflects the intrinsic difficulty of translating one-shot delimited continuations to asymmetric coroutines.

\section{One-Shot Effect Handlers as Asymmetric Coroutines}\label{chap:EFFoneToDELone}

Since Plotkin and Pretnar's proposal~\cite{plotkin2009handlers}, effect handlers have been actively studied in recent
years.  This section extends our results to effect handlers, namely, we establish macro-expressibility of one-shot
effect handlers in terms of asymmetric coroutines.  Since macro-translations are composable, it suffices to show that
one-shot effect handlers are macro-expressible by one-shot delimited-control operators.

\subsection{The Calculus for One-Shot Effect Handlers}
\label{subsec:eff}

\begin{figure}[tb]
  \setlength{\parindent}{0cm}%
  \setlength{\arraycolsep}{4pt}%
  \renewcommand{\arraystretch}{0.9}%
\begin{tabular}{rcll}
  $V,W$ & ::= & $\ldots$ &  \textsf{value}\\
& \textbar\  & $l_H\quad(l \in \syntacticset{L}_{\mathbf{E}})$ & continuation label \\
  $M,N$ & ::= & $\ldots$ &  \textsf{computation}\\
& \textbar\  & $\opcall{op}{V}\quad(\op{op} \in \syntacticset{O})$ & operation call \\
& \textbar\  & $\handle{H}{M}$ & handle \\
& \textbar\  & $\throw{V}{W}$ & throw \\
  $H$ & ::= & $\handler{x}{M_{\mathrm{ret}}}{(\app{\op{op}_i}{\app{p_i}{k_i}} \mapsto M_i)_i}$ & \textsf{handler}\footnotemark \\
\end{tabular}\\[1em]
  \begin{minipage}[t]{0.63\columnwidth}
    \begin{tabular}{rrcl}
      \textsf{pure frame} & $\mathcal{P}$ & ::= & \ldots \\
      \textsf{computational frame} & $\mathcal{F}$ & ::= & \ldots \textbar\  $\handle{H}{\hole}$ \\

    \end{tabular}
  \end{minipage}
  \begin{minipage}[t]{0.3\columnwidth}
    \begin{tabular}{rrcl}
      \textsf{pure context} & $\mathcal{H}$ & ::= & \ldots \\
      \textsf{evaluation context} & $\mathcal{C}$ & ::= & \ldots \\
    \end{tabular}
  \end{minipage}
\caption{Syntax of $\eff$}
\label{fig:eff_syntax}
\end{figure}\footnotetext{Note that $\var{x}$ is bound in $M_{\mathrm{ret}}$, and $\var{p_i}$ and $\var{k_i}$ are bound in their respective $M_i$.}

Figure~\ref{fig:eff_syntax} presents the syntax of $\eff$, the calculus
for one-shot effect handlers.
$\syntacticset{L}_{\mathbf{E}}$ denotes a countable set of labels.
A continuation label $l_H$ is a pair of its name $l \in \syntacticset{L}_{\mathbf{E}}$
and a handler $H$, and serves as a runtime representation of a one-shot continuation delimited by the handler $H$.
The term $\opcall{op}{V}$ is an operation invocation with an argument $V$,
where the operation $\op{op}$ is an element of a countable set $\syntacticset{O}$.
The term $\handle{H}{M}$ installs a handler $H$ and evaluates $M$ under it.
The term $\throw{V}{W}$ invokes the continuation represented by $V$, which
should be a continuation label.

We assume that every handler handles all operations that appear in programs.
This is not an essential restriction, 
as any handler can be rewritten to satisfy this condition.\footnote{For each unhandled operation $\op{op}$, we add a clause $\opcall{op}{\app{p}{k}} \mapsto \seq{r}{\opcall{op}{p}}{\throw{k}{\var{r}}}$ to the handler.}
$\eff$-Phrases are the values and computations, and $\eff$-Programs are the computations without continuation labels.

Similarly to $\del$, captured continuations can be invoked at most once
during program execution, hence handling the operation $\op{E}$ under the following handler raises a runtime error:
\[
   H \equiv \left\{
    \begin{array}{@{}l@{}}
      \return{x} \mapsto \return{x}, \\
      \opcall{E}{\app{\var{p}}{\var{k}}} \mapsto \letin{a}{\throw{k}{1}}{\letin{b}{\throw{k}{2}}{\return{\vpair{a}{b}}}}
    \end{array}
  \right\}
\]
$\eff$ uses a store to detect one-shotness,
which is a partial function:
$\theta : \syntacticset{L}_{\mathbf{E}}\times\mathsf{handler}\;\rightharpoonup\;\mathsf{computation}\sqcup\{\nil\}$.

A configuration $C$ is a pair of an $\eff$-computation $M$ and a store $\theta$, or the error state $\bot$.
The beta reduction rules $\rightarrow_{\mathbf{E}}^{\beta}$ on configurations are defined in Figure~\ref{fig:eff_beta}.
\begin{figure}[tb]
  \centering
  \setlength{\extrarowheight}{10pt}
  \begin{tabular}{cl}
    $(\mam)$ & $\inferrule{\redbetaM{M}{M'}}{\redbetaEp{M}{\theta}{M'}{\theta}}$ \\
    $(\mathrm{ret})$ &
    $\inferrule
       {H \equiv \handler{x}{M_{\mathrm{ret}}}{\ldots}}
       {\redbetaEp{\handle{H}{\paren{\return{V}}}}{\theta}{M_{\mathrm{ret}}[V/x]}{\theta}}$ \\
    $(\mathrm{op})$ &
    $\inferrule
    {
      H \equiv \handler{x}{M_{\mathrm{ret}}}{(\app{\op{op}_i}{\app{p_i}{k_i}} \mapsto M_i)_i} \\
      \text{$l_H$ is a fresh label}
    }
    { 
      \config{\handle{H}{\paren{\plug{H}{\app{\op{op}_j}{V}}}}}{\theta} \\\\
      \rightarrow^{\mathbf{E}}_{\beta} \config{M_j[V/p_j, l_H/k_j]}{\theta[l_H \coloneq \abs{x}{\handle{H}{\plug{H}{\return{x}}}}]}
      }$ \\
    $(\mathrm{throw})$ &
    $\inferrule
    {
      \theta(l_H) = \abs{x}{\handle{H}{\plug{H}{\return{x}}}}
    }
    {
      \redbetaEp{\throw{l_H}{V}}{\theta}{\handle{H}{\plug{H}{\return{V}}}}{\theta[l_H \coloneq \nil]}
    }$ \\
    $(\mathrm{fail})$ &
    $\inferrule
    {
      \theta(l_H) = \nil
    }
    {
      \config{\throw{l_H}{V}}{\theta} \rightarrow^{\beta}_{\mathbf{E}} \bot
    }$
  \end{tabular}
  \caption{Beta Reduction Rules of $\eff$}
  \label{fig:eff_beta}
\end{figure}
The reduction of $\eff$ is defined as follows:
\[
  \inferrule{\redbetaEp{M}{\theta}{M'}{\theta'}}{\redEp{\plug{C}{M}}{\theta}{\plug{C}{M'}}{\theta'}} \hspace {4em} \inferrule{\redbetaE{\config{M}{\theta}}{\bot}}{\redE{\config{\plug{C}{M}}{\theta}}{\bot}}
\]
Evaluation of an $\eff$-program is defined by: $\eval{\eff}{M} \defines V$ if there exists a store $\theta$ such that $\redEclos{\config{M}{\emptyset}}{\config{\return{V}}{\theta}}$.
$\eval{\eff}{M}$ is a well-defined partial function, since $\arrE$ is deterministic.

\subsection{Macro-Translation from \texorpdfstring{$\eff$}{EFFone} to \texorpdfstring{$\del$}{DELone}}
\label{sec:efftodel_macro_translation}

We present a translation from $\eff$ to $\del$ in Figure~\ref{fig:efftodel}, which is based on the translation by Forster et al.~\cite{forster2019expressive}.
\begin{figure}[tb]
  \centering
  \[
  \begin{array}{rll}
    \mt{\opcall{op}{V}} & \defines & \shift{k}{\abs{h}{\app{\force{h}}{\inj{\op{op}}{\vpair{\mt{V}}{\thunk{\abs{y}{\app{\throw{k}{y}}{h}}}}}}}} \\
    \mt{\throw{V_1}{V_2}} & \defines & \app{\force{\mt{V_1}}}{\mt{V_2}} \\
    \mt{\handle{H}{M}} & \defines & \app{\dollart{\mt{M}}{H^{\mathrm{ret}}}}{\thunk{H^{\mathrm{ops}}}} \\
                        & \mathrm{where} & (\handler{x}{M_{\mathrm{ret}}}{\ldots})^{\mathrm{ret}} = x. \abs{\underscore}{\mt{M_{\mathrm{ret}}}} \\
                        & & (\handler{x}{M_{\mathrm{ret}}}{(\app{\op{op}_i}{\app{p_i}{k_i}} \mapsto M_i)_i})^{\mathrm{ops}} \\
                        & &\quad= \abs{c}{\scase{c}{\op{op}_i}{\vpair{p_i}{k_i}}{\mt{M_i}}}
  \end{array}
  \]
  \caption{Translation from $\eff$ to $\del$}
  \label{fig:efftodel}
\end{figure}

\begin{theorem}\label{thm:efftodel}
  $\eff$ is macro-expressible in $\del$.
\end{theorem}

\begin{proof}
  First, this translation maps $\eff$-Programs to $\del$-Programs since it introduces no continuation labels.
  The translation does not alter $\mam$ constructors, so it homomorphically acts on them.
  Also, Figure~\ref{fig:efftodel} clearly shows that the program constructs peculiar to $\eff$ are expressed by syntactic abstractions of $\del$.
  We prove the preservation of semantics in Appendix~\ref{app:EFFoneToDELone}.
\end{proof}
From Theorem~\ref{thm:efftodel} together with
Theorem~\ref{thm:deltoac} and the compositionality of
macro-translations, we get the following theorem.
\begin{theorem}\label{thm:efftoac}
$\eff$ is macro-expressible in $\ac$.
\end{theorem}

\subsection{Macro-Translation
  from \texorpdfstring{$\del$}{DELone}
  to \texorpdfstring{$\eff$}{EFFone}
}
We can also show the macro-expressibility of $\del$ to $\eff$, using Forster et al.'s translation for the multi-shot $\mathit{shift}_0/\mathit{dollar}$.
See Appendix~\ref{app:DELoneToEFFone} for the proof.

\begin{theorem}\label{thm:deltoeff}
  Assume that $\op{shift0}$ is contained in the set of operations $\syntacticset{O}$.
  Then, the following translation is a macro-translation from $\del$ to $\eff$.%
  \[
  \begin{array}{rll}
    \mt{\shift{k}{M}} & \defines & \opcall{shift0}{\thunk{\abs{k}{\mt{M}}}} \\
    \mt{\throw{V}{W}} & \defines & \throw{\mt{V}}{\mt{W}} \\
    \mt{\dollar{M}{x}{N}} & \defines & \handle{\handler{x}{\mt{N}}{\opcall{shift0}{\app{p}{k}} \mapsto \app{\force{p}}{k}}}{\mt{M}}
  \end{array}
  \]  
\end{theorem}

\section{One-Shot Effect Handlers Cannot Macro-Express Asymmetric Coroutines}\label{chap:nonexistent}

When asymmetric coroutines macro-express one-shot effect handlers, it
is natural to ask whether the converse direction also holds.
In this section, we prove that the converse direction does not hold.

To show this, we introduce $\reflang$, a calculus with ML-style reference cells, defined in Figure~\ref{fig:ref_syntax}.
The definition of $\reflang$-frames is omitted since it is the same as that of $\mam$-frames.
\begin{figure}[tb]
  \begin{minipage}[t]{0.40\columnwidth}
    \begin{tabular}[t]{@{}l@{\hspace{0.5em}}l@{\hspace{1em}}l@{}}
      $V,W$ & $::= \ldots$ & \textsf{value} \\
            & \textbar\ $l\in\syntacticset{L}_{\mathbf{R}}$ & reference cell
    \end{tabular}
  \end{minipage}%
    \begin{minipage}[t]{0.38\columnwidth}
    \begin{tabular}[t]{@{}l@{\hspace{0.5em}}l@{\hspace{1em}}l@{}}
      $M,N$ & $::= \ldots$ & \textsf{computation} \\
            & \textbar\ $\create{V}$ & create
    \end{tabular}%
  \end{minipage}%
    \begin{minipage}[t]{0.2\columnwidth}
    \begin{tabular}[t]{@{}l@{\hspace{1em}}l@{}}
      \textbar\ $\set{V}{W}$ & set \\
      \textbar\ $\get{V}$ & get
    \end{tabular}%
  \end{minipage}%
\caption{Syntax of $\reflang$}
\label{fig:ref_syntax}
\end{figure}
$\reflang$-Phrases are the values and the computations of $\reflang$, and $\reflang$-Programs are the values and the computations which contain no reference cells.
As in the previous calculi, we use stores and configurations: a store is a partial function that maps reference cells to \textit{values}, and a configuration is a pair of a $\reflang$-computation and a store.
Figure~\ref{fig:ref_beta} presents the semantics of $\reflang$.

\begin{figure}[tb]
  \begin{minipage}[c]{0.65\columnwidth}
    \begin{tabular}{cl}
    $(\mam)$ & $\inferrule{\redbetaM{M}{M'}}{\redbetaRp{M}{\theta}{M'}{\theta}}$ \\
    
    $(\mathrm{create})$ &
    $\inferrule
    {
      \text{$l$ is a fresh reference cell}
    }
    {
      \redbetaRp{\create{V}}{\theta}{\return{l}}{\theta[l := V]}
    }$ \\
    
    $(\mathrm{set})$ &
    $\inferrule
    {
      l \in \dom{\theta}
    }
    {
      \redbetaRp{\set{l}{V}}{\theta}{\return{\unit}}{\theta[l := V]}
    }$ \\

    $(\mathrm{get})$ &
    $\inferrule
    {
      \theta(l) = V
    }
    {
      \redbetaRp{\get{l}}{\theta}{\return{V}}{\theta}}$
  \end{tabular}
  \end{minipage}%
  \begin{minipage}[c]{0.3\columnwidth}
    \fbox{$\inferrule
      {
        \redbetaRp{M}{\theta}{M'}{\theta'}
      }
      {
        \redRp{\plug{C}{M}}{\theta}{\plug{C}{M'}}{\theta'}
      }$}
  \end{minipage}%
  \caption{Semantics of $\reflang$}
  \label{fig:ref_beta}
\end{figure}

First, we prove that $\eff$ cannot macro-express $\reflang$:

\begin{theorem}\label{thm:reftoeff_nonexist}
  There is no valid macro-translation from $\reflang$ to $\eff$.
\end{theorem}

\begin{proof}[Proof sketch]
  Suppose there is a macro-translation from $\reflang$ to $\eff$, and consider the following term $M$ in $\reflang$.
  \[
      M \defines
      \left(\begin{array}{l}
        \letin{r}{\create{\inj{A}{\unit}}}
        {\\\letin{i}{\get{r}}
        {\\\letin{\underscore}{\set{r}{\inj{B}{\unit}}}
        {\\\letin{k}{\get{r}}
        {\\\return{\vpair{i}{k}}}}}}
      \end{array}\right) \mapsto
    \mt{M} = 
    \left(\begin{array}{l}
      \letin{r}{\app{\trconst{create}}{\paren{\inj{A}{\unit}}}}
      {\\\letin{i}{\app{\trconst{get}}{r}}
      {\\\letin{\underscore}{\app{\app{\trconst{set}}{r}}{\paren{\inj{B}{\unit}}}}
      {\\\letin{k}{\app{\trconst{get}}{r}}
      {\\\return{\vpair{i}{k}}}}}}
    \end{array}\right)
  \]%
  In $\reflang$, $M$ evaluates to $\vpair{\inj{A}{\unit}}{\inj{B}{\unit}}$.
  In $\eff$, however, $\mt{M}$ ought to evaluate to $\vpair{\inj{A}{\unit}}{\inj{A}{\unit}}$.
  This is because a macro-translation must be local and compositional, which prevents $\mt{M}$ from being enclosed by any handler in $\eff$, forcing the two evaluations of $\app{\mt{\mathbf{get}}}{\var{r}}$ in $\mt{M}$ to yield the same value.
  See Appendix~\ref{app:reftoeff_nonexist} for the complete proof.
\end{proof}

On the other hand, $\ac$ can macro-express $\reflang$:
\begin{theorem}\label{thm:reftoac}
  $\reflang$ is macro-expressible in $\ac$.
\end{theorem}
\begin{proof}
  The following translation is a valid macro-translation from $\reflang$ to $\ac$.\footnote{$\refcell{\cdot}$ is defined in Figure~\ref{fig:deltoac}.}
  \[
    \begin{array}{rll}
      \mt{\create{V}} & \defines & \create{\refcell{\mt{V}}} \\
      \mt{\set{V}{W}} & \defines & \resume{\mt{V}}{\paren{\inj{Set}{\mt{W}}}} \\
      \mt{\get{V}} & \defines & \resume{\mt{V}}{\paren{\inj{Get}{\unit}}} %
    \end{array}
  \]
  We prove the details in Appendix~\ref{app:reftoac}.
\end{proof}

\begin{corollary}\label{cor:nonexistent}
  $\eff$ cannot macro-express $\ac$.
\end{corollary}
\begin{proof}
  Suppose that a macro-translation from $\ac$ to $\eff$ exists.
  Then we have a macro-translation from $\reflang$ to $\eff$ by Theorem~\ref{thm:reftoac},
  which contradicts Theorem~\ref{thm:reftoeff_nonexist}.
\end{proof}

Corollary~\ref{cor:nonexistent} is counter-intuitive, since effect
handlers are considered a universal tool for expressing various
computational effects, such as coroutines.
This gap arose due to the strictness of macro-translations, which
adheres to Felleisen's original notion.
We conjecture that, under a suitably relaxed definition of
macro-translations, $\eff$ can macro-express $\reflang$ and $\ac$.
One possible relaxation is to allow macro-translations to insert a
fixed context at the root of each translated term in order to represent global effects.
Such relaxed macro-translations would establish that $\reflang$ and $\ac$ are
macro-expressible in $\eff$, while preserving the compositionality of
the translations.
We leave the formalization and verification of this idea for future work.

\section{Conclusion}\label{chap:conclusion}

This study investigated the relative macro-expressiveness of one-shot control operators, resulting in several key findings:
(1) One-shot delimited-control operators and one-shot effect handlers can be macro-expressed by asymmetric coroutines, and
(2) The converse direction does not hold.
Previously, the former was considered trivial; however, we spotted a gap in the literature and fixed the problem by devising the counter mechanism.
As our results demonstrate, establishing such a connection requires
a precise mathematical analysis.

To our knowledge, this work is the first study to conduct a systematic comparison of the expressiveness of one-shot control operators.
James and Sabry claimed that the one-shot yield operator can be seen as a one-shot variant of delimited-control operators~\cite{yield2011}, but they did not provide a formal justification.
Forster et al.\ studied multi-shot control operators -- effect handlers, monadic reflection, and delimited-control operators -- from operational and denotational semantics, and analyzed their macro-expressibility with and without types~\cite{forster2019expressive}.
Following their approach but adopting a purely operational viewpoint, we have revealed certain aspects of one-shot control operators that cannot be directly derived from prior multi-shot results.
Meanwhile, earlier work by de Moura and Ierusalimschy examined the expressiveness of symmetric coroutines, asymmetric coroutines, and one-shot subcontinuations in the presence of mutable states~\cite{moura2009revisiting}.
In contrast, we study the calculi without mutable states, which clarifies the expressive power of control operators.

There are a number of directions for future work.
Proposing a type system for coroutines and proving that our translation preserves types would be an important extension to our results.
Although Anton and Thiemann proposed a type system for coroutines~\cite{anton2010typing}, it remains unclear whether their system aligns with our translations.
Introducing affine types to statically guarantee that each continuation may be invoked at most once, in the spirit of linearly used continuations~\cite{berdine2002linear}, is another interesting future topic.

\subsection*{Acknowledgments}
We would like to thank
anonymous reviewers for their constructive comments.
This work was supported by
JSPS KAKENHI Grant Number JP23K24819.

\bibliographystyle{splncs04}
\bibliography{refs}

\appendix

\section{Definition of Conservative Extensions}\label{app:conservative_extensions}
In this appendix, we describe the formal definition of \textit{conservative extensions} by Felleisen \cite{felleisen1991expressive}.
\begin{definition}
  A language $\mathscr{L}$ is a conservative extension of $\mathscr{L'}$ if and only if:
  \begin{enumerate}
  \item the constructors of $\mathscr{L'}$ is a proper subset of the constructors of $\mathscr{L}$
    with the difference being $\{F_1, \ldots, F_n\}$;
  \item The set of $\mathscr{L'}$-Phrases is a full subset of $\mathscr{L}$-Phrases
    which do not contain any constructors in $\{F_1, \ldots, F_n\}$;
  \item The set of $\mathscr{L'}$-Programs is a full subset of $\mathscr{L}$-Programs
    which do not contain any constructors in $\{F_1, \ldots, F_n\}$; and
  \item if an $\mathscr{L'}$-Program $M$ evaluates to $V$ in $\mathscr{L'}$, then also in $\mathscr{L}$, $M$ evaluates to $V$.
  \end{enumerate}
\end{definition}

Informally, a conservative extension of $\mathscr{L}$ is a language extension which includes only terms generated from a finite number of additional constructors.

\section{Supplementary Proofs for Chapter~\ref{chap:DELoneToAC}}\label{app:DELoneToAC}

\subsection{Preservation of Well-Formedness}\label{app:DELoneToAC:well-formedness}

Syntactically, a coroutine $l$ can execute a computation that contains itself as an \emph{active label}:
\[
  \labeledc{l}{\paren{\labeledc{l}{M}}}.
\]
However, we demonstrate that $\ac$-programs cannot reach such states by formalizing the notion of \textit{well-formedness} and proving that it is preserved under reduction.

\begin{definition}
  For any $\ac$-expression $E$, we define $\activelabels{E}$ as the set of all active labels appearing in $E$.
\end{definition}

\begin{definition}\label{def:deltoac:wellformed}
  An $\ac$-configuration $\config{M}{\theta}$ is \textit{well-formed} if
  \begin{enumerate}
  \item $\wellformed{M}$ holds,
  \item $\activelabels{V} = \emptyset$ for all $V \in \im{\theta}$, and
  \item for all $l \in \activelabels{M}$, $\theta(l) = \nil$,
  \end{enumerate}
  where $\wellformed{M}$ is a predicate on $\ac$-computations inductively defined as follows:
  \begin{enumerate}
  \item
    $
    \inferrule
    {\activelabels{V} \cup \activelabels{M} = \emptyset}
    {\wellformed{\pcase{V}{x_1}{x_2}{M}}}$
  \item
    $\inferrule
    {\activelabels{V} \cup (\cup_i \activelabels{M_i}) = \emptyset}
    {\wellformed{\scase{V}{L_i}{x_i}{M_i}}}$

  \item
    $\inferrule
    {\activelabels{V} = \emptyset}
    {\wellformed{\force{V}}}$

  \item
    $\inferrule
    {\activelabels{V} = \emptyset}
    {\wellformed{\return{V}}}$

  \item
    $\inferrule
    {\wellformed{M} \\ \activelabels{N} = \emptyset}
    {\wellformed{\seq{x}{M}{N}}}$

  \item
    $\inferrule
    {\activelabels{M} = \emptyset}
    {\wellformed{\abs{x}{M}}}$

  \item
    $\inferrule
    {\wellformed{M} \\ \activelabels{N} = \emptyset}
    {\wellformed{\app{M}{V}}}$

  \item
    $\inferrule
    {\activelabels{M} \cup \activelabels{N} = \emptyset}
    {\wellformed{\cpair{M}{N}}}$

  \item
    $\inferrule
    {\wellformed{M}}
    {\wellformed{\prj{i}{M}}}$

  \item
    $\inferrule
    {\wellformed{M} \\ l \notin \activelabels{M}}
    {\wellformed{\labeledc{l}{M}}}$

  \item
    $\inferrule
    {\activelabels{V} = \emptyset}
    {\wellformed{\create{V}}}$

  \item
    $\inferrule
    {\activelabels{V} \cup \activelabels{W} = \emptyset}
    {\wellformed{\resume{V}{W}}}$

  \item
    $\inferrule
    {\activelabels{V} = \emptyset}
    {\wellformed{\yield{V}}}$
  \end{enumerate}
\end{definition}

\begin{proposition}\label{prop:deltoac:wellformed_preservation}
  Suppose that $\config{\plug{C}{M}}{\theta}$ is well-formed and $\redbetaAC{\config{M}{\theta}}{\config{M'}{\theta'}}$.
  Then, $\config{\plug{C}{M'}}{\theta'}$ is also well-formed.
\end{proposition}
\begin{proof}
  By case analysis on $\redbetaAC{\config{M}{\theta}}{\config{M'}{\theta'}}$.
  
  \noindent\textbf{Case} $(\mam)$:
  \begin{caseindent}
    \noindent\textbf{SubCase} $(F)$:
    \begin{caseindent}
      Suppose that $\redbetaAC{\config{M}{\theta}}{\config{M'}{\theta'}}$ is
      \[
        \redbetaAC{\config{\seq{x}{\return{V}}{M}}{\theta}}{\config{M[V/x]}{\theta'}}.
      \]
    Since $\seq{x}{\return{V}}{M}$ is well-formed, $V$ and $M$ do not contain any active labels, implying $\activelabels{M[V/x]} = \emptyset$.
    By straightforward induction on $\context{C}$, we conclude that $\config{\plug{C}{M[V/x]}}{\theta}$ is well-formed.
    \end{caseindent}

    \noindent~The remaining subcases are similar.
  \end{caseindent}

  \noindent\textbf{Case} $(\mathrm{create})$:
  \begin{caseindent}
    Suppose that $\redbetaAC{\config{M}{\theta}}{\config{M'}{\theta'}}$ is
    \[
      \inferrule
      {
        \text{$l$ is a fresh label}
      }
      {
        \redbetaACp{\create{V}}{\theta}{\return{l}}{\theta[l := V]}
      }.
    \]
    Since $l$ is fresh, $l \notin \activelabels{\context{C}}$.
    The well-formedness of $\create{V}$ ensures $\activelabels{V} = \emptyset$.
    By induction on $\context{C}$, we obtain that $\config{\plug{C}{\return{l}}}{{\theta[l := V]}}$ is well-formed.
  \end{caseindent}

  \noindent\textbf{Case} $(\mathrm{resume})$:
  \begin{caseindent}
    Suppose that $\redbetaAC{\config{M}{\theta}}{\config{M'}{\theta'}}$ is
    \[
      \inferrule
      {
        l \in \dom{\theta} \\ \theta(l) \neq \nil
      }
      {
        \redbetaACp{\resume{l}{V}}{\theta}{\labeledc{l}{(\app{\force{\theta(l)}}{V})}}{\theta[l := \nil]}
      }.
    \]
    If $l$ is active in a well-formed configuration $\config{\plug{C}{\resume{l}{V}}}{\theta}$, it follows that $\theta(l) = \nil$, but this is a contradiction.
    Therefore, $l$ is distinct in $\config{\plug{C}{\labeledc{l}{(\app{\force{\theta(l)}}{V})}}}{\theta[l := \nil]}$ and mapped to $\nil$ by $\theta[l := \nil]$.
    From the well-formedness of $\config{\plug{C}{\resume{l}{M}}}{\theta}$, we have that $\activelabels{\theta(l)} = \emptyset$ and $\activelabels{V} = \emptyset$.
    Then, it is easy to check that this configuration is well-formed by induction on $\context{C}$.
  \end{caseindent}

  \noindent\textbf{Case} $(\mathrm{fail})$:
  \begin{caseindent}
    $(\mathrm{fail})$ does not occur, as $\config{\plug{C}{M}}{\theta}$ reduces to $\config{\plug{C}{M'}}{\theta'}$, not $\bot$.
  \end{caseindent}

  \noindent\textbf{Case} $(\mathrm{ret})$:
  \begin{caseindent}
    Immediate from the definition of well-formedness.
  \end{caseindent}

  \noindent\textbf{Case} $(\mathrm{yield})$:
  \begin{caseindent}
    Suppose $\redbetaAC{\config{M}{\theta}}{\config{M'}{\theta'}}$ is
    \[
      \redbetaACp{\labeledc{l}{\context{H}[\yield{V}]}}{\theta}{\return{V}}{\theta[l := \thunk{\abs{y}{\context{H}[\return{y}]}}]}
    \]
    $l$ is no longer active in $\config{\plug{C}{\return{V}}}{\theta[l := \thunk{\abs{y}{\context{H}[\return{y}]}}]}$, while other active labels remain distinct and are mapped to $\nil$ by $\theta[l := \thunk{\abs{y}{\context{H}[\return{y}]}}]$.
    Moreover, from the well-formedness of $\plug{H}{\yield{V}}$, we obtain that $\context{H}$ and $V$ does not contain any active labels.
    Therefore, by induction on $\context{C}$, we see that $\config{\plug{C}{\return{V}}}{\theta[l := \thunk{\abs{y}{\context{H}[\return{y}]}}]}$ is well-formed.
  \end{caseindent}
  This completes the proof.
\end{proof}

\subsection{Definition of Simulation Relation}\label{app:DELoneToAC:contextual_relation}

\begin{definition}
  For any $l \in \syntacticset{L}_{\ac}$, $\ac$-store $\tau$, and natural number $i$, we define $\getnum{l, \tau} = i$ when
  \[
    \tau(l) = \refcell{i}.
  \]
\end{definition}

\begin{definition}\label{def:rsimuek}
  Let $\eta$ be a partial function from $\syntacticset{L}_{\del}$ to
  $\syntacticset{L}_{\ac} \times \syntacticset{L}_{\ac} \times
  \mathbb{N}$, $\kappa$ be a partial function from
  $\syntacticset{L}_{\ac}$ to $\syntacticset{L}_{\ac}$, and
  $\mathrm{Conf}_{\del}$ and $\mathrm{Conf}_{\ac}$ be the sets of
  configurations of $\del$ and $\ac$, respectively. We inductively
  define a binary relation $\rsimuek{\eta}{\kappa}$ over
  $\mathrm{Conf}_{\del} \times \mathrm{Conf}_{\ac}$, which is
  parameterized by $\eta$ and $\kappa$, as follows:

  \begin{enumerate}
  \item
    $
    \simuek
    {\config{\pcase{V}{x_1}{x_2}{M}}{\theta}}
    {\config{\mte{\pcase{V}{x_1}{x_2}{M}}{\eta}}{\tau}}
    {\eta}{\kappa}
    $,
  \item
    $
    \simuek
    {\config{\scase{V}{L_i}{x_i}{M_i}}{\theta}}
    {\config{\mte{\scase{V}{L_i}{x_i}{M_i}}{\eta}}{\tau}}
    {\eta}{\kappa}
    $,
  \item $\simuek{\config{\force{V}}{\theta}}{\config{\mte{\force{V}}{\eta}}{\tau}}{\eta}{\kappa}$,
  \item $\simuek{\config{\return{V}}{\theta}}{\config{\mte{\return{V}}{\eta}}{\tau}}{\eta}{\kappa}$,
  \item
    $
    \simuek{\config{\abs{x}{M}}{\theta}}{\config{\mte{\abs{x}{M}}{\eta}}{\tau}}{\eta}{\kappa}
    $,
  \item
    $
    \simuek{\config{\cpair{M_1}{M_2}}{\theta}}{\config{\mte{\cpair{M_1}{M_2}}{\eta}}{\tau}}{\eta}{\kappa}
    $,
  \item
    $
    \simuek{\config{\shift{k}{M}}{\theta}}{\config{\mte{\shift{k}{M}}{\eta}}{\tau}}{\eta}{\kappa}
    $,
  \item
    $
    {
      \simuek
      {\config{\dollar{M_1}{x}{M_2}}{\theta}}
      {\config{\mte{\dollar{M_1}{x}{M_2}}{\eta}}{\tau}}
      {\eta}{\kappa}
    }$,
  \item
    $
    \simuek
    {\config{\throw{V}{W}}{\theta}}
    {
      \config
      {
        \mte{\throw{V}{W}}{\eta}
      }
      {\tau}
    }
    {\eta}{\kappa}
    $,
  \item
    $l\inferrule
    {
      \simuek{\config{M_1}{\theta}}{\config{N_1}{\tau}}{\eta}{\kappa}
    }
    {
      \simuek{\config{\seq{x}{M_1}{M_2}}{\theta}}{\config{\seq{x}{N_1}{\mte{M_2}{\eta}}}{\tau}}{\eta}{\kappa}
    }$,
  \item
    $\inferrule
    {
      \simuek{\config{M}{\theta}}{\config{N}{\tau}}{\eta}{\kappa}
    }
    {
      \simuek{\config{\app{M}{V}}{\theta}}{\config{\app{N}{\mte{V}{\eta}}}{\tau}}{\eta}{\kappa}
    }$,
  \item
    $\inferrule
    {
      \simuek{\config{M}{\theta}}{\config{N}{\tau}}{\eta}{\kappa}
    }
    {
      \simuek{\config{\prj{i}{M}}{\theta}}{\config{\prj{i}{N}}{\tau}}{\eta}{\kappa}
    }$, and
  \item
    $\inferrule
    {
      \simuek{\config{M_1}{\theta}}{\config{N_1}{\tau}}{\eta}{\kappa} \\
      \getnum{\kappa(m), \tau} = i \\
      \tau(m) = \nil \\
      \eta^{-1}(m, \kappa(m), \_) \subseteq \theta^{-1}(\nil)
    }
    {
      \simuek
      {\config{\dollar{M_1}{x}{M_2}}{\theta}}
      {\config
        {\begin{array}{l}
        \letin{res}{\labeledc{m}{(\letin{x}{N_1}{\app{{\return{\thunk{\abs{\underscore}{\mte{M_2}{\eta}}}}}}})}}
        {\\\app{\force{\var{res}}}{\var{(m, \kappa(m), i)}}}
        \end{array}}{\tau}}{\eta}{\kappa}}$.
\end{enumerate}
\end{definition}

We refer to the pairs of evaluation contexts and stores as \textit{contextual configurations} and write $\mathrm{CConf}_{\del}$ and $\mathrm{CConf}_{\ac}$ for the sets consisting of them.
\begin{definition}\label{def:rsimuekc}
  Let $\eta$ be a partial function from $\syntacticset{L}_{\del}$ to
  $\syntacticset{L}_{\ac} \times \syntacticset{L}_{\ac} \times
  \mathbb{N}$, $\kappa$ be a partial function from
  $\syntacticset{L}_{\ac}$ to $\syntacticset{L}_{\ac}$, and
  $\mathrm{CConf}_{\del}$ and $\mathrm{CConf}_{\ac}$ be the sets of
  contextual configurations of $\del$ and $\ac$, respectively.  We
  inductively define a binary relation $\rsimuekc{\eta}{\kappa}$ over
  $\mathrm{CConf}_{\del} \times \mathrm{CConf}_{\ac}$, which is
  parameterized by $\eta$ and $\kappa$, as follows:

  \begin{enumerate}
  \item
    $
    \simuekc
    {\config{\hole{}}{\theta}}
    {\config{\hole{}}{\tau}}
    {\eta}{\kappa}
    $,
  \item
    $\inferrule
    {
      \simuekc{\config{\context{C}}{\theta}}{\config{\context{D}}{\tau}}{\eta}{\kappa}
    }
    {
      \simuekc{\config{\seq{x}{\plug{C}{}}{M_2}}{\theta}}{\config{\seq{x}{\plug{D}{}}{\mte{M_2}{\eta}}}{\tau}}{\eta}{\kappa}
    }$,
  \item
    $\inferrule
    {
      \simuekc{\config{\context{C}}{\theta}}{\config{\context{D}}{\tau}}{\eta}{\kappa}
    }
    {
      \simuekc{\config{\app{\paren{\plug{C}{}}}{V}}{\theta}}{\config{\app{\paren{\plug{D}{}}}{\mte{V}{\eta}}}{\tau}}{\eta}{\kappa}
    }$,
  \item
    $\inferrule
    {
      \simuekc{\config{\context{C}}{\theta}}{\config{\context{D}}{\tau}}{\eta}{\kappa}
    }
    {
      \simuekc{\config{\prj{i}{\plug{C}{}}}{\theta}}{\config{\prj{i}{\plug{D}{}}}{\tau}}{\eta}{\kappa}
    }$,
  \item
    $\inferrule
    {
      \simuekc{\config{\context{C}}{\theta}}{\config{\context{D}}{\tau}}{\eta}{\kappa} \\
      \getnum{\kappa(m), \tau} = i \\
      \tau(m) = \nil \\
      \eta^{-1}(m, \kappa(m), \_) \subseteq \theta^{-1}(\nil) \\
    }
    {
      \simuekc
      {\config{\dollar{\plug{C}{}}{x}{M_2}}{\theta}}
      {\config
        {\begin{array}{l}
          \letin{res}{\labeledc{m}{(\letin{x}{\plug{D}{}}{\app{{\return{\thunk{\abs{\underscore}{\mte{M_2}{\eta}}}}}}})}}
          {\\\app{\force{\var{res}}}{\var{(m, \kappa(m), i)}}}
        \end{array}}{\tau}}{\eta}{\kappa}}$.
\end{enumerate}
\end{definition}

\begin{definition}\label{def:deltoac:simulation}
  For any configurations of $\del$ and $\ac$, say $C$ and $D$, $\simu{C}{D}$ means that $C \equiv D \equiv \bot$, or
  there exist $\del$-computation $M$, $\ac$-computation $N$, $\del$-store $\theta$, $\ac$-store $\tau$, and partial functions $\eta : \syntacticset{L}_{\del} \rightharpoonup \syntacticset{L}_{\ac} \times \syntacticset{L}_{\ac} \times \mathbb{N}$ and $\kappa : \syntacticset{L}_{\ac} \rightharpoonup \syntacticset{L}_{\ac}$ such that the following conditions are satisfied:
  \begin{enumerate}
  \item $C = \config{M}{\theta}$.
  \item $D = \config{N}{\tau}$.
  \item $\simuek{C}{D}{\eta}{\kappa}$.
  \item $D$ is well-formed.%
  \item\label{def:deltoac:simulation:IC_start} $\kappa$ is injective.
  \item $\dom{\kappa} \cap \im{\kappa} = \emptyset$.
  \item $\dom{\kappa} \sqcup \im{\kappa} = \dom{\tau}$.
  \item $\dom{\eta} = \dom{\theta}$.
  \item For any $l \in \dom{\eta}$, if $\eta(l) = (z, zc, i)$, then $z \in \dom{\kappa}$ and
    $\kappa(z) = zc$.
  \item For any $l \in \dom{\eta}$, if $\theta(l) = \nil$, then
    $\getnum{\mathrm{pr}_2(\eta(l)), \tau} > \mathrm{pr}_3(\eta(l))$.
  \item\label{def:deltoac:simulation:IC_end} For any $l \in \dom{\eta}$, if $\theta(l) = \abs{y}{\dollar{\plug{H}{\return{y}}}{x}{M'}}$,
    then
    \begin{enumerate}
    \item $l$ is not contained in $M'$,
    \item $\getnum{\mathrm{pr}_2(\eta(l)), \tau} = \mathrm{pr}_3(\eta(l))$,
    \item for any $l' \in \dom{\theta}$, if $l' \neq l$ and
      $\mathrm{pr}_1(\eta(l')) = \mathrm{pr}_1(\eta(l))$, then
      $\theta(l') = \nil$, and
    \item $\tau(\mathrm{pr}_1(\eta(l))) = \thunk{\abs{y}{\seq{x}{\plug{\mte{H}{\mathrm{\eta}}}{\return{x}}}{\return{\thunk{\abs{\underscore}{\mte{M'}{\eta}}}}}}}$.
    \end{enumerate}
  \end{enumerate}
\end{definition}
We define that $\theta$, $\tau$, $\eta$, and $\tau$ satisfy the \textit{invariant conditions} (IC) if the conditions \ref{def:deltoac:simulation:IC_start} through \ref{def:deltoac:simulation:IC_end} in Definition~\ref{def:deltoac:simulation} are fulfilled.
Since well-formedness is preserved under reduction (Proposition~\ref{prop:deltoac:wellformed_preservation}), we leave it implicit unless necessary.

\subsection{Proof of Theorem~\ref{thm:deltoac_simulation}}\label{app:DELoneToAC:simulation}

In this section, we give a proof of Theorem.~\ref{thm:deltoac_simulation}.
First, we extend the translation $\mte{\cdot{}}{\eta}$ on frames and contexts.
For example, we define $\mte{\seq{x}{\plug{C}{}}{M}}{\eta}$ as $\seq{x}{\paren{\plug{\mte{C}{\eta}}{}}}{\mte{M}{\eta}}$.

We present several auxiliary lemmas for proving Theorem~\ref{thm:deltoac_simulation}.

\begin{lemma}\label{lem:deltoac:subst}
  For any $\eta : \syntacticset{L}_{\del} \rightharpoonup \syntacticset{L}_{\ac} \times \syntacticset{L}_{\ac} \times \mathbb{N}$, $\del$-computation $M$ with free variables $x_1, \ldots, x_n$, and $\del$-values $V_1, \ldots, V_n$,
  \[
    \mte{M}{\eta}\left[\mte{V_1}{\eta}/x_1, \ldots, \mte{V_n}{\eta}/x_n\right] \equiv \mte{M\left[V_1/x_1, \ldots, V_n/x_n\right]}{\eta}
  \]
  holds~\footnote{We use $\equiv$ to denote syntactic equality}.
\end{lemma}
\begin{proof}
  We prove a more general form of this lemma.

  Suppose $E$ is a $\del$-computation or a $\del$-value with free variables.
  Then, we aim to show that
  \[
    \mte{E}{\eta}\left[\mte{V_1}{\eta}/x_1, \ldots, \mte{V_n}{\eta}/x_n\right] \equiv \mte{E\left[V_1/x_1, \ldots, V_n/x_n\right]}{\eta}.
  \]
  by induction on $E$.
  Suppose first that $E$ is a $\del$-value.
  Then $E$ has six possible forms.
  If $E$ is a variable $x$ and $x \equiv x_i$ for some $i$, then
  \[
    \mte{x_i}{\eta}\left[\mte{V_1}{\eta}/x_1, \ldots, \mte{V_n}{\eta}/x_n\right] \equiv \mte{V_i}{\eta} \equiv \mte{x_i\left[V_1/x_1, \ldots, V_n/x_n\right]}{\eta}.
  \]
  If $x$ is not any of $x_1, \ldots, x_n$,
  \[
    \mte{x}{\eta}\left[\mte{V_1}{\eta}/x_1, \ldots, \mte{V_n}{\eta}/x_n\right] \equiv x \equiv \mte{x\left[V_1/x_1, \ldots, V_n/x_n\right]}{\eta}.
  \]
  If $E \equiv \unit$ or $E \equiv l$, the result is immediate.
  Suppose that $E \equiv \vpair{V}{W}$.
  By the induction hypothesis, we see that
  \[
    \mte{V}{\eta}\left[\mte{V_1}{\eta}/x_1, \ldots, \mte{V_n}{\eta}/x_n\right] \equiv \mte{V\left[V_1/x_1, \ldots, V_n/x_n\right]}{\eta}
  \]
  and
  \[
    \mte{W}{\eta}\left[\mte{V_1}{\eta}/x_1, \ldots, \mte{V_n}{\eta}/x_n\right] \equiv \mte{W\left[V_1/x_1, \ldots, V_n/x_n\right]}{\eta}.
  \]
  Using these fact, it follows that
  \begin{align*}
    \mte{\vpair{V}{W}}{\eta}\left[\mte{V_1}{\eta}/x_1, \ldots, \mte{V_n}{\eta}/x_n\right]
    &\equiv \vpair{\mte{V}{\eta}}{\mte{W}{\eta}}\left[\mte{V_1}{\eta}/x_1, \ldots, \mte{V_n}{\eta}/x_n\right] \\
    &\equiv \vpair{\mte{V}{\eta}\left[\mte{V_1}{\eta}/x_1, \ldots, \mte{V_n}{\eta}/x_n\right]}{\mte{W}{\eta}\left[\mte{V_1}{\eta}/x_1, \ldots, \mte{V_n}{\eta}/x_n\right]} \\
    &\equiv \vpair{\mte{V\left[V_1/x_1, \ldots, V_n/x_n\right]}{\eta}}{\mte{W\left[V_1/x_1, \ldots, V_n/x_n\right]}{\eta}} \\
    &\equiv \mte{\vpair{V\left[V_1/x_1, \ldots, V_n/x_n\right]}{W\left[V_1/x_1, \ldots, V_n/x_n\right]}}{\eta} \\
    &\equiv \mte{\vpair{V}{W}\left[V_1/x_1, \ldots, V_n/x_n\right]}{\eta}.
  \end{align*}
  The other cases are similar.

  When $E$ is a $\del$-value, we can prove it similarly.
  If $E$ is a term unique to $\del$, the calculation is more tangled but follows essentially the same steps as for the other constructors.
\end{proof}

\begin{lemma}\label{lem:deltoac:transd-correspondence}
  For any $\del$-computation $M$, $\del$-store $\theta$, $\ac$-store $\tau$, and partial functions $\eta : \syntacticset{L}_{\del} \rightharpoonup \syntacticset{L}_{\ac} \times \syntacticset{L}_{\ac} \times \mathbb{N}$ and $\kappa : \syntacticset{L}_{\ac} \rightharpoonup \syntacticset{L}_{\ac}$,
  \[
    \simuek{\config{M}{\theta}}{\config{\mte{M}{\eta}}{\tau}}{\eta}{\kappa}
  \]
  holds.
\end{lemma}
\begin{proof}
  We prove this lemma by induction on $M$, but give only one case since the other cases follow the similar steps.
  Suppose that $M \equiv \paren{\app{M'}{V}}$ for some $\del$-computation $M'$ and $\del$-value $V$.
  Then, using the induction hypothesis, we obtain $\simuek{\config{M'}{\theta}}{\config{\mte{M'}{\eta}}{\tau}}{\eta}{\kappa}$.
  Thus, it immediately follows that $\simuek{\config{\app{M'}{V}}{\theta}}{\config{\app{\mte{M'}{\eta}}{\mte{V}{\eta}}}{\tau}}{\eta}{\kappa}$.
\end{proof}

  \begin{lemma}\label{lem:deltoac:purec_simuek}
    Let $\context{H}$ be a pure $\del$-context and $\context{E}$ an $\ac$-context.
    Then, for any $\theta$, $\tau$, $\eta$, and $\kappa$, $\simuekc{\config{\context{H}}{\theta}}{\config{\context{E}}{\tau}}{\eta}{\kappa}$ if and only if $\context{E} \equiv \mte{\context{H}}{\eta}$.
  \end{lemma}
  \begin{proof}
    Both necessity and sufficiency can be proved by straightforward induction on $\simuekc{\config{\context{H}}{\theta}}{\config{\context{G}}{\tau}}{\eta}{\kappa}$ and $\context{H}$, respectively.
  \end{proof}

  \begin{lemma}\label{lem:deltoac:purec_insertion}
    For any pure context $\context{H}$, if $\simuek{\config{M}{\theta}}{\config{N}{\tau}}{\eta}{\kappa}$, then $\simuek{\config{\plug{H}{M}}{\theta}}{\config{\plug{\mte{H}{\eta}}{N}}{\tau}}{\eta}{\kappa}$.
  \end{lemma}
  \begin{proof}
    By straightforward induction on $\context{H}$.
  \end{proof}

  \begin{lemma}\label{lem:deltoac:purec_decomp}
    Let $\context{H}$ be a pure context of $\del$.
    Then, if $\simuek{\config{\plug{H}{M}}{\theta}}{\config{N}{\tau}}{\eta}{\kappa}$, there exist a $\ac$-computation $N'$ such that
    \begin{gather*}
      N \equiv \plug{\mte{H}{\eta}}{N'}, \\
      \simuek{\config{\plug{H}{M}}{\theta}}{\config{\plug{\mte{H}{\eta}}{N'}}{\tau}}{\eta}{\kappa}, \\
      \simuek{\config{M}{\theta}}{\config{N'}{\tau}}{\eta}{\kappa}.
    \end{gather*}
  \end{lemma}
  \begin{proof}
    By straightforward induction on $\context{H}$.
  \end{proof}

  \begin{lemma}\label{lem:deltoac:empty_context}
    Suppose that $\simuek{\config{M}{\theta}}{\config{N}{\tau}}{\eta}{\kappa}$ and $\simuek{\config{M}{\theta}}{\config{\plug{D}{N}}{\tau}}{\eta}{\kappa}$.
    Then, $\context{D} \equiv []$.
  \end{lemma}
  \begin{proof}
    By induction on the derivation of $\simuek{\config{M}{\theta}}{\config{N}{\tau}}{\eta}{\kappa}$.
    First, suppose that $N \equiv \mte{M}{\eta}$.
    Then, we see that $\plug{D}{N} \equiv \mte{M}{\eta}$.
    Therefore, we obtain $\context{D} \equiv []$.

    For the inductive step, we only show interesting cases for brevity.
    We assume that the derivation of $\simuek{\config{M}{\theta}}{\config{N}{\tau}}{\eta}{\kappa}$ is
    \[
      \inferrule
      {
        \simuek{\config{M_1}{\theta}}{\config{N_1}{\tau}}{\eta}{\kappa}
      }
      {
        \simuek{\config{M \equiv \paren{\seq{x}{M_1}{M_2}}}{\theta}}{\config{N \equiv \paren{\seq{x}{N_1}{\mte{M_2}{\eta}}}}{\tau}}{\eta}{\kappa}
      }.
    \]
    Suppose that $\context{D} \neq []$.
    Then, there is a context $\context{D'}$ such that $\context{D} \equiv \paren{\seq{x}{\plug{D'}{}}{\mte{M_2}{\eta}}}$ and $\simuek{\config{M_1}{\theta}}{\config{\plug{D'}{N} \paren{\equiv \plug{D'}{\seq{x}{N_1}{\mte{M_2}{\eta}}}}}{\tau}}{\eta}{\kappa}$.
    Applying the inductive hypothesis to this, we obtain
    \[
      \plug{D'}{\seq{x}{\hole}{\mte{M_2}{\eta}}} \equiv [],
    \]
    which is a contradiction.
    
    Next, we assume that the derivation of $\simuek{\config{M}{\theta}}{\config{N}{\tau}}{\eta}{\kappa}$ is
    \[
      \inferrule
      {
        \simuek{\config{M_1}{\theta}}{\config{N_1}{\tau}}{\eta}{\kappa} \\
        \getnum{\kappa(m), \tau} = i \\
        \tau(m) = \nil \\
        \eta^{-1}(m, \kappa(m), \_) \subseteq \theta^{-1}(\nil)
      }
      {
        \simuek
        {\config{M \equiv \dollar{M_1}{x}{M_2}}{\theta}}
        {\config
          {N \equiv \paren{\begin{array}{l}
            \letin{res}{\labeledc{m}{(\letin{x}{N_1}{\app{{\return{\thunk{\abs{\underscore}{\mte{M_2}{\eta}}}}}}})}}
            {\\\app{\force{\var{res}}}{\var{(m, \kappa(m), i)}}}
          \end{array}}}{\tau}}{\eta}{\kappa}}.
    \]
    Suppose that $\context{D} \neq []$.
    Then, there is a context $\context{D'}$ such that
    \[
      \context{D} \equiv \paren{\begin{array}{l}
            \letin{res}{\labeledc{m}{(\letin{x}{\plug{D'}{}}{\app{{\return{\thunk{\abs{\underscore}{\mte{M_2}{\eta}}}}}}})}}
            {\\\app{\force{\var{res}}}{\var{(m, \kappa(m), i)}}}
          \end{array}}
    \]
    and $\simuek{\config{M_1}{\theta}}{\config{\plug{D'}{N}}{\tau}}{\eta}{\kappa}$.
    This is because, if the derivation of $\simuek{\config{M}{\theta}}{\config{\plug{D}{N}}{\tau}}{\eta}{\kappa}$ is
    \[
      \simuek{\config{\dollar{M_1}{x}{M_2}}{\theta}}{\config{\plug{D}{N} \equiv \mte{\dollar{M_1}{x}{M_2}}{\eta}}{\tau}}{\eta}{\kappa},
    \]
    $\mte{\dollar{M_1}{x}{M_2}}{\eta}$ contains a labeled computation $\labeledc{m}{()}$, which is not in the image of the macro-translation.
    Therefore, from the inductive hypothesis, we obtain
    \[
      \plug{D'}{\begin{array}{l}
        \letin{res}{\labeledc{m}{(\letin{x}{[]}{\app{{\return{\thunk{\abs{\underscore}{\mte{M_2}{\eta}}}}}}})}}
        {\\\app{\force{\var{res}}}{\var{(m, \kappa(m), i)}}}
      \end{array}} \equiv [],
    \]
    which is a contradiction.
    The other cases are similar.
  \end{proof}

  \begin{lemma}\label{lem:deltoac:factor_context}
    Suppose that $\simuek{\config{\plug{C}{M}}{\theta}}{\config{\plug{D}{N}}{\tau}}{\eta}{\kappa}$ and $\simuek{\config{M}{\theta}}{\config{N}{\tau}}{\eta}{\kappa}$.
    Then, $\simuekc{\config{\context{C}}{\theta}}{\config{\context{D}}{\tau}}{\eta}{\kappa}$.
  \end{lemma}
  \begin{proof}
    By induction on $\context{C}$.
    The base case follows directly from Lemma~\ref{lem:deltoac:empty_context}.
    Suppose that $C \equiv \paren{\seq{x}{\plug{C'}{}}{M'}}$ for some $\context{C'}$ and $M'$.
    Then, the derivation of $\simuek{\config{\plug{C}{M}}{\theta}}{\config{\plug{D}{N}}{\tau}}{\eta}{\kappa}$ is
    \[
      \inferrule
      {
        \simuek{\config{\plug{C'}{M}}{\theta}}{\config{\plug{D'}{N}}{\tau}}{\eta}{\kappa}
      }
      {
        \simuek{\config{\seq{x}{\plug{C'}{M}}{M'}}{\theta}}{\config{\seq{x}{\plug{D'}{N}}{\mte{M'}{\eta}}}{\tau}}{\eta}{\kappa}
      }.
    \]
    Hence, we have $\simuek{\config{\plug{C'}{M}}{\theta}}{\config{\plug{D'}{N}}{\tau}}{\eta}{\kappa}$.
    From the induction hypothesis, we obtain
    \[
      \simuek{\config{\context{C}'}{\theta}}{\config{\context{D}'}{\tau}}{\eta}{\kappa},
    \]
    which implies that $\simuek{\config{\context{C}}{\theta}}{\config{\context{D}}{\tau}}{\eta}{\kappa}$.

    We present another non-trivial case: $C \equiv \dollar{\plug{C'}{}}{x}{M'}$.
    In this case, we argue that the derivation of $\simuek{\config{\plug{C}{M}}{\theta}}{\config{\plug{D}{N}}{\tau}}{\eta}{\kappa}$ is
    \[
      \inferrule
      {
        \simuek{\config{\plug{C'}{M}}{\theta}}{\config{\plug{D'}{N}}{\tau}}{\eta}{\kappa} \\
        \getnum{\kappa(m), \tau} = i \\
        \tau(m) = \nil \\
        \eta^{-1}(m, \kappa(m), \_) \subseteq \theta^{-1}(\nil)
      }
      {
        \simuek
        {\config{\dollar{\plug{C'}{M}}{x}{M'}}{\theta}}
        {\config
          {\begin{array}{l}
            \letin{res}{\labeledc{m}{(\letin{x}{\plug{D'}{N}}{\app{{\return{\thunk{\abs{\underscore}{\mte{M'}{\eta}}}}}}})}}
            {\\\app{\force{\var{res}}}{\var{(m, \kappa(m), i)}}}
          \end{array}}{\tau}}{\eta}{\kappa}}.
    \]
    Otherwise, $\plug{D}{N} \equiv \mte{\dollar{\plug{C'}{}}{x}{M'}}{\eta}$ implies that
    \[
      \plug{D}{N} \equiv \left(\begin{array}{l}
          \letin{z}{\create{\thunk{\abs{\underscore}{\letin{x}{\mte{\plug{C'}{M}}{\eta}}{\return{\thunk{\abs{\underscore}{\mte{M'}{\eta}}}}}}}}}
          {\\\letin{zc}{\app{\force{\var{ref}}}{\inj{Zero}{\unit}}}
          {\\\letin{res}{\resume{z}{\unit}}
           {\\\app{\force{\var{res}}}{(\var{z}, \var{zc}, \inj{Zero}{\unit})}}}}
        \end{array}\right)
    \]
    and, by the definition of contexts,
    \begin{gather*}
      \context{D} \equiv \left(\begin{array}{l}
          \letin{z}{\hole{}}
          {\\\letin{zc}{\app{\force{\var{ref}}}{\inj{Zero}{\unit}}}
          {\\\letin{res}{\resume{z}{\unit}}
           {\\\app{\force{\var{res}}}{(\var{z}, \var{zc}, \inj{Zero}{\unit})}}}}
      \end{array}\right), \\
      N \equiv \create{\thunk{\abs{\underscore}{\letin{x}{\mte{\plug{C'}{M}}{\eta}}{\return{\thunk{\abs{\underscore}{\mte{M'}{\eta}}}}}}}};
    \end{gather*}
    however, $\create{V}$ does not appear in the right hand side of $\rsimuek{\eta}{\kappa}$, which contradicts $\simuek{\config{M}{\theta}}{\config{N}{\tau}}{\eta}{\kappa}$.
    Thus, we obtain $\simuek{\config{\plug{C'}{M}}{\theta}}{\config{\plug{D'}{N}}{\tau}}{\eta}{\kappa}$ and by the induction hypothesis, $\simuek{\config{\context{C}'}{\theta}}{\config{\context{D}'}{\tau}}{\eta}{\kappa}$, yielding that $\simuek{\config{\context{C}}{\theta}}{\config{\context{D}}{\tau}}{\eta}{\kappa}$.
    This completes the proof.
  \end{proof}

  \begin{lemma}\label{lem:deltoac:factor_computation}
    Suppose that $\simuek{\config{\plug{C}{M}}{\theta}}{\config{\plug{D}{N}}{\tau}}{\eta}{\kappa}$
    and $\simuekc{\config{\context{C}}{\theta}}{\config{\context{D}}{\tau}}{\eta}{\kappa}$.
    Then, $\simuek{\config{M}{\theta}}{\config{N}{\tau}}{\eta}{\kappa}$.
  \end{lemma}
  \begin{proof}
    By induction on the derivation of $\simuekc{\config{\context{C}}{\theta}}{\config{\context{D}}{\tau}}{\eta}{\kappa}$.
  \end{proof}

  \begin{proposition}\label{prop:deltoac:decomp}
    Suppose that
    $\simuek{\config{\plug{C}{M}}{\theta}}{\config{N}{\tau}}{\eta}{\kappa}$
    and that $\theta$, $\tau$, $\eta$, and $\kappa$ satisfy the IC.
    Then, there exist an evaluation context $\context{D}$, a
    computation $N'$, an extension of $\tau'$, and a partial function
    $\kappa'$ such that
    \begin{gather*}
      \config{N}{\tau} \arrAC^{*} \config{\plug{D}{N'}}{\tau'}, \\
      \config{\context{C}}{\theta} \rsimuekc{\eta}{\kappa'} \config{\context{D}}{\tau'}, \\
      \config{M}{\theta} \rsimuek{\eta}{\kappa'} \config{N'}{\tau'}, \\
      \config{\plug{C}{M}}{\theta} \rsimuek{\eta}{\kappa'} \config{\plug{D}{N'}}{\tau'},
    \end{gather*}
    where $\tau'$ is a finite extension of $\tau$, $\kappa'$ is a finite extension of $\kappa$, and $\theta$, $\tau'$, $\eta$, and $\kappa'$ satisfy the IC.
    Moreover, if $M$ is a redex, so is $N'$.
  \end{proposition}
  \begin{proof}
    We prove this by induction on the number of computational frames that constitute $\context{C}$.
    The base case follows from Lemma~\ref{lem:deltoac:purec_decomp}.

    For the inductive step, we first show that we may suppose without loss of generality that $\context{C} \equiv \dollar{\plug{C'}{}}{x}{M'}$, where the outermost frame is computational.
    If, instead, $\context{C} \equiv \plug{P}{\plug{C'}{}}$ for some pure frame $\context{P}$, then we can decompose $\context{C}$ and $\context{C}'$ as
    \[
      \context{C}  \equiv \plug{P}{\plug{H}{\dollar{\plug{C''}{}}{x}{M'}}}
      \quad\text{and}\quad
      \context{C'} \equiv \plug{H}{\dollar{\plug{C''}{}}{x}{M'}},
    \]
    where $\context{C}''$ is a possibly computational context and $\context{H}$ is a pure context.
    First, by applying the argument from the base case, we obtain
    \begin{gather*}
      N \equiv \plug{\paren{\mte{\plug{P}{\context{H}}}{\eta}}}{N'} \\
      \simuek{\config{\dollar{\plug{C''}{M}}{x}{M'}}{\theta}}{\config{N'}{\tau}}{\eta}{\kappa}, \\
      \simuek{\config{\plug{P}{\plug{H}{\dollar{\plug{C''}{M}}{x}{M'}}}}{\theta}}{\config{\plug{\mte{\plug{P}{\context{H}}}{\eta}}{N'}}{\tau}}{\eta}{\kappa}.
    \end{gather*}
    Then, by treating $\dollar{\plug{C''}{}}{x}{M'}$ as $C$, we reduce this case to one in which the outermost frame of $\context{C}$ is computational, and we obtain
    \[
      \config{N'}{\tau} \arrAC^{*} \config{\plug{D'}{N''}}{\tau'} \quad\text{and}\quad
      \config{\dollar{\plug{C''}{M}}{x}{M'}}{\theta} \rsimuek{\eta}{\kappa'} \config{\plug{D'}{N''}}{\tau'},
    \]
    where $\theta$, $\tau'$, $\eta$, and $\kappa'$ satisfy the IC.
    Note that if $M$ is a redex, then so is $N''$.
    Applying Lemma~\ref{lem:deltoac:purec_simuek} implies $\config{\plug{P}{\plug{H}{}}}{\theta} \rsimuekc{\eta}{\kappa''} \config{\mte{\plug{P}{\plug{H}{}}}{\eta}}{\tau'}$.
    Finally, from Lemma~\ref{lem:deltoac:purec_insertion}, we conclude
    \[
      \simuek{\config{\plug{P}{\plug{H}{\dollar{\plug{C''}{M}}{x}{M'}}}}{\theta}}{\config{\plug{\mte{\plug{P}{\plug{H}{}}}{\eta}}{\plug{D'}{N''}}}{\tau''}}{\eta}{\kappa''},
    \]
    yielding the desired result.

    Now we return to the scenario where $C \equiv \dollar{\plug{C'}{}}{x}{M'}$.
    According to the definition of $\rsimuek{\eta}{\kappa}$, the derivation tree of $\simuek{\config{\dollar{\plug{C'}{M}}{x}{M'}}{\theta}}{\config{N}{\tau}}{\eta}{\kappa}$ has two possible forms:

    \begin{caseindent}
      \noindent \textbf{Case 1}:

      \[
        \inferrule
        {
          \simuek{\config{\plug{C'}{M}}{\theta}}{\config{N'}{\tau}}{\eta}{\kappa} \\
          \getnum{\kappa(m), \tau} = i \\
          \tau(m) = \nil \\
          \eta^{-1}(m, \kappa(m), \_) \subseteq \theta^{-1}(\nil) \\
        }
        {
          \simuek
          {\config{\dollar{\plug{C'}{M}}{x}{M'}}{\theta}}
          {\config{N \equiv \paren{\begin{array}{l}
            \letin{res}{\labeledc{m}{(\letin{x}{N'}{\app{{\return{\thunk{\abs{\underscore}{\mte{M'}{\eta}}}}}}})}}
        {\\\app{\force{\var{res}}}{\var{(m, \kappa(m), i)}}}
          \end{array}}}{\tau}}{\eta}{\kappa}
        }
      \]
      Then, by the induction hypothesis, we obtain
      \begin{gather}
        \label{lem:deltoac:decomp:N_prime_reduction}
        \config{N'}{\tau} \arrAC^{*} \config{\plug{D}{N''}}{\tau'}, \\
        \label{lem:deltoac:decomp:simuekc_C_prime_D}
        \config{\context{C'}}{\theta} \rsimuekc{\eta}{\kappa'} \config{\context{D}}{\tau'}, \\
        \config{M}{\theta} \rsimuek{\eta}{\kappa'} \config{N''}{\tau'}, \\
        \label{lem:deltoac:decomp:simuek}
        \config{\plug{C'}{M}}{\theta} \rsimuek{\eta}{\kappa'} \config{\plug{D}{N''}}{\tau'},
      \end{gather}
      where $\tau'$ is a finite extension of $\tau$, $\kappa'$ is a finite extension of $\kappa$, and $\theta$, $\tau'$, $\eta$, and $\kappa'$ satisfy the IC.
      Note that $N''$ is a redex if $M$ is also a redex.
      From (\ref{lem:deltoac:decomp:N_prime_reduction}), then, we have
      \[
        \config{N}{\tau} \arrAC^{*} \config{\paren{\begin{array}{l}
            \letin{res}{\labeledc{m}{(\letin{x}{\plug{D}{N''}}{\app{{\return{\thunk{\abs{\underscore}{\mte{M'}{\eta}}}}}}})}}
        {\\\app{\force{\var{res}}}{\var{(m, \kappa(m), i)}}}
          \end{array}}}{\tau'}.
      \]
      Since $\tau'$ and $\kappa'$ are finite extensions, it follows that $\getnum{\kappa'(m), \tau'} = i$, $\tau'(m) = \nil $, and $\eta^{-1}(m, \kappa'(m), \_) \subseteq \theta^{-1}(\nil)$.
      Thus, by (\ref{lem:deltoac:decomp:simuekc_C_prime_D}), we obtain
      \[
        \simuekc
          {\config{\dollar{\plug{C'}{}}{x}{M'}}{\theta}}
          {\config{\begin{array}{l}
            \letin{res}{\labeledc{m}{(\letin{x}{\plug{D}{}}{\app{{\return{\thunk{\abs{\underscore}{\mte{M'}{\eta}}}}}}})}}
            {\\\app{\force{\var{res}}}{\var{(m, \kappa'(m), i)}}}
          \end{array}}{\tau'}}{\eta}{\kappa'},
      \]
      and by (\ref{lem:deltoac:decomp:simuek}),
      \[
        \simuek
          {\config{\dollar{\plug{C'}{M}}{x}{M'}}{\theta}}
          {\config{\begin{array}{l}
            \letin{res}{\labeledc{m}{(\letin{x}{\plug{D}{N''}}{\app{{\return{\thunk{\abs{\underscore}{\mte{M'}{\eta}}}}}}})}}
            {\\\app{\force{\var{res}}}{\var{(m, \kappa'(m), i)}}}
          \end{array}}{\tau'}}{\eta}{\kappa'},
      \]
      which completes the proof for the current case. \\

      \noindent \textbf{Case 2}:
      \[
        \simuek
        {\config{\dollar{M_1}{x}{M_2}}{\theta}}
        {\config{N \equiv \mte{\dollar{M_1}{x}{M_2}}{\eta}}{\tau}}
        {\eta}{\kappa}
      \]
      $\config{N}{\tau}$ evaluates to
      \[
        \config{N \equiv \paren{\begin{array}{l}
          \letin{res}{\labeledc{m}{(\letin{x}{N'}{\app{{\return{\thunk{\abs{\underscore}{\mte{M'}{\eta}}}}}}})}}
          {\\\app{\force{\var{res}}}{\var{(m, mc, i)}}}
        \end{array}}}{\tau'},
      \]
      where $\tau'$ is $\tau[\var{m}\defines\nil, \var{mc}\defines\refcell{\inj{Zero}{\unit}}]$.
      Let $\kappa'$ be $\kappa[m \defines mc]$.
      Then, if $\theta$, $\tau'$, $\eta$, and $\kappa'$ satisfy the IC, this case is reducible to the previous case.
      Therefore, it remains to verify the IC.
      \begin{itemize}
      \item[(4)] $\kappa'$ is injective since $mc$ is taken as a fresh label.

      \item[(5)] $\dom{\kappa'} = \dom{\kappa} \sqcup \{\var{m}\}$ and $\im{\kappa'} = \im{\kappa} \sqcup \{\var{mc}\}$, so
        \begin{align*}
          \dom{\kappa'} \cap \im{\kappa'} &= (\dom{\kappa} \cap \im{\kappa}) \cup (\dom{\kappa} \cap \{\var{mc}\}) \cup
                                            (\{\var{m}\} \cap \im{\kappa}) \cup (\{\var{m}\} \cap \{\var{mc}\}) \\
                                          &= \emptyset,
        \end{align*}
        since $\var{m}$ and $\var{mc}$ are taken as fresh labels.

      \item[(6)] With regard to $\dom{\tau'}$,
        \begin{align*}
          \dom{\tau'} &= \dom{\tau} \sqcup \{\var{m}, \var{mc}\} \\
                      &= (\dom{\kappa} \sqcup \im{\kappa}) \sqcup \{\var{m}, \var{mc}\} \\
                      &= (\dom{\kappa} \sqcup \{\var{m}\}) \sqcup (\im{\kappa} \sqcup \{\var{mc}\}) \\
                      &= \dom{\kappa'} \sqcup \im{\kappa'}.
        \end{align*}
      \end{itemize}
      The other conditions remain true since $\theta$, $\tau$, $\eta$, $\kappa$ satisfy the IC and $\theta$ and $\eta$ are not extended.
    \end{caseindent}
    This completes the proof.
  \end{proof}

  \begin{proposition}\label{prop:deltoac:beta_sim}
    Suppose that $\simu{C}{D}$ and $\redbetaD{C}{C'}$, then there exists an $\ac$-configuration $D'$ such that $\redACplus{D}{D'}$ and $\simu{C'}{D'}$.
  \end{proposition}
  \begin{proof}
    Suppose that $C \equiv \config{M}{\theta}$.
    By the definition of $\sim$, there exist $N$, $\tau$, $\eta$, and $\kappa$ such that
    \begin{gather*}
      D \equiv \config{N}{\tau}, \\
      \simuek{C}{\config{N}{\tau}}{\eta}{\kappa},
    \end{gather*}
    where $D$ is well-formed and $\theta$, $\tau$, $\eta$, and $\kappa$ satisfy the IC.
    We analyze each case based on the definition of $\redbetaD{C}{C'}$.

    \noindent \textbf{Case} $(\mam)$:
    \begin{caseindent}
      In this case, we present only one case since the other cases are similar.
      Suppose that
      \[
        M \equiv \paren{\pcase{(V_1 ,V_2)}{x_1}{x_2}{M'}}.
      \]
      Then, $C \equiv \config{M}{\theta}$ evaluates to $C' \equiv \config{\paren{M'[V_1/x_1, V_2/x_2]}}{\theta}$.
      By the definition of $\rsimuek{\eta}{\kappa}$, we obtain
      \begin{align*}
        N &\equiv \paren{\mte{\pcase{(V_1 ,V_2)}{x_1}{x_2}{M'}}{\eta}} \\
           &\equiv \paren{\pcase{(\mte{V_1}{\eta}, \mte{V_2}{\eta})}{\mte{x_1}{\eta}}{\mte{x_2}{\eta}}{\mte{M'}{\eta}}}.
      \end{align*}
      $\config{N}{\tau}$ is reduced in one step to
      \[
        \config{\mte{M'}{\eta}[\mte{V_1}{\eta}/x_1, \mte{V_2}{\eta}/x_2]}{\tau} \equiv \config{\mte{M'[V_1/x_1, V_2/x_2]}{\eta}}{\tau},
      \]
      where the equation follows from Lemma~\ref{lem:deltoac:subst}.
      Then, Lemma~\ref{lem:deltoac:transd-correspondence} implies that
      \[
        \simuek{\config{M'[V_1/x_1, V_2/x_2]}{\theta}}{\config{\mte{M'[V_1/x_1, V_2/x_2]}{\eta}}{\tau}}{\eta}{\kappa},
      \]
      and Proposition~\ref{prop:deltoac:wellformed_preservation} implies that the right-hand-side configuration is well-formed.
      Since $\theta$, $\tau$, $\eta$, and $\kappa$ satisfy the IC, it follows that
      \[
        \simu{\config{M'[V_1/x_1, V_2/x_2]}{\theta}}{\config{\mte{M'[V_1/x_1, V_2/x_2]}{\eta}}{\tau}},
      \]
      which completes the proof for this case.\\
    \end{caseindent}

        \noindent \textbf{Case} $(\mathrm{throw})$:
    \begin{caseindent}
      Suppose that $M \equiv \paren{\throw{l}{V}}$ and $\theta(l) = \paren{\abs{y}{\dollar{\plug{H}{\return{y}}}{x}{M}}}$.
      Then, $\config{M}{\theta}$ evaluates to
      \[
        \config{\dollar{\plug{H}{\return{V}}}{x}{M}}{\theta'},
      \]
      where $\theta' = \theta[l \defines \nil]$.
      By the definition of $\rsimuek{\eta}{\kappa}$, we obtain
      \begin{align*}
        N &\equiv \mte{\throw{l}{V}}{\eta} \\
          &\equiv \left(\begin{array}{l}
          \mathbf{case}\;\mte{l}{\eta}\;\mathbf{of}\;\{\\
          \quad((\var{z}, \var{zc}), \var{i}) \mapsto\\
          \quad\quad\mathbf{let}\;\var{j} = \app{\force{\var{get}}}{\var{zc}}\;\mathbf{in} \\
          \quad\quad\mathbf{let}\;\var{b} = \app{\app{\force{\var{compare}}}{i}}{j} \\
          \quad\quad\mathbf{case}\;\var{b}\;\mathbf{of}\;\{\\
          \quad\quad\quad\paren{\inj{True}{\unit}} \mapsto\\
          \quad\quad\quad\quad\mathbf{let}\;\var{i'} = \app{\force{\var{incr}}}{i}\;\mathbf{in}\\
          \quad\quad\quad\quad\mathbf{let}\;\unit = \app{\app{\force{\var{set}}}{\var{zc}}}{i'}\;\mathbf{in}\\
          \quad\quad\quad\quad\mathbf{let}\;\var{res} = \app{\app{\mathbf{resume}}{\var{z}}}{\mte{V}{\eta}}\;\mathbf{in}\\
          \quad\quad\quad\quad\app{\force{\var{res}}}{((\var{z}, \var{zc}), \var{i'})} \\
          \quad\quad\quad\paren{\inj{False}{\unit}} \mapsto \force{\var{fail}}\\
          \quad\quad\} \\
          \}
        \end{array} \right).
      \end{align*}
      Let $m = \pri{l}$ and $i = \getnum{\prii{l}, \tau}$, then, since $\theta$, $\tau$ $\eta$, and $\kappa$ satisfy the IC, we obtain
      \begin{gather*}
        \prii{l} = \kappa(l), \\
        i = \priii{l}, \\
        \tau(m) = \thunk{\abs{y}{\seq{x}{\plug{\mte{H}{\eta}}{\return{y}}}{\return{\thunk{\abs{\underscore}{\mte{M}{\eta}}}}}}}.
      \end{gather*}
      Using these facts, the evaluation of $\config{N}{\tau}$ proceeds as follows:
      \begin{align*}
        \config{N}{\tau}
        &\arrAC^{*}
          \config
          {\paren{\begin{array}{l}
            \mathbf{let}\;\var{i'} = \app{\force{\var{incr}}}{i}\;\mathbf{in}\\
            \mathbf{let}\;\unit = \app{\app{\force{\var{set}}}{\var{\kappa(m)}}}{i'}\;\mathbf{in}\\
            \mathbf{let}\;\var{res} = \app{\app{\mathbf{resume}}{m}}{\mte{V}{\eta}}\;\mathbf{in} \\
            \app{\force{\var{res}}}{((\var{m}, \var{\kappa(m)}), \var{i'})}
          \end{array}}}
          {\tau} \\
        &\arrAC^{*}
          \config
          {\paren{
          \begin{array}{l}
            \mathbf{let}\;\var{res} = \app{\app{\mathbf{resume}}{m}}{\mte{V}{\eta}}\;\mathbf{in} \\
            \app{\force{\var{res}}}{((m, \kappa(m)), i+1)}
          \end{array}
          }}
          {\tau[\kappa(m) \defines \refcell{i+1}]} \\
        &\arrAC
          \config
          {\paren{
          \begin{array}{l}
            \mathbf{let}\;\var{res} = \labeledc{m}{\paren{\seq{x}{\plug{\mte{H}{\eta}}{\return{\mte{V}{\eta}}}}{\return{\thunk{\abs{\underscore}{\mte{M}{\eta}}}}}}}\;\mathbf{in} \\
            \app{\force{\var{res}}}{((m, \kappa(m)), i+1)}
          \end{array}
          }}
          {\tau'}
      \end{align*}
      where $\tau' \defines \tau[m \defines \nil, \kappa(m) \defines \refcell{i+1}]$.
      It is straightforward to check that $\plug{\mte{H}{\eta}}{\return{\mte{V}{\eta}}} = \mte{\plug{H}{\return{V}}}{\eta}$.
      Thus, by Lemma~\ref{lem:deltoac:transd-correspondence}, we obtain
      \[
        \simuek{\config{\plug{\mte{H}{\eta}}{\return{\mte{V}{\eta}}}}{\theta'}}{\config{\mte{\plug{H}{\return{V}}}{\eta}}{\tau'}}{\eta}{\kappa'}.
      \]
      Now, in order to derive
      \[
        \simuek
        {\config{\dollar{\plug{H}{\return{V}}}{x}{M}}{\theta'}}
        {\config
          {\paren{
          \begin{array}{l}
            \mathbf{let}\;\var{res} = \labeledc{m}{\paren{\seq{x}{\mte{\plug{H}{\return{V}}}{\eta}}{\return{\thunk{\abs{\underscore}{\mte{M}{\eta}}}}}}}\;\mathbf{in} \\
            \app{\force{\var{res}}}{((m, \kappa(m)), i+1)}
          \end{array}
          }}
        {\tau'}}
        {\eta}{\kappa},
      \]
      we shall check that the other premises hold.
      \begin{itemize}
      \item
        $\getnum{\kappa(m), \tau'} = i + 1$ since $\tau'(\kappa(m)) = \refcell{i+1}$.

      \item
        Obviously, $\tau'(m) = \nil$.

      \item
        Let $l'$ be a continuation label of $\del$ and suppose that $\pri{\eta(l')} = m$ and $\prii{\eta(l')} = \kappa(m)$.
        We shall prove that $\theta'(l') = \nil$.
        If $l' = l$, then $\theta'(l) = \nil$.
        Otherwise, since $\theta$, $\tau$, $\eta$, and $\kappa$ satisfy the IC, $\theta(l) \neq \nil$ implies that $\theta(l') = \nil$; hence we obtain $\theta'(l') = \nil$.

      \end{itemize}
      Thus, we obtain
      \[
        \simuek
        {\config{\dollar{\plug{H}{\return{V}}}{x}{M}}{\theta'}}
        {\config
          {\paren{
          \begin{array}{l}
            \mathbf{let}\;\var{res} = \labeledc{m}{\paren{\seq{x}{\mte{\plug{H}{\return{V}}}{\eta}}{\return{\thunk{\abs{\underscore}{\mte{M}{\eta}}}}}}}\;\mathbf{in} \\
            \app{\force{\var{res}}}{((m, \kappa(m)), i+1)}
          \end{array}
          }}
        {\tau'}}
        {\eta}{\kappa}.
      \]
      Furthermore, Proposition~\ref{prop:deltoac:wellformed_preservation} shows that the right-hand-side configuration is well-formed.
      Finally, we check that the IC is preserved through the evaluation. 
      However, we present only non-trivial conditions for brevity.
      \begin{itemize}
      \item[6.] $\dom{\tau'} = \dom{\tau} = \dom{\kappa} \sqcup \im{\kappa}$.
        
      \item[7.] $\dom{\eta} = \dom{\theta} = \dom{\theta'}$.
        
      \item[9.]
        Suppose that $\theta'(l') = \nil$.
        If $l' \equiv l$, then $\getnum{\prii{\eta(l)}, \tau'} = i+1 > i = \priii{\eta(l)}$.
        Otherwise, the previous IC implies that $\getnum{\prii{\eta(l')}, \tau} > \priii{\eta(l')}$.
        If $\prii{\eta(l')} = \kappa(m)$, then $\getnum{\kappa(m), \tau'} = i+1 > i = \getnum{\kappa(m), \tau'} > \priii{\eta(l')}$.
        If $\prii{\eta(l')} \neq \kappa(m)$, then the previous IC yields that $\prii{\eta(l')} \neq m$.
        Hence, $\prii{\eta(l')} \in \dom{\tau}$, yielding $\getnum{\prii{\eta(l')}, \tau'} = \getnum{\prii{\eta(l')}, \tau} > \priii{\eta(l')}$.

      \item[10.]
        Suppose that $\theta'(l') \neq \nil$.
        Then, we obtain $l' \neq l$ and $\theta'(l') = \theta(l') = \abs{y}{\dollar{\plug{G}{\return{y}}}{x}{S}}$ for some $\context{G}$ and $S$.
        \begin{itemize}
        \item[(a)]
          The previous IC yields that $l'$ does not appear in $S$.
          
        \item[(b)] 
          We shall argue that $\prii{\eta(l')} \neq \kappa(m)$.
          Suppose otherwise.
          Then, the previous IC implies that $\pri{\eta(l')} = m$ and $\theta(l') = \nil$, which is a contradiction.
          Hence, $\getnum{\prii{\eta(l')}, \tau'} = \getnum{\prii{\eta(l')}, \tau} = \priii{\eta(l)}$.
          
        \item[(c)]
          Let $l'' \in \dom{\theta'}$ and suppose that $l'' \neq l'$ and $\pri{\eta(l'')} = \pri{\eta(l')}$.
          If $l'' \equiv l$, then the previous IC yields that $\theta(l') = \nil$, which contradicts the assumption.
          Hence, $l'' \neq l$, which immediately implies that $\theta'(l'') = \theta(l'') = \nil$.
          Note that the last equation follows from the previous IC.
          
        \item[(d)]
          This directly follows from the previous IC since $\pri{\eta(l')} \neq m$ and $\prii{\eta(l')} \neq \kappa(m)$.
        \end{itemize}
      \end{itemize}
      This completes the proof for the current case. \\
    \end{caseindent}

    \noindent \textbf{Case} $(\mathrm{fail})$:
    \begin{caseindent}
      Suppose that $M \equiv \paren{\throw{l}{V}}$ and $\theta(l) = \nil$.
      Then, $\config{M}{\theta}$ evaluates to $\bot$.
      By the definition of $\rsimuek{\eta}{\kappa}$, we obtain
      \begin{align*}
        N &\equiv \mte{\throw{l}{V}}{\eta} \\
          &\equiv \left(\begin{array}{l}
          \mathbf{case}\;\mte{l}{\eta}\;\mathbf{of}\;\{\\
          \quad((\var{z}, \var{zc}), \var{i}) \mapsto\\
          \quad\quad\mathbf{let}\;\var{j} = \app{\force{\var{get}}}{\var{zc}}\;\mathbf{in} \\
          \quad\quad\mathbf{let}\;\var{b} = \app{\app{\force{\var{compare}}}{i}}{j} \\
          \quad\quad\mathbf{case}\;\var{b}\;\mathbf{of}\;\{\\
          \quad\quad\quad\paren{\inj{True}{\unit}} \mapsto\\
          \quad\quad\quad\quad\mathbf{let}\;\var{i'} = \app{\force{\var{incr}}}{i}\;\mathbf{in}\\
          \quad\quad\quad\quad\mathbf{let}\;\unit = \app{\app{\force{\var{set}}}{\var{zc}}}{i'}\;\mathbf{in}\\
          \quad\quad\quad\quad\mathbf{let}\;\var{res} = \app{\app{\mathbf{resume}}{\var{z}}}{\mte{V}{\eta}}\;\mathbf{in}\\
          \quad\quad\quad\quad\app{\force{\var{res}}}{((\var{z}, \var{zc}), \var{i'})} \\
          \quad\quad\quad\paren{\inj{False}{\unit}} \mapsto \force{\var{fail}}\\
          \quad\quad\} \\
          \}
        \end{array} \right).
      \end{align*}
      Let $m = \pri{l}$ and $j = \getnum{\prii{l}, \tau}$, then, since $\theta$, $\tau$ $\eta$, and $\kappa$ satisfy the IC, we obtain
      \[
        \prii{l} = \kappa(l) \quad\text{and}\quad j > \priii{l}.
      \]
      Using this, the evaluation of $\config{N}{\tau}$ proceeds as follows:
      \begin{align*}
        \config{N}{\tau}
        &\arrAC^{*}
          \config
          {\force{\var{fail}}}
          {\tau} \\
        &\arrAC^{*}
          \bot.
      \end{align*}
      Hence, we obtain $\simu{\bot}{\bot}$.
      This complete the proof. \\
    \end{caseindent}

    Now, the remaining cases are for $(\mathrm{ret})$ and $(\mathrm{yield})$, where the computation $M$ is enclosed within a dollar term, i.e., $M \equiv \dollar{M_1}{x}{M_2}$.
    In such instances, there are two possible derivations for $\simuek{\config{M}{\theta}}{\config{N}{\tau}}{\eta}{\kappa}$:
    \[
      \simuek
      {\config{\dollar{M_1}{x}{M_2}}{\theta}}
      {\config{N \equiv \paren{\mte{\dollar{M_1}{x}{M_2}}{\eta}}}{\tau}}
      {\eta}{\kappa},
    \]
    or
    \[
      \inferrule
      {
        \simuek{\config{M_1}{\theta}}{\config{N_1}{\tau}}{\eta}{\kappa} \\
        \getnum{\kappa(m), \tau} = i \\
        \tau(m) = \nil \\
        \eta^{-1}(m, \kappa(m), \_) \subseteq \theta^{-1}(\nil) \\
      }
      {
        \simuek
        {\config{\dollar{M_1}{x}{M_2}}{\theta}}
        {\config
          {N \equiv \paren{\begin{array}{l}
            \letin{res}{\labeledc{m}{(\letin{x}{N_1}{\app{{\return{\thunk{\abs{\underscore}{\mte{M_2}{\eta}}}}}}})}}
            {\\\app{\force{\var{res}}}{\var{(m, \kappa(m), i)}}}
          \end{array}}}{\tau}}{\eta}{\kappa}}.
    \]
    Then, by applying Proposition~\ref{prop:deltoac:decomp} with $\context{C}[\dollar{\hole}{x}{M_2}]$ as the instance of $\context{C}$, the former case reduces to the latter.
    Thus, for the remaining cases, we restrict our attention to the situation where
    \[
      N \equiv \paren{\begin{array}{l}
        \letin{res}{\labeledc{m}{(\letin{x}{N_1}{\app{{\return{\thunk{\abs{\underscore}{\mte{M_2}{\eta}}}}}}})}}
        {\\\app{\force{\var{res}}}{\var{(m, \kappa(m), i)}}}
      \end{array}}.
    \]

    \noindent \textbf{Case} $(\mathrm{ret})$:
    \begin{caseindent}

      Suppose that $M \equiv \dollar{\return{V}}{x}{M'}$.
      Then, $\config{M}{\theta}$ is reduced to $\config{M'[V/x]}{\theta}$.
      As noted previously, we shall assume without loss of generality that the derivation of $\simuek{\config{M}{\theta}}{\config{N}{\tau}}{\eta}{\kappa}$ is
      \[
        \inferrule
        {
          \simuek{\config{\return{V}}{\theta}}{\config{\return{\mte{V}{\eta}}}{\tau}}{\eta}{\kappa} \\
          \getnum{\kappa(m), \tau} = i \\
          \tau(m) = \nil \\
          \eta^{-1}(m, \kappa(m), \_) \subseteq \theta^{-1}(\nil) \\
        }
        {
          \simuek
          {\config{\dollar{\return{V}}{x}{M'}}{\theta}}
          {\config
            {\begin{array}{l}
              \letin{res}{\labeledc{m}{(\letin{x}{\return{\mte{V}{\eta}}}{\app{{\return{\thunk{\abs{\underscore}{\mte{M'}{\eta}}}}}}})}}
              {\\\app{\force{\var{res}}}{\var{(m, \kappa(m), i)}}}
            \end{array}}{\tau}}{\eta}{\kappa}}
      \]
      $\config{N}{\tau}$ is evaluated as follows.
      \begin{align*}
        \config{N}{\tau}
        &\arrAC \config{\paren{\app{\force{\thunk{\abs{\underscore}{\mte{M'}{\eta}[\mte{V}{\eta}/x]}}}}{\var{(m, \kappa(m), i)}}}}{\tau} \\
        &\arrAC \config{\mte{M'}{\eta}[\mte{V}{\eta}/x]}{\tau} \\
        &\equiv \config{\mte{M'[V/x]}{\eta}}{\tau} & \text{by Lemma~\ref{lem:deltoac:subst}}.
      \end{align*}
      From Lemma~\ref{lem:deltoac:transd-correspondence}, we obtain
      \[
        \simuek{\config{M'[V/x]}{\theta}}{\config{\mte{M'[V/x]}{\eta}}{\tau}}{\eta}{\kappa},
      \]
      where Proposition~\ref{prop:deltoac:wellformed_preservation} implies that the configuration on the right-hand side  is well-formed.
      Since $\theta$, $\tau$, $\eta$, and $\kappa$ satisfy the IC, we conclude that
      \[
        \simu{\config{M'[V/x]}{\theta}}{\config{\mte{M'[V/x]}{\eta}}{\tau}}.
      \]\\
    \end{caseindent}

    \noindent \textbf{Case} $(\mathrm{shift})$:
    \begin{caseindent}
      Suppose that $M \equiv \dollar{\plug{H}{\shift{k}{M_1}}}{x}{M_2}$.
      Then, $\config{M}{\theta}$ evaluates to
      \[
        \config{M_1[l/k]}{\theta'},
      \]
      where $l$ is a fresh continuation label and $\theta'$ is $\theta[l \defines \abs{y}{\dollar{\plug{H}{\return{y}}}{x}{M_2}}]$.
      Conducting case analysis implies that the derivation tree of $\simuek{\config{\dollar{\plug{H}{\shift{k}{M_1}}}{x}{M_2}}{\theta}}{\config{N}{\tau}}{\eta}{\kappa}$ is
      \[
        \inferrule
        {
          \simuek{\config{\plug{H}{\shift{k}{M_1}}}{\theta}}{\config{N_1}{\tau}}{\eta}{\kappa} \\
          \getnum{\kappa(m), \tau} = i \\
          \tau(m) = \nil \\
          \eta^{-1}(m, \kappa(m), \_) \subseteq \theta^{-1}(\nil) \\
        }
        {
          \simuek
          {\config{\dollar{\plug{H}{\shift{k}{M_1}}}{x}{M_2}}{\theta}}
          {\config
            {N \equiv \paren{\begin{array}{l}
              \letin{res}{\labeledc{m}{(\letin{x}{N_1}{\app{{\return{\thunk{\abs{\underscore}{\mte{M_2}{\eta}}}}}}})}}
              {\\\app{\force{\var{res}}}{\var{(m, \kappa(m), i)}}}
            \end{array}}}{\tau}}{\eta}{\kappa}}.
      \]
      Then, applying Lemma~\ref{lem:deltoac:purec_decomp} to $\simuek{\config{\plug{H}{\shift{k}{M_1}}}{\theta}}{\config{N_1}{\tau}}{\eta}{\kappa}$ implies that there is an $\ac$-computation $N_1'$ such that
      \[
        N_1 \equiv \plug{\mte{H}{\eta}}{N_1'} \quad\text{and}\quad \simuek{\config{\shift{k}{M_1}}{\theta}}{\config{N_1'}{\tau}}{\eta}{\kappa}.
      \]
      Moreover, by performing case inversion on $\simuek{\config{\shift{k}{M_1}}{\theta}}{\config{N_1'}{\tau}}{\eta}{\kappa}$, we obtain
      \[
        N_1' \equiv \mte{\shift{k}{M_1}}{\eta} \equiv \yield{\thunk{\abs{k}{\mte{M_1}{\eta}}}}.
      \]
      Thus, the evaluation of $\config{N}{\tau}$ is as follows:
      \begin{align*}
        \config{N}{\tau}
        &\arrAC^{*}
          \config
          {\paren{\app{\force{\thunk{\abs{k}{\mte{M_1}{\eta}}}}}{\var{(m, \kappa(m), i)}}}}
          {\tau'} \\
        &\arrAC^{*}
          \config
          {\mte{M_1}{\eta}[\var{(m, \kappa(m), i)}/k]}
          {\tau'},
      \end{align*}
      where
      \[
        \tau' \defines \tau\left[\var{m} \defines \thunk{\abs{y}{\seq{x}{\plug{\mte{H}{\eta}}{\return{y}}}{\return{\thunk{\abs{\underscore}{\mte{M_2}{\eta}}}}}}}\right].
      \]
      Thus, defining $\eta'$ to be $\eta[l \defines \var{(m, \kappa(m), i)}]$ yields
      \[
        \mte{M_1}{\eta}[\var{(m, \kappa(m), i)}/k] \equiv \mte{M_1}{\eta'}[\var{(m, \kappa(m), i)}/k] \equiv \mte{M_1}{\eta'}[\eta'(l)/k] \equiv \mte{M_1[l/k]}{\eta'},
      \]
      where the first equation follows since $M_1$ does not contain $l$, which is taken as a fresh label, and the last equation follows from Lemma~\ref{lem:deltoac:subst}.
      Therefore by Lemma~\ref{lem:deltoac:transd-correspondence}, we have
      \[
        \simuek{\config{M_1[l/k]}{\theta'}}{\config{\mte{M_1[l/k]}{\eta'}}{\tau'}}{\eta'}{\kappa}.
      \]
      Moreover, Proposition~\ref{prop:deltoac:wellformed_preservation} shows that $\arrAC^{*}\config{\mte{M_1}{\eta}[\var{(m, \kappa(m), i)}/k]}{\tau'}$ is well-formed.
      Then, it remains to show that $\theta'$, $\tau'$, $\eta'$, and $\kappa$ satisfy the IC.
      We verify only the non-trivial conditions here; the remaining conditions hold automatically as $\theta$, $\tau$, $\eta$, and $\kappa$ already satisfy the IC.
      \begin{itemize}
      \item[6.]
        $\dom{\kappa} \sqcup \im{\kappa} = \dom{\tau} = \dom{\tau'}$. %

      \item[7.]
        $\dom{\eta'} = \dom{\eta} \cup \{l\} = \dom{\theta} \cup \{l\} = \dom{\theta'}$.

      \item[8.]
        Let $l' \in \dom{\eta'}$ and suppose that $\eta'(l') = (\var{z}, \var{zc}, i)$.
        If $l' \in \dom{\eta}$, then we immediately see that $\var{z} \in \dom{\kappa}$ and $\kappa(\var{z}) = \var{zc}$.
        If $l' \equiv l$, then, $l \in \dom{\theta}$, $(\var{z}, \var{zc}, i) = (m, \kappa(m), i)$, and $m \in \dom{\kappa}$ since
        \[
          \dom{\tau} \subset \dom{\kappa} \quad\text{and}\quad m \in \dom{\tau}.
        \]

      \item[9.]
        Suppose that $l' \in \dom{\eta'}$ and $\theta(l') = \nil$.
        Then, since $l' \in \dom{\theta}$,
        \[
          \getnum{\prii{\eta'(l')}, \tau'} = \getnum{\prii{\eta(l')}, \tau} < \priii{\eta(l')} < \priii{\eta'(l')}.
        \]

      \item[10.]
        Suppose that $l' \in \dom{\eta'}$ and $\theta'(l') = \abs{y}{\dollar{\plug{G}{\return{y}}}{x}{S}}$.
        First, assume that $l' \in \dom{\theta}$.
        Then, from the previous IC, it follows that $l'$ is not contained in $S$ and that
        \[
          \getnum{\prii{\eta'(l')}, \tau'} = \getnum{\prii{\eta(l')}, \tau'} = \getnum{\prii{\eta(l')}, \tau} = \priii{\eta(l')} = \priii{\eta'(l')},
        \]
        where the second equation holds since $\tau'$ is an extension of $\tau$ and $\prii{\eta(l')} \in \dom{\tau}$.
        Next, suppose that $l'' \in \dom{\theta'}$, $l'' \neq l'$, and $\pri{\eta(l'')} = \pri{\eta(l')}$.
        Then, if $l'' \in \dom{\theta}$, it immediately follows that
        \[
          \theta'(l'') = \theta(l'') = \nil.
        \]
        If, instead, $l'' \in \dom{\theta'}\setminus\dom{\theta}$, i.e., $l'' = l$, by one of the premises for deriving $\simuek{\config{S}{\theta}}{\config{N'}{\tau}}{\eta}{\kappa}$, we see that
        \[
          \eta^{-1}(m, \kappa(m), \_) \subseteq \theta^{-1}(\nil).
        \]
        Moreover, since $\pri{\eta(l')} = \pri{\eta(l)} = m$, we also obtain $\prii{\eta(l')} = \kappa(m)$.
        Thus, we see that $\theta(l') = \nil$, but this is contradiction.
        Finally, we have $\tau'(\eta'(l')) = \tau'(\eta(l')) = \tau(\eta(l')) = \abs{y}{\dollar{\plug{\mte{G}{\eta}}{\return{y}}}{x}{\mte{S}{\eta}}}$.
        Here, $\context{G}$ and $S$ do not contain $l$ since it is taken as a fresh label.
        Hence, $\mte{\context{G}}{\eta} = \mte{\context{G}}{\eta'}$ and $\mte{S}{\eta} = \mte{S}{\eta'}$, which implies
        \[
          \tau'(\eta'(l')) = \abs{y}{\dollar{\plug{\mte{G}{\eta'}}{\return{y}}}{x}{\mte{S}{\eta'}}}.
        \]

        Now, assume that $l' = l$.
        Since, $l$ is taken as a fresh label, it does not appear in $M_2$.
        Then, the second condition is ensured by the one of premises of $\simuek{\config{M}{\theta}}{\config{N'}{\tau}}{\eta}{\kappa}$:
        \[
          \getnum{\kappa(m), \tau} = i
        \]
        The second condition can be also verified as in the case in which $l' \in \dom{\theta}$.
        As for the last condition, we see that
        \begin{align*}
          \tau'(\pri{\eta(l')}) &= \thunk{\abs{y}{\seq{x}{\plug{\mte{H}{\eta}}{\return{y}}}{\return{\thunk{\abs{\underscore}{\mte{M_2}{\eta}}}}}}} \\
                                         &= \thunk{\abs{y}{\seq{x}{\plug{\mte{H}{\eta'}}{\return{y}}}{\return{\thunk{\abs{\underscore}{\mte{M_2}{\eta'}}}}}}}.
        \end{align*}
      \end{itemize}
      This completes the proof for the current case.
    \end{caseindent}
\end{proof}

  \begin{proposition}\label{prop:deltoac:sim_withc}
    Suppose that
    \begin{gather*}
      \simuekc{\config{\context{C}}{\theta}}{\config{\context{D}}{\tau}}{\eta}{\kappa}, \\
      \simuek{\config{M}{\theta}}{\config{N}{\tau}}{\eta}{\kappa}, \\
      \simuek{\config{\plug{C}{M}}{\theta}}{\config{\plug{D}{N}}{\tau}}{\eta}{\kappa}, \\
      \redbetaD{\config{M}{\theta}}{\config{M'}{\theta'}},
    \end{gather*}
    where $\config{\plug{D}{N}}{\tau}$ is well-formed and $\theta$, $\tau$, $\eta$, and $\kappa$ satisfy the IC.
    Then, there exist an $\ac$-computation $N'$, an $\ac$-store $\tau'$, partial functions $\eta'$ and $\kappa'$ such that
    \begin{gather*}
      \redACplus{\config{N}{\tau}}{\config{N'}{\tau'}}, \\
      \simuek{\config{M'}{\theta'}}{\config{N'}{\tau'}}{\eta'}{\kappa'}, \\
      \simuek{\config{\plug{C}{M'}}{\theta'}}{\config{\plug{D}{N'}}{\tau'}}{\eta'}{\kappa'}.
    \end{gather*}
    Moreover, from Proposition~\ref{prop:deltoac:wellformed_preservation}, $\config{\plug{D}{N'}}{\tau'}$ is well-formed.
  \end{proposition}
  \begin{proof}
    We prove by induction on the derivation of $\simuekc{\config{\context{C}}{\theta}}{\config{\context{D}}{\tau}}{\eta}{\kappa}$.
    The base case follows immediately from Lemma~\ref{prop:deltoac:beta_sim}.

    For the inductive step, we suppose, without loss of generality, that the outermost frame of $C$ is computational; i.e., $\context{C} \equiv \dollar{\plug{C'}{}}{x}{M_2}$, which is justified by Lemma~\ref{lem:deltoac:purec_insertion} and Lemma~\ref{lem:deltoac:purec_decomp}.
    Suppose that $\simuek{\config{\dollar{\plug{C'}{M_1}}{x}{M_2}}{\theta}}{\config{\plug{D}{N}}{\tau}}{\eta}{\kappa}$, $\simuek{\config{M_1}{\theta}}{\config{N}{\tau}}{\eta}{\kappa}$, and $\redbetaD{\config{M_1}{\theta}}{\config{M_1'}{\theta'}}$.
    As noted in the proof of Lemma~\ref{prop:deltoac:beta_sim}, or alternatively by applying Proposition~\ref{prop:deltoac:decomp}, we shall assume that the derivation of $\simuek{\config{\dollar{\plug{C'}{M_1}}{x}{M_2}}{\theta}}{\config{\plug{D}{N}}{\tau}}{\eta}{\kappa}$ is
    \[
      \inferrule
      {
        \simuek{\config{\plug{C'}{M_1}}{\theta}}{\config{\plug{D'}{N}}{\tau}}{\eta}{\kappa} \\
        \getnum{\kappa(m), \tau} = i \\
        \tau(m) = \nil \\
        \eta^{-1}(m, \kappa(m), \_) \subseteq \theta^{-1}(\nil) \\
      }
      {
        \simuek
        {\config{\dollar{\plug{C'}{M_1}}{x}{M_2}}{\theta}}
        {\config
          {\paren{\begin{array}{l}
            \letin{res}{\labeledc{m}{(\letin{x}{\plug{D'}{N}}{\app{{\return{\thunk{\abs{\underscore}{\mte{M_2}{\eta}}}}}}})}}
            {\\\app{\force{\var{res}}}{\var{(m, \kappa(m), i)}}}
          \end{array}}}{\tau}}{\eta}{\kappa}}
    \]
    Moreover, from Lemma~\ref{lem:deltoac:factor_context}, we obtain $\simuekc{\config{\context{C'}}{\theta}}{\config{\context{D'}}{\tau}}{\eta}{\kappa}$.
    Applying the induction hypothesis to $\simuek{\config{\plug{C'}{M_1}}{\theta}}{\config{\plug{D'}{N}}{\tau}}{\eta}{\kappa}$ implies that there exist $N'$, $\tau'$, $\eta'$, and $\kappa'$ such that
    \begin{gather*}
      \redACplus{\config{N}{\tau}}{\config{N'}{\tau'}} \\
      \simuek{\config{\plug{C'}{M_1'}}{\theta'}}{\config{\plug{D'}{N'}}{\tau'}}{\eta'}{\kappa'},
    \end{gather*}
    where $\theta'$, $\tau'$, $\eta'$, and $\kappa'$ satisfy the IC.
    Then, we shall show that
    \[
    \inferrule
    {
      \simuek{\config{\plug{C'}{M_1'}}{\theta'}}{\config{\plug{D'}{N'}}{\tau'}}{\eta'}{\kappa'} \\
      \getnum{\kappa'(m), \tau} = i \\
      \tau'(m) = \nil \\
      \eta'^{-1}(m, \kappa'(m), \_) \subseteq \theta'^{-1}(\nil) \\
    }
    {
      \simuek{\config{\dollar{\plug{C'}{M_1'}}{x}{M_2}}{\theta'}}{\config{\paren{\begin{array}{l}
            \letin{res}{\labeledc{m}{(\letin{x}{\plug{D'}{N'}}{\app{{\return{\thunk{\abs{\underscore}{\mte{M_2}{\eta'}}}}}}})}}
            {\\\app{\force{\var{res}}}{\var{(m, \kappa(m), i)}}}
          \end{array}}}{\tau'}}{\eta'}{\kappa'}}
    \]
    is derivable by checking that that $\mte{M_2}{\eta'} \equiv \mte{M_2}{\eta}$ and that the premises other than $\simuek{\config{\plug{C'}{M_1'}}{\theta'}}{\config{\plug{D'}{N'}}{\tau'}}{\eta'}{\kappa'}$ are also satisfied.

    Here, we conduct case analysis on $\redbetaD{\config{M_1}{\theta}}{\config{M_1'}{\theta'}}$ as follows.
    
    \noindent\textbf{Case} $(\mam)$:
    \begin{caseindent}
      In this case, none of $\theta$, $\tau$, $\eta$, and $\kappa$ is updated.
      Hence, it is trivial to check that the premises are preserved. \\
    \end{caseindent}

    \noindent\textbf{Case } $(\mathrm{throw})$:
    \begin{caseindent}
      Suppose that $\redbetaD{\config{M_1}{\theta}}{\config{M_1'}{\theta'}}$ is
      \[
        \redbetaD{\config{\throw{l}{V}}{\theta}}{\config{\dollar{\plug{H}{\return{V}}}{x}{S}}{\theta'}},
      \]
      where
      \begin{align*}
        \theta(l) &= \paren{\abs{y}{\dollar{\plug{H}{\return{y}}}{x}{S}}}, \\
        \theta' &\defines \theta[l \defines \nil].
      \end{align*}
      From the proof of Lemma~\ref{prop:deltoac:beta_sim}, we obtain
      \begin{gather*}
        N \equiv \mte{\throw{l}{V}}{\eta}, \\
        N' \equiv \paren{
          \begin{array}{l}
            \mathbf{let}\;\var{res} = \labeledc{n}{\paren{\seq{x}{\plug{\mte{H}{\eta}}{\return{\mte{V}{\eta}}}}{\return{\thunk{\abs{\underscore}{\mte{S}{\eta}}}}}}}\;\mathbf{in} \\
            \app{\force{\var{res}}}{((n, \kappa(n)), j+1)}
          \end{array}
        }, \\
        \simuek{\config{M_1'}{\theta'}}{\config{N'}{\tau'}}{\eta}{\kappa},
      \end{gather*}
      where
      \begin{gather*}
        m \neq n \quad(\text{by Proposition~\ref{prop:deltoac:wellformed_preservation}}), \\
        \tau' \defines \tau\left[n \defines \nil, \kappa(n) \defines \refcell{j+1}\right], \\
      \end{gather*}
      and $\theta'$, $\tau'$, $\eta$, and $\kappa$ satisfy the IC.
      Then, we shall check that the premises are satisfied as follows:
      \begin{itemize}
      \item
        $\getnum{\kappa(m), \tau'} = i$ since $m \neq n$ and $\tau'(\kappa(m)) = \refcell{i}$.
        
      \item
        $\tau'(m) = \tau(m) = \nil$.
        
      \item
        Let $l' \in \dom{\eta}$ and suppose that $\pri{\eta(l')} = m$ and $\prii{\eta(l')} = \kappa(m)$.
        Then, since $m \neq n$, we see that $l' \neq l$ and $\theta'(l') = \theta(l') = \nil$, where the last equation follows from the previous IC.
        
      \end{itemize}       
    \end{caseindent}

    \noindent\textbf{Case} $(\mathrm{fail})$:
    \begin{caseindent}
      $(\mathrm{fail})$ is not applicable here, as $\config{M}{\theta}$ reduces to a configuration $\config{M'}{\theta'}$, not $\bot$.
    \end{caseindent}

    Now, the remaining cases for $(\mathrm{ret})$ and $(\mathrm{yield})$.
    Both of them assumes that the computation $M_1$ is enclosed within a dollar term: $M_1 \equiv \dollar{S_1}{x}{S_2}$.
    As in the proof of Proposition~\ref{prop:deltoac:beta_sim}, we shall assume that the the derivation of $\simuek{\config{M_1}{\theta}}{\config{N}{\tau}}{\eta}{\kappa}$ is
    \[
      \inferrule
      {
        \simuek{\config{S_1}{\theta}}{\config{N_1}{\tau}}{\eta}{\kappa} \\
        \getnum{\kappa(n), \tau} = n \\
        \tau(n) = \nil \\
        \eta^{-1}(n, \kappa(n), \_) \subseteq \theta^{-1}(\nil) \\
      }
      {
        \simuek
        {\config{\dollar{S_1}{x}{S_2}}{\theta}}
        {\config
          {N \equiv \paren{\begin{array}{l}
            \letin{res}{\labeledc{m}{(\letin{x}{N_1}{\app{{\return{\thunk{\abs{\underscore}{\mte{S_2}{\eta}}}}}}})}}
            {\\\app{\force{\var{res}}}{\var{(n, \kappa(n), j)}}}
          \end{array}}}{\tau}}{\eta}{\kappa}},
    \]
    not
    \[
      \simuek{\config{\dollar{S_1}{x}{S_2}}{\theta}}{\config{\mte{\dollar{S_1}{x}{S_2}}{\eta}}{\tau}}{\eta}{\kappa}.
    \]
    This is because, the evaluation from $\config{\mte{\dollar{S_1}{x}{S_2}}{\eta}}{\tau}$ to
    \[
      \config
          {N \equiv \paren{\begin{array}{l}
            \letin{res}{\labeledc{m}{(\letin{x}{N_1}{\app{{\return{\thunk{\abs{\underscore}{\mte{S_2}{\eta}}}}}}})}}
            {\\\app{\force{\var{res}}}{\var{(n, \kappa(n), j)}}}
          \end{array}}}{\tau}
    \]
    preserves not only the IC, but also the premises of
    \[
    \inferrule
    {
      \simuek{\config{\plug{C'}{M_1'}}{\theta'}}{\config{\plug{D'}{N'}}{\tau'}}{\eta'}{\kappa'} \\
      \getnum{\kappa'(m), \tau} = i \\
      \tau'(m) = \nil \\
      \eta'^{-1}(m, \kappa'(m), \_) \subseteq \theta'^{-1}(\nil) \\
    }
    {
      \simuek{\config{\dollar{\plug{C'}{M_1'}}{x}{M_2}}{\theta'}}{\config{\paren{\begin{array}{l}
            \letin{res}{\labeledc{m}{(\letin{x}{\plug{D'}{N'}}{\app{{\return{\thunk{\abs{\underscore}{\mte{M_2}{\eta'}}}}}}})}}
            {\\\app{\force{\var{res}}}{\var{(m, \kappa(m), i)}}}
          \end{array}}}{\tau'}}{\eta'}{\kappa'}},
    \]
    as we have proven in the \textbf{Case 1} in the proof of Proposition~\ref{prop:deltoac:decomp}.
    
    \noindent\textbf{Case} $(\mathrm{ret})$:
    \begin{caseindent}
      Suppose that $\redbetaD{\config{M_1}{\theta}}{\config{M_1'}{\theta'}}$ is
      \begin{gather*}
        \redbetaD{\config{M_1 \equiv \dollar{\return{V}}{x}{M'}}{\theta}}{\config{M'[V/x]}{\theta}}.
      \end{gather*}
      From the proof of Lemma~\ref{prop:deltoac:beta_sim}, we see that none of $\theta$, $\tau$, $\eta$, and $\kappa$ is updated, which implies that the premises are preserved.\\
    \end{caseindent}

    \noindent\textbf{Case} $(\mathrm{shift})$:
    \begin{caseindent}
      Suppose that $\redbetaD{\config{M_1}{\theta}}{\config{M_1'}{\theta'}}$ is
      \begin{gather*}
        \redbetaD{\config{\dollar{\plug{H}{\shift{k}{S_1}}}{x}{S_2}}{\theta}}{\config{S_1[l/k]}{\theta'}},
      \end{gather*}
      where $l$ is a fresh continuation label, and $\theta'$ is $\theta[l \defines \abs{y}{\dollar{\plug{H}{\return{y}}}{x}{S_2}}]$.
      From the proof of Lemma~\ref{prop:deltoac:beta_sim}, we obtain
      \begin{gather*}
        N \equiv \paren{\begin{array}{l}
              \letin{res}{\labeledc{n}{\paren{\letin{x}{\plug{\mte{H}{\eta}}{\yield{\thunk{\abs{k}{\mte{S_1}{\eta}}}}}}{\app{{\return{\thunk{\abs{\underscore}{\mte{S_2}{\eta}}}}}}}}}}
              {\\\app{\force{\var{res}}}{\var{(n, \kappa(n), j)}}}
            \end{array}}, \\
        N' \equiv \mte{S_1}{\eta}[\var{(n, \kappa(n), j)}/k], \\
        \simuek{\config{M_1'}{\theta'}}{\config{N'}{\tau'}}{\eta'}{\kappa},
      \end{gather*}
      where
      \begin{gather*}
        \tau' \defines \tau\left[\var{n} \defines \thunk{\abs{y}{\seq{x}{\plug{\mte{H}{\eta}}{\return{y}}}{\return{\thunk{\abs{\underscore}{\mte{S_2}{\eta}}}}}}}\right], \\
        \eta' \defines \eta[l \defines \var{(n, \kappa(n), j)}],
      \end{gather*}
      and $\theta'$, $\tau'$, $\eta'$, and $\kappa$ satisfy the IC.
      First, since $\eta$ is extended on a fresh label $l$, we have that $\mte{M_2}{\eta'} \equiv \mte{M_2}{\eta}$.
      Moreover, the well-formedness implies that $m \neq n$.
      Then, we shall check that the premises are satisfied as follows:
      \begin{itemize}
      \item
        $\getnum{\kappa(m), \tau'} = \getnum{\kappa(m), \tau} = i$.
        
      \item
        $\tau'(m) = \tau(m) = \nil$
        
      \item
        Let $l' \in \dom{\eta'}$ and suppose that $\pri{\eta'(l')} = m$ and $\prii{\eta'(l')} = \kappa(m)$.
        Since $m \neq n$, we have $l' \neq l$ and $l \in \dom{\eta}$.
        Then, from the previous IC, we obtain $\theta(l) = \nil$.
        Therefore, $\theta'(l) = \theta(l) = \nil$.
      \end{itemize}
      Therefore, we obtain
      \[
      \simuek{\config{\dollar{\plug{C'}{M_1'}}{x}{M_2}}{\theta'}}{\config{\paren{\begin{array}{l}
            \letin{res}{\labeledc{m}{(\letin{x}{\plug{D'}{N'}}{\app{{\return{\thunk{\abs{\underscore}{\mte{M_2}{\eta'}}}}}}})}}
            {\\\app{\force{\var{res}}}{\var{(m, \kappa(m), i)}}}
          \end{array}}}{\tau'}}{\eta'}{\kappa'},
      \]
    \end{caseindent}
  \end{proof}

  Finally, we give the proof of Theorem~\ref{thm:deltoac_simulation}.

  \begin{proof}[Proof of Theorem~\ref{thm:deltoac_simulation}]
    Suppose that $\simu{\config{\plug{C}{M}}{\theta}}{D}$ and $\redbetaD{\config{M}{\theta}}{C'}$.
    Then, by definition, there exist $N$ and $\tau$ such that $D \equiv \config{N}{\tau}$, which is well-formed.
    Moreover, by Proposition~\ref{prop:deltoac:decomp}, there exist $\context{D}$, $N'$, $\tau'$, and $\kappa'$ such that
      \begin{gather*}
        \config{N}{\tau} \arrAC^{*} \config{\plug{D}{N'}}{\tau'}, \\
        \simuekc{\config{\mathcal{C}}{\theta}}{\config{\mathcal{D}}{\tau'}}{\eta}{\kappa'}, \\
        \simuek{\config{M}{\theta}}{\config{N'}{\tau'}}{\eta}{\kappa'},
      \end{gather*}
      where $\theta$, $\tau'$, $\eta$, and $\kappa'$ satisfy the IC, and Proposition~\ref{prop:deltoac:wellformed_preservation} implies that ${\config{\plug{D}{N'}}{\tau'}}$ is well-formed.
    Then, we consider two cases for the derivation of $\redbetaD{\config{M}{\theta}}{C'}$.

    \noindent \textbf{Case 1}:
    \[
      \redbetaD{\config{M}{\theta}}{\bot}
    \]
    \begin{caseindent}
      In this case, the reduction rule $(\mathrm{fail})$ is applied to $\config{M}{\theta}$ since it is the only rule that introduces $\bot$.
      Therefore, from Lemma~\ref{prop:deltoac:beta_sim}, we obtain $\redbetaD{\config{N'}{\tau'}}{\bot}$, which implies $\redbetaD{\config{\plug{D}{N'}}{\tau'}}{\bot}$, completing this case.
    \end{caseindent}

    \noindent \textbf{Case 2}:
    \[
      \redbetaD{\config{M}{\theta}}{\config{M'}{\theta'}}
    \]
    \begin{caseindent}
      This case follows directly by Lemma~\ref{prop:deltoac:sim_withc}.
    \end{caseindent}
    This completes the proof.
  \end{proof}

  \subsection{Proposition for proving strong macro-expressibility}\label{sec:app_deltoac:strong}

  To establish the strong macro-expressibility of $\del$ in $\ac$, we need to prove the following proposition:

  \begin{proposition}\label{prop:deltoac:strong}
    For any $\del$-program $M$, if $\eval{\ac}{\mt{M}}$ terminates, then $\eval{\del}{M}$ also terminates.
  \end{proposition}

  Before proving this proposition, we show four lemmas.

  \begin{lemma}\label{lem:deltoac:eff_well_formed}
    We say an $\del$-configuration $\config{M}{\theta}$ is well-formed
    if for any continuation label $l$ that appears in $M$,
    $\theta(l)$ is defined, and this property is preserved under
    reduction.
  \end{lemma}
  \begin{proof}
    By straightforward case analysis on $M$.
  \end{proof}

  \begin{lemma}\label{lem:deltoac:diverge}
    Suppose that $\eval{\del}{M}$ diverges, i.e., the reduction sequence of $M$ is infinite.
    Then $\eval{\ac}{\mt{M}}$ also diverges.
  \end{lemma}
  \begin{proof}
    By applying Theorem~\ref{thm:deltoac_simulation} repeatedly, we also obtain an infinite reduction sequence of $\mt{M}$.
    Since the reduction of $\ac$ is deterministic, this implies that $\eval{\ac}{\mt{M}}$ also diverges.
  \end{proof}

  \begin{lemma}\label{lem:deltoac:stuck}
    Suppose that $\eval{\del}{M}$ gets stuck: there exists a term $M'$
    and a store $\theta$ such that
    $\redDclos{\config{M}{\emptyset}}{\config{M'}{\theta}}$,
    $M' \neq \return{V}$ for any value $V$, and there is no rule that
    can reduce $\config{M'}{\theta}$.  Then, $\eval{\ac}{\mt{M}}$ also
    gets stuck.
  \end{lemma}
  \begin{proof}
    It suffices to show the following statement:
    \begin{quote}
      Suppose that there exist $M$, $N$, $\theta$, $\tau$, $\eta$, and
      $\kappa$ such that
      $\simuek{\config{M}{\theta}}{\config{N}{\tau}}{\eta}{\kappa}$,
      $\config{M}{\theta}$ is well-formed, and $\theta$, $\tau$,
      $\eta$, and $\kappa$ satisfy the IC. If $\config{M}{\theta}$ is
      stuck, then $\config{N}{\tau}$ will also be stuck: there exist
      $N'$, $\tau'$, and $\kappa'$ such that
      \begin{itemize}
      \item $\redDclos{\config{N}{\tau}}{\config{N'}{\tau'}}$,
      \item $\simuek{\config{M}{\theta}}{\config{N'}{\tau'}}{\eta}{\tau'}$,
      \item $\theta$, $\tau'$, $\eta$, and $\tau'$ satisfy the IC, and
      \item $\config{N'}{\tau'}$ is stuck.
      \end{itemize}
    \end{quote}
    We prove by induction on
    $\simuek{\config{M}{\theta}}{\config{N}{\tau}}{\eta}{\kappa}$.
    There are twelve
    cases to consider,\footnote{$\simuek{\config{\return{V}}{\theta}}{\config{\mte{\return{V}}{\eta}}{\tau}}{\eta}{\kappa}$
      is excluded by the premise.} but we treat only five
    cases. The other cases follow by similar reasoning.
    
    \noindent\textbf{Case 1}:
    \begin{caseindent}
      \begin{gather*}
        \simuek
        {\config{\pcase{V}{x_1}{x_2}{M}}{\theta}}
        {\config{\mte{\pcase{V}{x_1}{x_2}{M}}{\eta}}{\tau}}
        {\eta}{\kappa}
      \end{gather*}
      Since the left-hand configuration is stuck, we know that
      $V \neq \vpair{V_1}{V_2}$ for any values $V_1$ and $V_2$.
      Consequently, $\mte{V}{\eta}$ also cannot be of the form
      $\vpair{W_1}{W_2}$ for any values $W_1$ and $W_2$.
      Therefore, the configuration
      $\config{\mte{\pcase{V}{x_1}{x_2}{M}}{\eta}}{\tau}$ is stuck,
      and this concludes the current case.
    \end{caseindent}

    \noindent\textbf{Case 2}:
    \begin{caseindent}
      \[
        \inferrule
        {
          \simuek{\config{M_1}{\theta}}{\config{N_1}{\tau}}{\eta}{\kappa}
        }
        {
          \simuek{\config{\seq{x}{M_1}{M_2}}{\theta}}{\config{\seq{x}{N_1}{\mte{M_2}{\eta}}}{\tau}}{\eta}{\kappa}
        }
      \]
      By the induction hypothesis, there exist $N_1'$, $\tau'$, and
      $\kappa'$ such that
      $\redDclos{\config{N_1}{\tau}}{\config{N_1'}{\tau'}}$,
      $\simuek{\config{M_1}{\theta}}{\config{N_1'}{\tau'}}{\eta}{\kappa'}$,
      and $\config{N_1'}{\tau'}$ is stuck.
      Since $\config{\seq{x}{M_1}{M_2}}{\theta}$ is stuck, $M_1$
      cannot be of the form $\return{V}$ for any value $V$.
      Together with
      $\simuek{\config{M_1}{\theta}}{\config{N_1'}{\tau'}}{\eta}{\kappa'}$,
      this implies that $N_1 \neq \return{W}$ for any value $W$.
      Consequently, the configuration
      $\config{\seq{x}{N_1'}{\mte{M_2}{\eta}}}{\tau'}$ is also stuck,
      which completes the current case.
    \end{caseindent}

    \noindent\textbf{Case 3}:
    \begin{caseindent}
      \[
        \simuek
        {\config{\throw{V}{W}}{\theta}}
        {
          \config
          {
            \mte{\throw{V}{W}}{\eta}
          }
          {\tau}
        }
        {\eta}{\kappa}
      \]
      It follows that $V \neq l$ for any continuation label $l$, since
      if $V = l$ for some $l$, then by the well-formedness of
      $\config{\throw{V}{W}}{\theta}$, we know that
      $l \in \dom{\theta}$. However, this contradicts that
      $\config{\throw{V}{W}}{\theta}$ is stuck.
      This implies that $\mte{\throw{V}{W}}{\eta}$ is also stuck,
      which concludes the current case.
    \end{caseindent}

    \noindent\textbf{Case 4}:
    \begin{caseindent}
      \[
        \simuek
        {\config{\dollar{M_1}{x}{M_2}}{\theta}}
        {\config{\mte{\dollar{M_1}{x}{M_2}}{\eta}}{\tau}}
        {\eta}{\kappa}
      \]
      The right-hand configuration evaluates to
      \[
        \config{\paren{\begin{array}{l}
          \letin{res}{\labeledc{m}{(\letin{x}{N'}{\app{{\return{\thunk{\abs{\underscore}{\mte{M'}{\eta}}}}}}})}}
          {\\\app{\force{\var{res}}}{\var{(m, mc, i)}}}
        \end{array}}}{\tau'},
      \]
      which is reducible to the next case.
    \end{caseindent}

    \noindent\textbf{Case 5}:
    \begin{caseindent}
      \[
        \inferrule
        {
          \simuek{\config{M_1}{\theta}}{\config{N_1}{\tau}}{\eta}{\kappa} \\
          \getnum{\kappa(m), \tau} = i \\
          \tau(m) = \nil \\
          \eta^{-1}(m, \kappa(m), \_) \subseteq \theta^{-1}(\nil)
        }
        {
          \simuek
          {\config{\dollar{M_1}{x}{M_2}}{\theta}}
          {\config
            {\begin{array}{l}
              \letin{res}{\labeledc{m}{(\letin{x}{N_1}{\app{{\return{\thunk{\abs{\underscore}{\mte{M_2}{\eta}}}}}}})}}
              {\\\app{\force{\var{res}}}{\var{(m, \kappa(m), i)}}}
            \end{array}}{\tau}}{\eta}{\kappa}}
      \]
      By the induction hypothesis, there exist $N_1'$, $\tau'$, and
      $\kappa'$ such that
      $\redDclos{\config{N_1}{\tau}}{\config{N_1'}{\tau'}}$,
      $\simuek{\config{M_1}{\theta}}{\config{N_1'}{\tau'}}{\eta}{\kappa'}$,
      and $\config{N_1'}{\tau'}$ is stuck.
      Since $\config{\seq{x}{M_1}{M_2}}{\theta}$ is stuck, $M_1$
      cannot be neither of the form $\plug{H}{\shift{k}{M_1'}}$ for
      any pure context $\context{H}$ and computation $M_1'$, nor of
      the form $\return{V}$ for any value $V$.
      Then, it follows (by induction on
      $\simuek{\config{M_1}{\theta}}{\config{N_1'}{\tau'}}{\eta}{\kappa'}$)
      that $N_1'$ cannot be neither of the form $\plug{P}{\yield{W}}$
      nor of the form $\return{W}$, for any pure context $\context{P}$
      and value $W$.
      Therefore, the configuration
      \[
        \config
        {\begin{array}{l}
          \letin{res}{\labeledc{m}{(\letin{x}{N_1'}{\app{{\return{\thunk{\abs{\underscore}{\mte{M_2}{\eta}}}}}}})}}
          {\\\app{\force{\var{res}}}{\var{(m, \kappa(m), i)}}}
        \end{array}}{\tau'}
      \]
      is also stuck, and this completes the current case.
    \end{caseindent}
  \end{proof}

  \begin{lemma}\label{lem:deltoac:bot}
    If $\eval{\del}{M}$ reaches the error state $\bot$, i.e., $\redDplus{\config{M}{\emptyset}}{\bot}$, then $\eval{\ac}{\mt{M}}$ also reaches $\bot$.
  \end{lemma}
  \begin{proof}
    This lemma directly follows from Theorem~\ref{thm:deltoac_simulation}.
  \end{proof}

  Finally, we give the proof of Proposition~\ref{prop:deltoac:strong}.

  \begin{proof}[Proof of Proposition~\ref{prop:deltoac:strong}]
    We prove the contrapositive of the proposition: if
    $\eval{\del}{M}$ does not terminate, $\eval{\ac}{\mt{M}}$ also
    does not terminate.
    There are three cases where $\eval{\del}{M}$ does not terminate:
    (1) $\eval{\del}{M}$ diverges; (2) $\eval{\del}{M}$ gets stuck;
    (3) $\eval{\del}{M}$ reaches $\bot$.
    In each case, Lemmas~\ref{lem:deltoac:diverge},
    \ref{lem:deltoac:stuck}, and \ref{lem:deltoac:bot} imply that
    $\eval{\ac}{\mt{M}}$ also diverges, gets stuck, or reaches $\bot$,
    respectively.
    Thus, $\eval{\ac}{\mt{M}}$ does not terminate, which completes the
    proof.
  \end{proof}

\section{Supplementary Proofs for Chapter~\ref{chap:EFFoneToDELone}}

\subsection{From \texorpdfstring{$\eff$}{EFFone} to \texorpdfstring{$\del$}{DELone}}\label{app:EFFoneToDELone}

To derive a simulation relation, we extend the macro-translation with $\eta$, a partial function which maps the continuation labels in $\eff$ onto those in $\del$:
\[
  \mte{l_H}{\eta} \defines \thunk{\abs{y}{\app{\throw{(\eta(l_H))}{y}}{\thunk{H^{\mathrm{ops}}_{\eta}}}}}
\]
$H^{\mathrm{ops}}_{\eta}$ denote the translation of operational clauses with respect to $\eta$.
Moreover, we derive a macro-translation on contexts by mapping holes to holes; we let $\mte{\context{K}}{\eta}$ denote the translation of a context $\context{K}$.

\begin{definition}
  For any $\eff$-configuration $C$ and $\del$-configuration $D$, $\simu{C}{D}$ is defined to hold if and only if $C \equiv D \equiv \bot$, or there exist an $\eff$-computation $M$, a $\del$-computation $N$, an $\eff$-store $\theta$, a $\del$-store $\tau$, and a partial function $\eta$ such that the following conditions are satisfied:
  \begin{enumerate}
  \item $C = \config{M}{\theta}$.
  \item $D = \config{N}{\tau}$.
  \item $N = \mte{M}{\eta}$.
  \item $\eta$ is injective.
  \item $\eta(\dom{\theta}) \subseteq \dom{\tau}$
  \item For any continuation label $l_H$ appearing in $M$, $\theta(l_H)$ is defined.
  \item For any continuation label $l_H \in \dom{\theta}$, if $\theta(l_H) = \nil$, then $\tau(\eta(l_H)) = \nil$. Otherwise, there exists a pure context $\mathcal{H}$ in $\eff$ such that
  \begin{align*}
  \theta(l_H) &= \abs{y}{\handle{H}{\plug{H}{\return{y}}}} \\
  \tau(\eta(l_H)) &= \abs{y}{\dollart{\mte{\plug{H}{\return{y}}}{\eta}}{H^{\mathrm{ret}}_{\eta}}}
  \end{align*}
\end{enumerate}
\end{definition}

Intuitively, $\simu{\config{M}{\theta}}{\config{N}{\tau}}$ means the correspondence between the two configurations: $M$ is translated into $N$ and the continuations stored in $\theta$ is translated into those in $\tau$.
We define that $\theta$, $\tau$, and $\eta$ satisfy the \emph{invariant conditions} (IC) if the conditions 4 through 6 in the definition are fulfilled.

\begin{lemma}\label{lem:EFFoneToDELone:property}
  \phantom{}
  \begin{enumerate}
  \item
    For any $\eff$-computation $M$, $\simu{\config{M}{\emptyset}}{\config{\mt{M}}{\emptyset}}$ holds.

  \item
    For any value $V$, $\simu{\config{\return{V}}{\theta}}{\config{N}{\tau}}$ implies that $N \equiv \return{\mte{V}{\eta}}$ for some $\eta$.
  \end{enumerate}
\end{lemma}
\begin{proof}
  By case analysis on $M$.
\end{proof}

\begin{lemma}\label{lem:eff_to_del_subst}
  Suppose that an $\eff$-computation $M$ has the free variables $x_1, \ldots x_n$, then we have $\mte{M[V_1/x_1,\ldots, V_n/x_n]}{\eta} = \mte{M}{\eta}[\mte{V_1}{\eta}/x_1, \ldots, \mte{V_n}{\eta}/x_n]$ for any values $V_1, \ldots V_n$ and $\eta$.
\end{lemma}
\begin{proof}
  By straightforward induction on $M$. For example, if $M \equiv \app{M_1}{M_2}$, then
  \begin{align*}
    \mte{M[V_1/x_1,\ldots, V_n/x_n]}{\eta} &\equiv \mte{(\app{M_1}{M_2})[V_1/x_1,\ldots, V_n/x_n]}{\eta} \\
    &= \app{\mte{M_1[V_1/x_1,\ldots, V_n/x_n]}{\eta}}{\mte{M_2[V_1/x_1,\ldots, V_n/x_n]}{\eta}} \\
    &= \app{(\mte{M_1}{\eta}[\mte{V_1}{\eta}/x_1, \ldots, \mte{V_n}{\eta}/x_n])}{(\mte{M_2}{\eta}[\mte{V_1}{\eta}/x_1, \ldots, \mte{V_n}{\eta}/x_n])} \\
                                           &= (\app{\mte{M_1}{\eta}}{\mte{M_2}{\eta}})[\mte{V_1}{\eta}/x_1, \ldots, \mte{V_n}{\eta}/x_n]
  \end{align*}
\end{proof}

\begin{lemma}\label{lem:eff_to_del_beta_sim}
  Suppose that $\simu{C}{D}$ and $\redbetaE{C}{C'}$, then there exists a $\del$ configuration $D'$ such that $\redDplus{D}{D'}$ and $\simu{C'}{D'}$.
\end{lemma}
\begin{proof}
  We prove by case analysis on $\redbetaE{C}{C'}$.
  For the sake of brevity, we consider a few non-trivial cases.

  \noindent \textbf{Case} $(\mathrm{op})$:
  \begin{caseindent}
    Suppose $C \equiv \config{\handle{H}{\paren{\plug{H}{\opcall{op}{V}}}}}{\theta}$ and
  \[
    C' \equiv \config{M_{\op{op}}[V/p, l_H/k]}{\theta[l_H \coloneq \abs{x}{\handle{H}{\plug{H}{\return{x}}}}]},
  \]
  where $H \equiv \handler{x}{M^{\mathrm{ret}}}{\ldots (\opcall{op}{\app{p}{k}} \mapsto M^{\op{op}}) \ldots}$ and $l_H$ is a fresh label.
  By the definition of $\simu{C}{D}$, we have
  \[
    D \equiv \config{\app{\dollart{\plug{\mte{\context{H}}{\eta}}{\shift{k}{\abs{h}{\app{\force{h}}{\inj{\op{op}}{\vpair{\mte{V}{\eta}}{\thunk{\abs{y}{\app{\throw{k}{y}}{h}}}}}}}}}}{H^{\mathrm{ret}}_{\eta}}}{\thunk{H^{\mathrm{ops}}_{\eta}}}}{\tau}
  \]
  for some $\eta$.
  The configuration $D$ evaluates to
  \[
    D' \defines \config{\mte{M_{\op{op}}}{\eta}\left[\mte{V}{\eta}/p, \thunk{\abs{y}{\app{\throw{l}{y}}{\thunk{H^{\mathrm{ops}}_{\eta}}}}}/k\right]}
    {\tau\left[ l \defines \abs{y}{\dollart{\plug{\mte{\context{H}}{\eta}}{\return{y}}}{H^{\mathrm{ret}}_{\eta}}} \right]}
  \]
  where $l$ is a fresh label.
  Let $\eta'$ be $\eta[ l_H \defines l ]$.
  By Lemma~\ref{lem:eff_to_del_subst}, it follows that
  \[
    \mte{M_{\op{op}}}{\eta}[\mte{V}{\eta}/p, \thunk{\abs{y}{\app{\throw{l}{y}}{\thunk{H^{\mathrm{ops}}_{\eta}}}}}/k] \equiv \mte{M_{\op{op}}[V/p, l_H/k]}{\eta'}.
  \]
  Note that $\mte{M_{\op{op}}}{\eta} \equiv \mte{M_{\op{op}}}{\eta'}$, $\mte{\context{H}}{\eta} \equiv \mte{\context{H}}{\eta'}$, $H^{\mathrm{ret}}_{\eta} \equiv H^{\mathrm{ret}}_{\eta'}$, and $H^{\mathrm{ops}}_{\eta} \equiv H^{\mathrm{ops}}_{\eta'}$ hold since $l_H$ does not appear in any of $M_{\op{op}}$, $\context{H}$, $H^{\mathrm{ret}}$, and $H^{\mathrm{ops}}_{\eta}$.
  Finally, we conclude that
  \[
    {\config{M_{\op{op}}[V/p, l_H/k]}{\theta[l_H \coloneq \abs{x}{\handle{H}{\plug{H}{\return{x}}}}]}} \sim D'.
  \]
  It is straightforward to check that the IC are satisfied with respect to the updated stores and $\eta'$.
  \end{caseindent}

  \noindent \textbf{Case} $(\mathrm{throw})$:
  \begin{caseindent}
    Suppose $C \equiv \config{\throw{l_H}{V}}{\theta}$ and $C' \equiv \config{\handle{H}{\plug{H}{\return{V}}}}{\theta[l_H \coloneq \nil]}$, where $\theta(l_H) = \abs{x}{\handle{H}{\plug{H}{\return{x}}}}$.
    By the definition of $\simu{C}{D}$, we obtain
    \[
      D \equiv \config{\app{\force{\thunk{\abs{y}{\app{\throw{(\eta(l_H))}{y}}{\thunk{H^{\mathrm{ops}}_{\eta}}}}}}}{\mte{V}{\eta}}}{\tau},
    \]
    for some $\eta$ and $\tau$ such that
    \[
      \tau(\eta(l_H)) = \abs{y}{\dollart{\mte{\plug{H}{\return{y}}}{\eta}}{H^{\mathrm{ret}}_{\eta}}}.
    \]
    Thus, $D$ evaluates to
    \[
      D' \defines \config{\app{\paren{\dollart{\mte{\plug{H}{\return{y}}}{\eta}}{H^{\mathrm{ret}}_{\eta}}[\mte{V}{\eta}/y]}}{\thunk{H^{\mathrm{ops}}_{\eta}}}}{\tau[\eta(l_H) := \nil]}.
    \]
    Next, from Lemma ~\ref{lem:eff_to_del_subst}, we obtain
    \[
      \dollart{\mte{\plug{H}{\return{y}}}{\eta}}{H^{\mathrm{ret}}_{\eta}}[\mte{V}{\eta}/y] \equiv \dollart{\mte{\plug{H}{\return{V}}}{\eta}}{H^{\mathrm{ret}}_{\eta}}.
    \]
    Therefore, we conclude that
    \[
    \config{\handle{H}{\plug{H}{\return{V}}}}{\theta[l_H \coloneq \nil]} \sim D'
    \]
    since
    \[
      \mte{\handle{H}{\plug{H}{\return{V}}}}{\eta} \equiv \dollart{\mte{\plug{H}{\return{V}}}{\eta}}{H^{\mathrm{ret}}_{\eta}}{\thunk{H^{\mathrm{ops}}_{\eta}}}.
    \]
    Note that the updated stores and $\eta$ satisfy the IC.
  \end{caseindent}

  \noindent \textbf{Case} $(\mathrm{fail})$:
  \begin{caseindent}
    Suppose $C \equiv \config{\throw{l_H}}{\theta}$, $\theta(l_H) = \nil$, and $\config{\throw{l_H}{V}}{\theta} \rightarrow^{\mathbf{E}}_{\beta} \bot$.
    By the definition of $\simu{C}{D}$, we have
    \[
      D \equiv \config{\app{\force{\thunk{\abs{y}{\app{\throw{(\eta(l_H))}{y}}{\thunk{H^{\mathrm{ops}}_{\eta}}}}}}}{\mte{V}{\eta}}}{\tau},
    \]
    where $\tau$ is a store such that $\tau(\eta(l_H)) = \nil$.
    Hence, the invocation of $\eta(l_H)$ fails and $D$ evaluates to $\bot$, which concludes the current case.
  \end{caseindent}
\end{proof}

\begin{lemma}\label{lem:eff_to_del_plug}
  $\mte{\plug{K}{M}}{\eta} \equiv \mte{\context{K}}{\eta}[\mte{M}{\eta}]$ for an arbitrary context $K$ and computation $M$.
\end{lemma}
\begin{proof}
  By straightforward induction on $K$.
\end{proof}

\begin{lemma}\label{lem:eff_to_del_decomp}
  If $\simu{\config{M}{\theta}}{\config{N}{\tau}}$ and $M$ can be decomposed into an evaluation context $\mathcal{C}$ and a redex $M'$, then there exist some $\eta$ such that $N \equiv \mte{M}{\eta}$ can be also decomposed into an evaluation context $\mte{\mathcal{C}}{\eta}$ and a redex $\mte{M'}{\eta}$.
\end{lemma}
\begin{proof}
  It is easy to check this by induction on the number of frames of $\mathcal{C}$. Note that the extended macro-translation maps frames to frames and redexes to redexes.
\end{proof}

\begin{proposition}[Simulation]\label{prop:EFFoneToDELone:sim}
  Suppose that $\simu{C}{D}$ and $\redE{C}{C'}$, then there exists a $\del$ configuration $D'$ such that $\redDplus{D}{D'}$ and $\simu{C'}{D'}$.
\end{proposition}
\begin{proof}
  Since $C$ is not in normal form, there exist $M$ and $\theta$ such that $C \equiv \config{M}{\theta}$, and according to the semantics of $\eff$, $M$ can be decomposed into an evaluation context $\mathcal{C}$ and a redex $M'$.
  By the definition of $\simu{C}{D}$, there exist $\eta$ and $\tau$ such that $D \equiv \config{\mte{M}{\eta}}{\tau}$, and $\theta$, $\tau$, and $\eta$ satisfy the IC.
  By Lemma~\ref{lem:eff_to_del_decomp}, $\mte{M}{\eta}$ can also be decomposed into $\mte{\mathcal{C}}{\eta}$ and a redex $\mte{M'}{\eta}$, namely $\mte{M}{\eta} = \mte{\mathcal{C}}{\eta}\left[\mte{M'}{\eta}\right]$.

  If $\redbetaE{\config{M}{\theta}}{\bot}$, then we have $\redE{(C \equiv \config{\plug{C}{M}}{\theta})}{\bot}$.
  By Lemma~\ref{lem:eff_to_del_beta_sim}, $\config{M'}{\tau}$ also evaluates to $\bot$.
  Hence, we conclude that $\redDplus{\config{\plug{D}{M'}}{\tau}}{\bot}$.

  Next, suppose $\redbetaEp{M'}{\theta}{M''}{\theta'}$.
  By Lemma~\ref{lem:eff_to_del_beta_sim}, we obtain $\redDplus{\config{\mte{M'}{\eta}}{\tau}}{\config{\mte{M''}{\eta'}}{\tau'}}$ and $\simu{\config{M''}{\theta'}}{\config{\mte{M''}{\eta'}}{\tau'}}$, where $\theta'$, $\tau'$, and $\eta'$ satisfy the IC.
  Moreover, $\mte{\context{C}}{\eta} \equiv \mte{\context{C}}{\eta'}$ holds since the labels in $\dom{\eta'}\setminus\dom{\eta}$ are fresh and do not appear in $\context{C}$.
  Therefore, we conclude that $\simu{\config{\plug{C}{M''}}{\theta'}}{\config{\mte{\plug{C}{M''}}{\eta'}}{\tau'}}$ by Lemma~\ref{lem:eff_to_del_plug}.
\end{proof}

To establish the strong macro-expressibility of $\eff$ in $\del$, we prove four lemmas.

\begin{lemma}\label{lem:efftodel:eff_well_formed}
  We say an $\eff$-configuration $\config{M}{\theta}$ is well-formed
  if for any continuation label $l_H$ that appears in $M$,
  $\theta(l)$ is defined, and this property is preserved under
  reduction.
\end{lemma}
\begin{proof}
  By straightforward case analysis on $M$.
\end{proof}

\begin{lemma}\label{lem:efftodel:diverge}
  Suppose that $\eval{\eff}{M}$ diverges, i.e., the reduction sequence of $M$ is infinite.
  Then $\eval{\del}{\mt{M}}$ also diverges.
\end{lemma}
\begin{proof}
  By applying Theorem~\ref{prop:EFFoneToDELone:sim} repeatedly, we also obtain an infinite reduction sequence of $\mt{M}$.
  Since the reduction of $\del$ is deterministic, this implies that $\eval{\del}{\mt{M}}$ also diverges.
\end{proof}

\begin{lemma}\label{lem:efftodel:stuck}
  Suppose that $\eval{\eff}{M}$ gets stuck: there exists a term $M'$
  and a store $\theta$ such that
  $\redEclos{\config{M}{\emptyset}}{\config{M'}{\theta}}$,
  $M' \neq \return{V}$ for any value $V$, and there is no rule that
  can reduce $\config{M'}{\theta}$.  Then, $\eval{\del}{\mt{M}}$ also
  gets stuck.
\end{lemma}
\begin{proof}
  It suffices to show the following statement:
  \begin{quote}
    Suppose that there exist $M$, $N$, $\theta$, $\tau$, and $\eta$
    such that $N = \mte{M}{\eta}$, $\config{M}{\theta}$ is well-formed, and $\theta$, $\tau$, and $\eta$
    satisfy the IC.
    If $\config{M}{\theta}$ is stuck, then $\config{N}{\tau}$ is
    also be stuck.
  \end{quote}
  We prove by induction on the structure of $M$.
  However, we treat only four cases. The other cases follow by similar reasoning.

  \noindent\textbf{Case 1}:
  \begin{caseindent}
    \[
      M \equiv \pcase{V}{x_1}{x_2}{M'}
    \]
    Since the $\config{M}{\theta}$ is stuck, we know that
    $V \neq \vpair{V_1}{V_2}$ for any values $V_1$ and $V_2$.
    Consequently, $\mte{V}{\eta}$ also cannot be of the form
    $\vpair{W_1}{W_2}$ for any values $W_1$ and $W_2$.
    Therefore, the configuration
    $\config{\mte{\pcase{V}{x_1}{x_2}{M}}{\eta}}{\tau}$ is stuck,
    and this concludes the current case.
  \end{caseindent}

  \noindent\textbf{Case 2}:
  \begin{caseindent}
    \[
      M \equiv \throw{V}{W}
    \]
    It follows that $V \neq l_H$ for any continuation label $l_H$,
    since if $V = l_H$ for some $l$, then by the well-formedness of
    $\config{\throw{V}{W}}{\theta}$, we know that
    $l_H \in \dom{\theta}$. However, this contradicts that
    $\config{\throw{V}{W}}{\theta}$ is stuck.
    This implies that $\mte{\throw{V}{W}}{\eta}$ is also stuck, which
    concludes the current case.
  \end{caseindent}

  \noindent\textbf{Case 3}:
  \begin{caseindent}
    \[
      M \equiv \seq{x}{M_1}{M_2}
    \]
    Since $\config{M}{\theta}$ is stuck, $M_1$
    cannot be of the form $\return{V}$ for any value $V$.
    By the induction hypothesis, we obtain that $N_1$ is stuck and
    that $N_1 \neq \return{W}$ for any value $W$.
    Consequently, the configuration
    $\config{\seq{x}{N_1}{\mte{M_2}{\eta}}}{\tau}$ is also stuck,
    which completes the current case.
  \end{caseindent}

  \noindent\textbf{Case 4}:
  \begin{caseindent}
    \[
      M \equiv \handle{H}{M'}
    \]
    By the definition of the translation, we obtain
    \[
      N \equiv \app{\dollart{\mte{M'}{\eta}}{H^{\mathrm{ret}}}}{\thunk{H^{\mathrm{ops}}}}.
    \]
    Since $\config{M}{\theta}$ is stuck, $M'$ cannot be neither of the
    form $\plug{H}{\app{\op{op}}{V}}$ nor of the form $\return{V}$ for
    any pure context $\context{H}$, operation symbol $\op{op}$, and
    value $V$.
    Then, it follows by induction on $M'$ that $\mte{M'}{\eta}$ cannot
    be neither of the form $\plug{P}{\shift{k}{N'}}$ nor of the form
    $\return{W}$, for any pure context $\context{P}$, computation
    $N'$, and value $W$.
    Therefore, the configuration $\config{N}{\tau}$ is also stuck, and
    this completes the current case.
\end{caseindent}
\end{proof}

\begin{lemma}\label{lem:efftodel:bot}
  If $\eval{\eff}{M}$ reaches the error state $\bot$, i.e., $\redEplus{\config{M}{\emptyset}}{\bot}$, then $\eval{\del}{\mt{M}}$ also reaches $\bot$.
\end{lemma}
\begin{proof}
  This lemma directly follows from Theorem~\ref{prop:EFFoneToDELone:sim}.
\end{proof}

\begin{corollary}[Preservation of semantics]\label{cor:eff_to_del_correctness}
  $\eval{\eff}{M}$ terminates if and only if $\eval{\del}{\mt{M}}$ terminates.
\end{corollary}
\begin{proof}
  We first show the ``only if'' direction.
  Suppose that
  $\redEplus{\config{M}{\emptyset}}{\config{\return{V}}{\theta}}$ for
  some $\theta$.
  Lemma~\ref{lem:EFFoneToDELone:property} yields that
  $\config{\mt{M}}{\emptyset}$.
  By applying Proposition~\ref{prop:EFFoneToDELone:sim} iteratively,
  there exists $\eta$ and $\tau$ such that
  $\redDplus{\config{\mt{M}}{\emptyset}}{\config{\return{\mte{V}{\eta}}}{\tau}}$.
  This implies that $\eval{\del}{\mt{M}} = V$.
  We then show the ``if'' direction.
  We prove the contrapositive of the proposition: if $\eval{\eff}{M}$
  does not terminate, $\eval{\del}{\mt{M}}$ also does not terminate.
  There are three cases where
  $\eval{\eff}{M}$ does not terminate: (1) $\eval{\eff}{M}$
  diverges; (2) $\eval{\eff}{M}$ gets stuck; (3) $\eval{\eff}{M}$
  reaches $\bot$.
  In each case, Lemmas~\ref{lem:efftodel:diverge},
  \ref{lem:efftodel:stuck}, and \ref{lem:efftodel:bot} imply that
  $\eval{\del}{\mt{M}}$ also diverges, gets stuck, or reaches $\bot$,
  respectively.
  Thus, $\eval{\del}{\mt{M}}$ does not terminate, which completes the
  proof.
\end{proof}

\subsection{From \texorpdfstring{$\del$}{DELone} to \texorpdfstring{$\eff$}{EFFone}}\label{app:DELoneToEFFone}

As in Appendix~\ref{app:EFFoneToDELone}, we extend the macro-translation with $\eta$, a partial function from labels in $\del$ to those in $\eff$, by
\[
  \mte{l}{\eta} := \eta(l),
\]
and derive the macro-translation on contexts $\mathcal{K} \mapsto \mte{\mathcal{K}}{\eta}$ by mapping holes to holes.

\begin{definition}
  For any configurations of $\del$ and $\eff$, denoted by $C$ and $D$, $\simu{C}{D}$ is defined to hold if and only if $C \equiv D \equiv \bot$, or there exist a $\del$-computation $M$, an $\eff$-computation $N$, a $\del$-store $\theta$, an $\eff$-store $\tau$, and a partial function $\eta$ such that the following conditions are satisfied:
  \begin{enumerate}
  \item $C = \config{M}{\theta}$.
  \item $D = \config{N}{\tau}$.
  \item $N = \mte{M}{\eta}$.
  \item $\eta(\dom{\theta}) \subseteq \dom{\tau}$.
  \item For any continuation label $l$ appearing in $M$, $\theta(l)$ is defined.
  \item For any continuation label $l \in \dom{\theta}$, if $\theta(l) = \nil$, then $\tau(\eta(l)) = \nil$. Otherwise, there exists a pure context $\mathcal{H}$ of $\del$ such that
    \begin{align*}
      \theta(l) &= \abs{y}{\dollar{\plug{H}{\return{y}}}{x}{N}}, \\
      \tau(\eta(l)) &= \abs{y}{\handle{\handler{x}{\mt{N}}{\opcall{shift0}{\app{p}{k}} \mapsto \app{\force{p}}{k}}}{\mte{\plug{H}{\return{y}}}{\eta}}}. \\
    \end{align*}
  \end{enumerate}
\end{definition}

We define that $\theta$, $\tau$, and $\eta$ satisfy the \emph{invariant conditions} (IC) if the conditions 4 through 6 in the definition are satisfied.

\begin{lemma}\label{lem:deltoeff:property}
  \phantom{}
  \begin{enumerate}
  \item
    For any $\del$-computation $M$, $\simu{\config{M}{\emptyset}}{\config{\mt{M}}{\emptyset}}$ holds.

  \item
    For any $\del$-value $V$, $\simu{\config{\return{V}}{\theta}}{\config{N}{\tau}}$ implies that $N \equiv \return{\mte{V}{\eta}}$ for some $\eta$.
  \end{enumerate}
\end{lemma}
\begin{proof}
  By case analysis on $M$.
\end{proof}

\begin{lemma}\label{lem:deltoeff:subst}
  Suppose that a $\del$-computation $M$ has the free variables $x_1, \ldots x_n$, then we have $\mte{M[V_1/x_1,\ldots, V_n/x_n]}{\eta} = \mte{M}{\eta}[\mte{V_1}{\eta}/x_1, \ldots, \mte{V_n}{\eta}/x_n]$ for any values $V_1, \ldots V_n$ and $\eta$.
\end{lemma}
\begin{proof}
  By straightforward induction on $M$.
\end{proof}

\begin{lemma}\label{lem:deltoeff:beta_sim}
  Suppose that $\simu{C}{D}$ and $\redbetaD{C}{C'}$, then there exists an $\eff$ configuration $D'$ such that $\redEplus{D}{D'}$ and $\simu{C'}{D'}$.
\end{lemma}
\begin{proof}
  We prove by case analysis on $\redbetaD{C}{C'}$.
  For the sake of brevity, we consider a few non-trivial cases.

  \noindent \textbf{Case} $(\mathrm{shift})$:
  \begin{caseindent}
    Suppose that
    $C \equiv \config{\dollar{\plug{H}{\shift{k}{M}}}{x}{N}}{\theta}$,
    $l$ is a fresh label in this configuration, and
    $C' \equiv \config{M[l/k]}{\theta'}$, where
    $\theta' \defines \theta[l \coloneq
    \abs{y}{\dollar{\plug{H}{\return{y}}}{x}{N}}]$.
    By the definition of $\simu{C}{D}$, we have for some $\eta$,
    \[
      D \equiv \config {\handle
        {H}
        {\plug{\mte{\context{H}}{\eta}}{\opcall{shift0}{\thunk{\abs{k}{\mte{M}{\eta}}}}}}}
      {\tau},
    \]
    where
    $H \equiv \handler{x}{\mte{N}{\eta}}{\opcall{shift0}{\app{p}{k}}
      \mapsto \app{\force{p}}{k}}$ and $\theta$, $\tau$, and $\eta$
    satisfy the IC.

    $D$ evaluates to the following configuration:
    \[
      D' \defines \config{\mte{M}{\eta}[m_H/k]}{\tau'},
    \]
    where $\tau' \defines \tau[m_H \defines \abs{y}{\handle{H}{\plug{\mte{\context{H}}{\eta}}{\return{y}}}}]$ and $m_H$ is a fresh label.
    Let $\eta'$ be $\eta[l \defines m_H]$.
    By Lemma\ref{lem:deltoeff:subst}, we obtain
    \[
      \mte{M}{\eta}[m_H/k] \equiv \mte{M[l/k]}{\eta'}.
    \]
    Moreover, since $m_H$ is took as a fresh label in $D$, we know that
    \begin{align*}
      & \abs{y}{\handle{H}{\plug{\mte{\context{H}}{\eta}}{\return{y}}}} \\
      &\equiv \abs{y}{\handle{H}{\plug{\mte{\context{H}}{\eta'}}{\return{y}}}} \\
      &\equiv \abs{y}{\handle{H}{\mte{\plug{H}{\return{y}}}{\eta'}}},
    \end{align*}
    and this implies that $\theta'$, $\tau'$, and $\eta'$ satisfy the
    IC.

    Therefore, we conclude that
    \[
      \simu{\config{M[l/k]}{\theta'}}{\config{\mte{M}{\eta}[m_H/k]}{\tau'}},
    \]
    which completes the current case.
  \end{caseindent}

  \noindent \textbf{Case} $(\mathrm{throw})$:
  \begin{caseindent}
    Suppose that $C \equiv \config{\throw{l}{V}}{\theta}$,
    $\theta(l) = \abs{y}{\dollar{\plug{H}{\return{y}}}{x}{N}}$, and
    $C' \equiv \config{\dollar{\plug{H}{\return{V}}}{x}{N}}{\theta[l \defines \nil]}$.
    By the definition of $\simu{C}{D}$, we obtain
    \[
      D \equiv \config{\throw{\eta(l)}{\mte{V}{\eta}}}{\tau},
    \]
    for some $\tau$ and $\eta$ such that
    \[
      \tau(\eta(l)) = \abs{y}{\handle{\handler{x}{\mt{N}}{\opcall{shift0}{\app{p}{k}} \mapsto \app{\force{p}}{k}}}{\mte{\plug{H}{\return{y}}}{\eta}}},
    \]
    where $\theta$, $\tau$, and $\eta$ satisfy the IC.

    $D$ evaluates to
    \begin{align*}
      D' &\defines \config{\handle{\handler{x}{\mt{N}}{\opcall{shift0}{\app{p}{k}} \mapsto \app{\force{p}}{k}}}{\plug{\mte{\context{H}}{\eta}}{\return{\mte{V}{\eta}}}}}{\tau[\eta(l) \defines \nil]} \\
         &\equiv \config{\handle{\handler{x}{\mt{N}}{\opcall{shift0}{\app{p}{k}} \mapsto \app{\force{p}}{k}}}{\mte{\plug{\context{H}}{\return{V}}}{\eta}}}{\tau[\eta(l) \defines \nil]}.
    \end{align*}
    It is trivial that $\theta[l \defines \nil]$, $\tau[\eta(l) \defines \nil]$, and $\eta$ satisfy the IC.
    Therefore, we obtain $\simu{C'}{D'}$, and this completes the current case.
  \end{caseindent}
\end{proof}

\begin{lemma}\label{lem:deltoeff:plug}
  $\mte{\plug{K}{M}}{\eta} \equiv \mte{\context{K}}{\eta}[\mte{M}{\eta}]$ for an arbitrary context $K$ and computation $M$.
\end{lemma}
\begin{proof}
  By straightforward induction on $K$.
\end{proof}

\begin{lemma}\label{lem:deltoeff:decomp}
  If $\simu{\config{M}{\theta}}{\config{N}{\tau}}$ and $M$ can be decomposed into an evaluation context $\mathcal{C}$ and a redex $M'$, then there exist some $\eta$ such that $N \equiv \mte{M}{\eta}$ can be also decomposed into an evaluation context $\mte{\mathcal{C}}{\eta}$ and a redex $\mte{M'}{\eta}$.
\end{lemma}
\begin{proof}
  It is easy to check this by induction on the number of frames of $\mathcal{C}$. Note that the extended macro-translation maps frames to frames and redexes to redexes.
\end{proof}

\begin{proposition}[Simulation]\label{prop:deltoeff:sim}
  Suppose that $\simu{C}{D}$ and $\redD{C}{C'}$, then there exists an $\eff$ configuration $D'$ such that $\redEplus{D}{D'}$ and $\simu{C'}{D'}$.
\end{proposition}
\begin{proof}
  Since $C$ is not in normal form, there exist $M$ and $\theta$ such that $C \equiv \config{M}{\theta}$ and according to the semantics of $\del$, $M$ can be decomposed into an evaluation context $\mathcal{C}$ and a redex $M'$.
  By the definition of $\simu{C}{D}$, there exist $\eta$ and $\tau$ such that $D \equiv \config{\mte{M}{\eta}}{\tau}$, and $\theta$, $\tau$, and $\eta$ satisfy the IC.
  By Lemma~\ref{lem:deltoeff:decomp}, $\mte{M}{\eta}$ can also be decomposed into $\mte{\mathcal{C}}{\eta}$ and a redex $\mte{M'}{\eta}$, namely $\mte{M}{\eta} = \mte{\mathcal{C}}{\eta}\left[\mte{M'}{\eta}\right]$.

  If $\redbetaD{\config{M}{\theta}}{\bot}$, then we have $\redD{(C \equiv \config{\plug{C}{M}}{\theta})}{\bot}$.
  By Lemma~\ref{lem:deltoeff:beta_sim}, $\config{M'}{\tau}$ also evaluates to $\bot$.
  Hence, we conclude that $\redEplus{\config{\plug{D}{M'}}{\tau}}{\bot}$.

  Next, suppose $\redbetaDp{M'}{\theta}{M''}{\theta'}$.
  By Lemma~\ref{lem:deltoeff:beta_sim}, we obtain $\redEplus{\config{\mte{M'}{\eta}}{\tau}}{\config{\mte{M''}{\eta'}}{\tau'}}$ and $\simu{\config{M''}{\theta'}}{\config{\mte{M''}{\eta'}}{\tau'}}$, where $\theta'$, $\tau'$, and $\eta'$ satisfy the IC.
  Moreover, $\mte{\context{C}}{\eta} \equiv \mte{\context{C}}{\eta'}$ holds since the labels in $\dom{\eta'}\setminus\dom{\eta}$ are fresh and do not appear in $\context{C}$.
  Therefore, we conclude that $\simu{\config{\plug{C}{M''}}{\theta'}}{\config{\mte{\plug{C}{M''}}{\eta'}}{\tau'}}$ by Lemma~\ref{lem:deltoeff:plug}.
\end{proof}

To establish the strong macro-expressibility of $\eff$ in $\del$, we prove four lemmas.

\begin{lemma}\label{lem:deltoeff:well_formed}
  We say a $\del$-configuration $\config{M}{\theta}$ is well-formed
  if for any continuation label $l$ that appears in $M$,
  $\theta(l)$ is defined, and this property is preserved under
  reduction.
\end{lemma}
\begin{proof}
  By straightforward case analysis on $M$.
\end{proof}

\begin{lemma}\label{lem:deltoeff:diverge}
  Suppose that $\eval{\del}{M}$ diverges, i.e., the reduction sequence of $M$ is infinite.
  Then $\eval{\eff}{\mt{M}}$ also diverges.
\end{lemma}
\begin{proof}
  By applying Theorem~\ref{prop:deltoeff:sim} repeatedly, we also obtain an infinite reduction sequence of $\mt{M}$.
  Since the reduction of $\eff$ is deterministic, this implies that $\eval{\eff}{\mt{M}}$ also diverges.
\end{proof}

\begin{lemma}\label{lem:deltoeff:stuck}
  Suppose that $\eval{\del}{M}$ gets stuck: there exists a term $M'$
  and a store $\theta$ such that
  $\redDclos{\config{M}{\emptyset}}{\config{M'}{\theta}}$,
  $M' \neq \return{V}$ for any value $V$, and there is no rule that
  can reduce $\config{M'}{\theta}$.  Then, $\eval{\eff}{\mt{M}}$ also
  gets stuck.
\end{lemma}
\begin{proof}
  It suffices to show the following statement:
  \begin{quote}
    Suppose that there exist $M$, $N$, $\theta$, $\tau$, and $\eta$
    such that $N = \mte{M}{\eta}$, $\config{M}{\theta}$ is well-formed, and $\theta$, $\tau$, and $\eta$
    satisfy the IC.
    If $\config{M}{\theta}$ is stuck, then $\config{N}{\tau}$ is
    also be stuck.
  \end{quote}
  We prove by induction on the structure of $M$.
  However, we treat only four cases. The other cases follow by similar reasoning.

  \noindent\textbf{Case 1}:
  \begin{caseindent}
    \[
      M \equiv \pcase{V}{x_1}{x_2}{M'}
    \]
    Since the $\config{M}{\theta}$ is stuck, we know that
    $V \neq \vpair{V_1}{V_2}$ for any values $V_1$ and $V_2$.
    Consequently, $\mte{V}{\eta}$ also cannot be of the form
    $\vpair{W_1}{W_2}$ for any values $W_1$ and $W_2$.
    Therefore, the configuration
    $\config{\mte{\pcase{V}{x_1}{x_2}{M}}{\eta}}{\tau}$ is stuck,
    and this concludes the current case.
  \end{caseindent}

  \noindent\textbf{Case 2}:
  \begin{caseindent}
    \[
      M \equiv \throw{V}{W}
    \]
    It follows that $V \neq l$ for any continuation label $l$,
    since if $V = l$ for some $l$, then by the well-formedness of
    $\config{\throw{V}{W}}{\theta}$, we know that
    $l \in \dom{\theta}$. However, this contradicts that
    $\config{\throw{V}{W}}{\theta}$ is stuck.
    This implies that $\mte{\throw{V}{W}}{\eta}$ is also stuck, which
    concludes the current case.
  \end{caseindent}

  \noindent\textbf{Case 3}:
  \begin{caseindent}
    \[
      M \equiv \seq{x}{M_1}{M_2}
    \]
    Since $\config{M}{\theta}$ is stuck, $M_1$
    cannot be of the form $\return{V}$ for any value $V$.
    By the induction hypothesis, we obtain that $N_1$ is stuck and
    that $N_1 \neq \return{W}$ for any value $W$.
    Consequently, the configuration
    $\config{\seq{x}{N_1}{\mte{M_2}{\eta}}}{\tau}$ is also stuck,
    which completes the current case.
  \end{caseindent}

  \noindent\textbf{Case 4}:
  \begin{caseindent}
    \[
      M \equiv \dollar{M_1}{x}{M_2}
    \]
    By the definition of the translation, we obtain
    \[
      N \equiv \handle{\handler{x}{\mte{M_2}{\eta}}{\opcall{shift0}{\app{p}{k}} \mapsto \app{\force{p}}{k}}}{\mte{M_1}{\eta}}.
    \]
    Since $\config{M}{\theta}$ is stuck, $M_1$ cannot be neither of
    the form $\plug{H}{\shift{k}{L}}$ nor of the form $\return{V}$ for
    any pure context $\context{H}$, computation $L$, and value $V$.
    Then, it follows by induction on $M_1$ that $\mte{M_1}{\eta}$
    cannot be neither of the form $\plug{P}{\opcall{op}{W}}$ nor of
    the form $\return{W}$, for any pure context $\context{P}$, and
    value $W$.
    Therefore, the configuration $\config{N}{\tau}$ is also stuck, and
    this completes the current case.
\end{caseindent}
\end{proof}

\begin{lemma}\label{lem:deltoeff:bot}
  If $\eval{\del}{M}$ reaches the error state $\bot$, i.e., $\redDplus{\config{M}{\emptyset}}{\bot}$, then $\eval{\eff}{\mt{M}}$ also reaches $\bot$.
\end{lemma}
\begin{proof}
  This lemma directly follows from Theorem~\ref{prop:EFFoneToDELone:sim}.
\end{proof}

\begin{corollary}[Preservation of semantics]\label{cor:deltoeff:correctness}
  $\eval{\del}{M}$ terminates if and only if $\eval{\eff}{\mt{M}}$ terminates.
\end{corollary}
\begin{proof}
  We first show the ``only if'' direction.
  Suppose that
  $\redDplus{\config{M}{\emptyset}}{\config{\return{V}}{\theta}}$ for
  some $\theta$.
  Lemma~\ref{lem:deltoeff:property} yields that
  $\simu{\config{M}{\emptyset}}{\config{\mt{M}}{\emptyset}}$.
  By applying Proposition~\ref{prop:deltoeff:sim} iteratively,
  there exists $\eta$ and $\tau$ such that
  $\redEplus{\config{\mt{M}}{\emptyset}}{\config{\return{\mte{V}{\eta}}}{\tau}}$.
  This implies that $\eval{\eff}{\mt{M}} = V$.
  We then show the ``if'' direction.
  We prove the contrapositive of the proposition: if $\eval{\del}{M}$
  does not terminate, $\eval{\eff}{\mt{M}}$ also does not terminate.
  There are three cases where
  $\eval{\del}{M}$ does not terminate: (1) $\eval{\del}{M}$
  diverges; (2) $\eval{\del}{M}$ gets stuck; (3) $\eval{\del}{M}$
  reaches $\bot$.
  In each case, Lemmas~\ref{lem:deltoeff:diverge},
  \ref{lem:deltoeff:stuck}, and \ref{lem:deltoeff:bot} imply that
  $\eval{\eff}{\mt{M}}$ also diverges, gets stuck, or reaches $\bot$,
  respectively.
  Thus, $\eval{\eff}{\mt{M}}$ does not terminate, which completes the
  proof.
\end{proof}

\section{Supplementary Proofs for Chapter~\ref{chap:nonexistent}}\label{app:nonexistent}

In this appendix, we provide the complete proof of Theorems~\ref{thm:reftoeff_nonexist} and \ref{thm:reftoac}.

\subsection{Proof of Theorem~\ref{thm:reftoeff_nonexist}}\label{app:reftoeff_nonexist}

\begin{proof}%
  \[
    M \defines
    \left(\begin{array}{l}
      \letin{r}{\create{\inj{A}{\unit}}}
      {\\\letin{i}{\get{r}}
      {\\\letin{\underscore}{\set{r}{\inj{B}{\unit}}}
      {\\\letin{k}{\get{r}}
      {\\\return{\vpair{i}{k}}}}}}
    \end{array}\right) \mapsto
  \left(\begin{array}{l}
    \letin{r}{\app{\trconst{create}}{\paren{\inj{A}{\unit}}}}
    {\\\letin{i}{\app{\trconst{get}}{r}}
    {\\\letin{\underscore}{\app{\app{\trconst{set}}{r}}{\paren{\inj{B}{\unit}}}}
    {\\\letin{k}{\app{\trconst{get}}{r}}
    {\\\return{\vpair{i}{k}}}}}}
  \end{array}\right)
  \]
  
  Consider the above $\reflang$-Program $M$.
  In $\reflang$, $M$ evaluates to $\vpair{\inj{A}{\unit}}{\inj{B}{\unit}}$.
  Suppose that we have a macro-translation $\mt{\cdot}$ from $\reflang$ to $\eff$.
  Then the evaluation of the translated program $\mt{M}$ terminates with a value $\vpair{\inj{A}{\unit}}{\inj{B}{\unit}}$.
  This implies that there exist a value $X$ and store $\theta$ such that
\[
  \redEplus{\config{\plug{C}{\app{\trconst{create}}{\paren{\inj{A}{\unit}}}}}{\emptyset}}{\config{\plug{C}{\return{X}}}{\theta}}
\]
  where
  \[
    \mathcal{C} =
      \left(\letin{r}{\hole}
      {\letin{i}{\app{\trconst{get}}{r}}
      {\letin{\underscore}{\app{\trconst{set}}{\app{r}{\paren{\inj{B}{\unit}}}}}
      {\letin{k}{\app{\trconst{get}}{r}}
      {\return{\vpair{i}{k}}}}}}\right).
  \]

  We claim that the two evaluations of $\app{\trconst{get}}{r}$ (with r being substituted for X) yield the same value $\inj{A}{\unit}$.
  The first $\app{\trconst{get}}{X}$ is evaluated under the configuration $\config{\plug{D}{\app{\trconst{get}}{X}}}{\theta}$
  where
  \[
    \mathcal{D} = \left(
      \letin{i}{\hole}
      {\letin{\underscore}{\app{\trconst{set}}{\app{X}{\paren{\inj{B}{\unit}}}}}
        {\letin{k}{\app{\trconst{get}}{X}}
          {\return{\vpair{i}{k}}}}}\right).
  \]
  Since the reduction of $\mt{M}$ terminates, there exists a store $\theta'$ such that
  \begin{align*}
    \redEplus{\config{\plug{D}{\app{\trconst{get}}{X}}}{\theta}}{\config{\plug{D}{\return{\paren{\inj{A}{\unit}}}}}{\theta'}}
  \end{align*}
  because $\mt{M}$ evaluates to $\vpair{\inj{A}{\unit}}{\inj{B}{\unit}}$.
  We shall show that a non-local computation which goes beyond $\app{\trconst{get}}{X}$ does not happen in this evaluation.
  But this is trivial; since the evaluation context $\mathcal{D}$ does not have a handler that encloses the hole,
  there is no way to capture a raised operation in $\mathcal{D}$.
  All operations invoked in the evaluation should be caught within $\app{\trconst{get}}{X}$;
  otherwise, the whole computation would not terminate normally.
  
  This implies that the evaluation of $\plug{E}{\app{\trconst{get}}{X}}$ also yields $\inj{A}{\unit}$\footnote{To be precise, the second reduction of $\app{\trconst{get}}{X}$ is identical to the first one modulo the labels of continuations.} where
  \[
    \mathcal{E} = \left(
      \letin{k}{\hole}
      {\return{\vpair{\inj{A}{\unit}}{k}}}\right).
  \]  
  Therefore, $\mt{M}$ evaluates to $\vpair{\inj{A}{\unit}}{\inj{A}{\unit}}$.

  Consider the following program $L$:
  \[
    L \defines \paren{\begin{array}{l}
      \mathbf{let}\;\var{r} = M\;\mathbf{in}\; \\
      \mathbf{case}\;r\;\mathbf{of}\;\{ \\
      \quad (\inj{A}{\unit}, \inj{B}{\unit}) \mapsto \return{\unit} \\
      \quad \underscore \mapsto  \app{\paren{\abs{x}{\app{\force{x}}{x}}}}{\thunk{\abs{x}{\app{\force{x}}{x}}}}\\
      \}
    \end{array}}
  \]
  In $\reflang$, $L$ evaluates to $\unit$.
  However, in $\eff$, the evaluation of $\mt{L}$ does not terminate, which is a contradiction.
  Therefore, there is no (weak) macro-translation from $\reflang$ to $\eff$.
\end{proof}

\subsection{Proof of Theorem~\ref{thm:reftoac}}\label{app:reftoac}

We extend the translation $M \mapsto \mt{M}$ by introducing a partial function $\eta$, which maps reference cells onto coroutine labels, as follows:
\[
  \mte{l}{\eta} \defines \eta(l).
\]

\begin{definition}
  A binary relation $\simu{\config{M}{\theta}}{\config{N}{\tau}}$ is defined to hold if and only if there exists $\eta$ such that the following conditions are satisfied:
\begin{enumerate}
\item $N = \mte{M}{\eta}$
\item $\eta$ is injective
\item $\eta(\dom{\theta}) \subseteq \dom{\tau}$
\item For any $l \in \dom{\theta}$, if $\theta(l) = V$, then
  \[
    \tau(\eta(l)) = \thunk{\abs{x}{\letin{y}{\return{x}}{\app{\app{\force{\var{loop}}}{\mte{V}{\eta}}}{y}}}}.
  \]
\end{enumerate}
\end{definition}
We define $\theta$, $\tau$, and $\eta$ satisfy the invariant conditions (IC) if the conditions 2 through 4 in the definition are fulfilled.

We prove the following important property of $\var{loop}$:
\begin{lemma}\label{lem:reftoac_loop}
  For any $\ac$ store $\theta$ and value $V$,
  \[
    \redACplus{\config{\app{\force{\var{loop}}}{V}}{\theta}}{\config{
        \app{\app{
            \left\{\
              \begin{array}{l}
                \lambda f. \lambda s. \lambda a.\;\mathbf{case}\;a\;\mathbf{of} \{\\
                \phantom{\lambda f.}\mathbf{inj}_{\mathrm{Get}}\;\unit \mapsto \letin{y}{\yield{s}}{\app{\app{\force{f}}{s}}{y}} \\
                \phantom{\lambda f.}\mathbf{inj}_{\mathrm{Set}}\;v \mapsto \letin{y}{\yield{\unit}}{\app{\app{\force{f}}{v}}{y}}\}
              \end{array}
\right\}!}{\var{loop}}}{V}
      }{\theta}}
  \]
\end{lemma}
\begin{proof}
  By straightforward calculation.
\end{proof}

Next, we establish a substitution property for the translation.
\begin{lemma}\label{lem:reftoac_subst}
  Suppose that a $\reflang$-computation $M$ has free variables $x_1, \ldots x_n$.
  Then for any values $V_1, \ldots V_n$ and store $\eta$, the following equality holds:
  \[
    \mte{M[V_1/x_1,\ldots, V_n/x_n]}{\eta} \equiv \mte{M}{\eta}[\mte{V_1}{\eta}/x_1, \ldots, \mte{V_n}{\eta}/x_n].
  \]
\end{lemma}
\begin{proof}
  By induction on $M$.
\end{proof}

\begin{lemma}\label{lem:reftoac_beta_sim}
  Suppose that $\simu{C}{D}$ and $\redbetaE{C}{C'}$, then there exists a $\del$ configuration $D'$ such that $\redDplus{D}{D'}$ and $\simu{C'}{D'}$.
\end{lemma}
\begin{proof}
  We prove by case analysis on $\redbetaE{C}{C'}$.
  For brevity, we focus on the reduction rules specific to $\reflang$.

  \noindent \textbf{Case} $(\mathrm{create})$:
  \begin{caseindent}
    Suppose that $C \equiv \config{\create{V}}{\theta}$ and $C' \equiv \config{\return{l}}{\theta[l \defines V]}$ for a fresh label $l$.
    By the definition of $\simu{C}{D}$, there exists $\eta$ such that
    \[
      D = \config{\create{\thunk{\abs{x}{\letin{y}{\return{x}}{\app{\app{\force{\var{loop}}}{\mte{V}{\eta}}}{y}}}}}}{\tau}.
    \]
    The configuration $D$ evaluates to $D' \equiv \config{\return{l'}}{\tau\left[ l' \defines \thunk{\abs{x}{\letin{y}{\return{x}}{\app{\app{\force{\var{loop}}}{\mte{V}{\eta}}}{y}}}}\right]}$.
    Let $\eta'$ be $\eta[l \defines l']$.
    By Lemma~\ref{lem:reftoac_subst}, we have $\return{l'} \equiv \mte{\return{l}}{\eta'}$.
    Therefore, we conclude that
    \[
      \simu{\config{\return{l}}{\theta[l \defines V]}}{\config{\mte{\return{l}}{\eta'}}{\tau\left[ l' \defines \thunk{\abs{x}{\letin{y}{\return{x}}{\app{\app{\force{\var{loop}}}{\mte{V}{\eta'}}}{y}}}}\right]}}.
    \]
    Note that $V$ does not contain $l$ since $l$ is a fresh label and the updated stores and $\eta'$ satisfy the IC.
  \end{caseindent}
 
  \noindent \textbf{Case} $(\mathrm{set})$:
  \begin{caseindent}
      Suppose that $C \equiv \config{\set{l}{V}}{\theta}$ and $C' \equiv \config{\return{\unit}}{\theta[l := V]}$ for some $l \in \dom{\theta}$.
  By the definition of $\simu{C}{D}$, there exists $\eta$ such that
  \[
    D \equiv \config{\resume{(\eta(l))}{\paren{\inj{Set}{\mte{V}{eta}}}}}{\tau}
  \]
  and
  \[
    \tau(\eta(l)) = \thunk{\abs{x}{\letin{y}{\return{x}}{\app{\app{\force{\var{loop}}}{\mte{\theta(l)}{\eta}}}{y}}}}.
  \]
  The reduction of $D$ proceeds as follows:
  \begin{align*}
    D &\equiv \config{\resume{(\eta(l))}{\paren{\inj{Set}{\mte{V}{\eta}}}}}{\tau} \\
      &\arrAC \left\langle
        \begin{array}{l}
          {\labeledcp{(\eta(l))}{\app{\force{\thunk{\abs{x}{\letin{y}{\return{x}}{\app{\app{\force{\var{loop}}}{\mte{\theta(l)}{\eta}}}{y}}}}}}{\paren{\inj{Set}{\mte{V}{\eta}}}}}}; \\
          {\tau[(\eta(l)) \defines \nil]}
        \end{array} \right\rangle \\
      &\arrAC^{+} \config{\labeledcp{(\eta(l))}{\app{\app{\force{\var{loop}}}{\mte{\theta(l)}{\eta}}}{\paren{\inj{Set}{\mte{V}{\eta}}}}}}{\tau[(\eta(l)) \defines \nil]} \\
      &\arrAC^{+} \left\langle
        \begin{array}{l}
          \labeledcp{(\eta(l))}{\begin{array}{l}
        \left\{\
        \begin{array}{l}
          \lambda f. \lambda s. \lambda a.\;\mathbf{case}\;a\;\mathbf{of} \{\\
          \phantom{\lambda f.}\mathbf{inj}_{\mathrm{Get}}\;\unit \mapsto \letin{y}{\yield{s}}{\app{\app{\force{f}}{s}}{y}} \\
          \phantom{\lambda f.}\mathbf{inj}_{\mathrm{Set}}\;v \mapsto \letin{y}{\yield{\unit}}{\app{\app{\force{f}}{v}}{y}}\}
        \end{array}
            \right\}! \\
            {\var{loop}}\;{\mte{\theta(l)}{\eta}}\;{\paren{\inj{Set}{\mte{V}{\eta}}}}\end{array}
            };\\
          {\tau[(\eta(l)) \defines \nil]}
        \end{array}
        \right\rangle \\
      &\arrAC^{+} \left\langle
        \begin{array}{l}
          \labeledcp{(\eta(l))}{{\begin{array}{l}
            \mathbf{case}\;\paren{\inj{Set}{\mte{V}{\eta}}}\;\mathbf{of} \{\\
            \phantom{\mathbf{ca}}\mathbf{inj}_{\mathrm{Get}}\;\unit \mapsto \letin{y}{\yield{\mte{\theta(l)}{\eta}}}{\app{\app{\force{\var{loop}}}{\mte{\theta(l)}{\eta}}}{y}} \\
            \phantom{\mathbf{ca}}\mathbf{inj}_{\mathrm{Set}}\;v \mapsto \letin{y}{\yield{\unit}}{\app{\app{\force{\var{loop}}}{v}}{y}}\}
          \end{array}}};\\
          {\tau[(\eta(l)) \defines \nil]}
        \end{array}
        \right\rangle \\
      &\arrAC \config{\labeledcp{(\eta(l))}{\letin{y}{\yield{\unit}}{\app{\app{\force{\var{loop}}}{\mte{V}{\eta}}}{y}}}}{\tau[(\eta(l)) \defines \nil]} \\
      &\arrAC \config{\return{\unit}}{\tau[(\eta(l)) \defines \thunk{\abs{x}{\letin{y}{\yield{\return{x}}}{\app{\app{\force{\var{loop}}}{\mte{V}{\eta}}}{y}}}}]}
        \end{align*}        
        Therefore, we obtain
        \[
          \simu{C'}{\config{\return{\unit}}{\tau[(\eta(l)) \defines \thunk{\abs{x}{\letin{y}{\yield{\return{x}}}{\app{\app{\force{\var{loop}}}{\mte{V}{\eta}}}{y}}}}]}},
        \]
        since the updated stores and $\eta$ satisfy the IC.
  \end{caseindent}

  \noindent \textbf{Case} $(\mathrm{get})$:
  \begin{caseindent}

  Suppose that $C \equiv \config{\get{l}}{\theta}$ and $C' \equiv \config{\return{V}}{\theta}$, where $l \in \dom{\theta}$ and $\theta(l) = V$.
  By the definition of $\simu{C}{D}$, there exists $\eta$ such that
  \[
    D \equiv \config{\resume{(\eta(l))}{\paren{\inj{Get}{\unit}}}}{\tau}
  \]
  and
  \[
    \tau(\eta(l)) = \thunk{\abs{x}{\letin{y}{\return{x}}{\app{\app{\force{\var{loop}}}{\mte{\theta(l)}{\eta}}}{y}}}}.
  \]
  The reduction of $D$ proceeds as follows:
    \begin{align*}
    D &\equiv \config{\resume{(\eta(l))}{\paren{\inj{Get}{\unit}}}}{\tau} \\
      &\arrAC \left\langle
        \begin{array}{l}
          {\labeledcp{(\eta(l))}{\app{\force{\thunk{\abs{x}{\letin{y}{\return{x}}{\app{\app{\force{\var{loop}}}{\mte{\theta(l)}{\eta}}}{y}}}}}}{\paren{\inj{Get}{\unit}}}}}; \\
          {\tau[(\eta(l)) \defines \nil]}
        \end{array} \right\rangle \\
      &\arrAC^{+} \config{\labeledcp{(\eta(l))}{\app{\app{\force{\var{loop}}}{\mte{\theta(l)}{\eta}}}{\paren{\inj{Get}{\unit}}}}}{\tau[(\eta(l)) \defines \nil]} \\
      &\arrAC^{+} \left\langle
        \begin{array}{l}
          \labeledcp{(\eta(l))}{\begin{array}{l}
        \left\{\
        \begin{array}{l}
          \lambda f. \lambda s. \lambda a.\;\mathbf{case}\;a\;\mathbf{of} \{\\
          \phantom{\lambda f.}\mathbf{inj}_{\mathrm{Get}}\;\unit \mapsto \letin{y}{\yield{s}}{\app{\app{\force{f}}{s}}{y}} \\
          \phantom{\lambda f.}\mathbf{inj}_{\mathrm{Set}}\;v \mapsto \letin{y}{\yield{\unit}}{\app{\app{\force{f}}{v}}{y}}\}
        \end{array}
            \right\}! \\
            {\var{loop}}\;{\mte{\theta(l)}{\eta}}\;{\paren{\inj{Get}{\unit}}}\end{array}
            };\\
          {\tau[(\eta(l)) \defines \nil]}
        \end{array}
        \right\rangle \\
      &\arrAC^{+} \left\langle
        \begin{array}{l}
          \labeledcp{(\eta(l))}{{\begin{array}{l}
            \mathbf{case}\;\paren{\inj{Get}{\unit}}\;\mathbf{of} \{\\
            \phantom{\mathbf{ca}}\mathbf{inj}_{\mathrm{Get}}\;\unit \mapsto \letin{y}{\yield{\mte{\theta(l)}{\eta}}}{\app{\app{\force{\var{loop}}}{\mte{\theta(l)}{\eta}}}{y}} \\
            \phantom{\mathbf{ca}}\mathbf{inj}_{\mathrm{Set}}\;v \mapsto \letin{y}{\yield{\unit}}{\app{\app{\force{\var{loop}}}{v}}{y}}\}
          \end{array}}};\\
          {\tau[(\eta(l)) \defines \nil]}
        \end{array}
        \right\rangle \\
      &\arrAC \config{\labeledcp{(\eta(l))}{\letin{y}{\mte{\theta(l)}{\eta}}{\app{\app{\force{\var{loop}}}{\mte{\theta(l)}{\eta}}}{y}}}}{\tau[(\eta(l)) \defines \nil]} \\
      &\arrAC \config{\return{\mte{\theta(l)}{\eta}}}{\tau[(\eta(l)) \defines \thunk{\abs{x}{\letin{y}{\yield{\return{x}}}{\app{\app{\force{\var{loop}}}{\mte{\theta(l)}{\eta}}}{y}}}}]}
        \end{align*}
  
        Therefore, we obtain
        \[
          \simu{C'}{\config{\return{\mte{\theta(l)}{\eta}}}{\tau[(\eta(l)) \defines \thunk{\abs{x}{\letin{y}{\yield{\return{x}}}{\app{\app{\force{\var{loop}}}{\mte{\theta(l)}{\eta}}}{y}}}}]}},
        \]
          since the updated stores and $\eta$ satisfy the IC.
      \end{caseindent}
     
      \end{proof}

We define a macro-translation on contexts by mapping holes to holes and let $\mte{\context{K}}{\eta}$ denote the translation of a context $\context{K}$.

\begin{lemma}\label{lem:reftoac_plug}
  $\mte{\plug{K}{M}}{\eta} \equiv \mte{\mathcal{K}}{\eta}[\mte{M}{\eta}]$ for an arbitrary context $K$ and computation $M$.
\end{lemma}
\begin{proof}
  By straightforward induction on $K$.
\end{proof}

\begin{lemma}\label{lem:reftoac_decomp}
  If $\simu{\config{M}{\theta}}{\config{N}{\tau}}$ and $M$ can be decomposed into an evaluation context $\mathcal{C}$ and a redex $M'$, then there exists some $\eta$ for which $N = \mte{M}{\eta}$ can be also decomposed into an evaluation context $\mte{\mathcal{C}}{\eta}$ and a redex $\mte{M'}{\eta}$ such that $N = \mte{\plug{C}{M'}}{\eta}$.
\end{lemma}
\begin{proof}
  It is easy to check this by induction on the number of frames of $\mathcal{C}$.
  Observe that the extended macro-translation maps frames to frames and redexes to redexes.
\end{proof}

\begin{proposition}[Simulation]\label{prop:reftoac_sim}
  Suppose that $\simu{C}{D}$ and $\redR{C}{C'}$, then there exists a $\ac$ configuration $D'$ such that $\redACplus{D}{D'}$ and $\simu{C'}{D'}$.
\end{proposition}
\begin{proof}
  By the definition of $\simu{C}{D}$, there exist $M$, $\theta$, $\tau$, and $\eta$ such that $C \equiv \config{M}{\theta}$, $D \equiv \config{\mte{M}{\eta}}{\tau}$, and $\theta$, $\tau$, and $\eta$ satisfy the IC.
  Since $C$ is not in normal form, $M$ can be decomposed into an evaluation context $\mathcal{C}$ and a redex $M'$.
  By Lemma~\ref{lem:reftoac_decomp}, $\mte{M}{\eta}$ can also be decomposed into $\mte{\mathcal{C}}{\eta}$ and a redex $\mte{M'}{\eta}$, namely, $\mte{M}{\eta} = \mte{\mathcal{C}}{\eta}\left[\mte{M'}{\eta}\right]$.
  
  Now, suppose $\redbetaRp{M'}{\theta}{M''}{\theta'}$.
  By Lemma~\ref{lem:reftoac_beta_sim}, there exist $\eta'$ and $\tau'$ such that $\redACplus{\config{\mte{M'}{\eta}}{\tau}}{\config{\mte{M''}{\eta'}}{\tau'}}$, $\simu{\config{M''}{\theta'}}{\config{\mte{M''}{\eta'}}{\tau'}}$, and $\theta'$, $\tau'$, and $\eta'$ satisfy the IC.
  Moreover, $\mte{\context{C}}{\eta} \equiv \mte{\context{C}}{\eta'}$ holds since the labels in $\dom{\eta'}\setminus\dom{\eta}$ are fresh and do not appear in $C$.
  Hence, we conclude that $\simu{\config{\plug{C}{M''}}{\theta'}}{\config{\mte{\plug{C}{M''}}{\eta'}}{\tau'}}$ by Lemma~\ref{lem:reftoac_plug}.
\end{proof}

\begin{lemma}\label{lem:reftoac_initial}
  For any $\reflang$-computation $M$, $\simu{\config{M}{\emptyset}}{\config{\mt{M}}{\emptyset}}$ holds.
\end{lemma}
\begin{proof}
  By induction on $M$.
\end{proof}

\begin{lemma}\label{lem:reftoac:well_formed}
  We say a $\reflang$-configuration $\config{M}{\theta}$ is well-formed
  if for any reference cell $l$ that appears in $M$,
  $\theta(l)$ is defined, and this property is preserved under
  reduction.
\end{lemma}
\begin{proof}
  By straightforward case analysis on $M$.
\end{proof}

\begin{lemma}\label{lem:reftoac:diverge}
  Suppose that $\eval{\reflang}{M}$ diverges, i.e., the reduction sequence of $M$ is infinite.
  Then $\eval{\ac}{\mt{M}}$ also diverges.
\end{lemma}
\begin{proof}
  By applying Theorem~\ref{prop:reftoac_sim} repeatedly, we also obtain an infinite reduction sequence of $\mt{M}$.
  Since the reduction of $\ac$ is deterministic, this implies that $\eval{\ac}{\mt{M}}$ also diverges.
\end{proof}

\begin{lemma}\label{lem:reftoac:stuck}
  Suppose that $\eval{\reflang}{M}$ gets stuck: there exists a term $M'$
  and a store $\theta$ such that
  $\redRclos{\config{M}{\emptyset}}{\config{M'}{\theta}}$,
  $M' \neq \return{V}$ for any value $V$, and there is no rule that
  can reduce $\config{M'}{\theta}$.  Then, $\eval{\ac}{\mt{M}}$ also
  gets stuck.
\end{lemma}
\begin{proof}
  It suffices to show the following statement:
  \begin{quote}
    Suppose that there exist $M$, $N$, $\theta$, $\tau$, and $\eta$
    such that $N = \mte{M}{\eta}$, $\config{M}{\theta}$ is well-formed, and $\theta$, $\tau$, and $\eta$
    satisfy the IC.
    If $\config{M}{\theta}$ is stuck, then $\config{N}{\tau}$ is
    also be stuck.
  \end{quote}
  We prove by induction on the structure of $M$.
  However, we treat only three cases. The other cases follow by similar reasoning.

  \noindent\textbf{Case 1}:
  \begin{caseindent}
    \[
      M \equiv \pcase{V}{x_1}{x_2}{M'}
    \]
    Since the $\config{M}{\theta}$ is stuck, we know that
    $V \neq \vpair{V_1}{V_2}$ for any values $V_1$ and $V_2$.
    Consequently, $\mte{V}{\eta}$ also cannot be of the form
    $\vpair{W_1}{W_2}$ for any values $W_1$ and $W_2$.
    Therefore, the configuration
    $\config{\mte{\pcase{V}{x_1}{x_2}{M}}{\eta}}{\tau}$ is stuck,
    and this concludes the current case.
  \end{caseindent}

  \noindent\textbf{Case 2}:
  \begin{caseindent}
    \[
      M \equiv \set{V}{W}
    \]
    It follows that $V \neq l$ for any continuation label $l$,
    since if $V = l$ for some $l$, then by the well-formedness of
    $\config{\throw{V}{W}}{\theta}$, we know that
    $l \in \dom{\theta}$. However, this contradicts that
    $\config{\throw{V}{W}}{\theta}$ is stuck.
    This implies that $\mte{\set{V}{W}}{\eta}$ is also stuck, which
    concludes the current case.
  \end{caseindent}

  \noindent\textbf{Case 3}:
  \begin{caseindent}
    \[
      M \equiv \seq{x}{M_1}{M_2}
    \]
    Since $\config{M}{\theta}$ is stuck, $M_1$
    cannot be of the form $\return{V}$ for any value $V$.
    By the induction hypothesis, we obtain that $N_1$ is stuck and
    that $N_1 \neq \return{W}$ for any value $W$.
    Consequently, the configuration
    $\config{\seq{x}{N_1}{\mte{M_2}{\eta}}}{\tau}$ is also stuck,
    which completes the current case.
  \end{caseindent}
\end{proof}

We now present the proof of Theorem~\ref{thm:reftoac}.

\begin{proof}[Complete proof of Theorem~\ref{thm:reftoac}]
  We prove that $\eval{\reflang}{M}$ terminates if and only if
  $\eval{\ac}{\mt{M}}$ terminates and show the ``only if'' direction
  first.
  Suppose that
  $\redRplus{\config{M}{\emptyset}}{\config{\return{V}}{\theta}}$ for
  some $\theta$.
  By Lemma~\ref{lem:reftoac_initial}, we obtain $\simu{\config{M}{\emptyset}}{\config{\mt{M}}{\emptyset}}$.
  By repeatedly applying Proposition~\ref{prop:reftoac_sim}, it follows that there exist $\eta$ and $\tau$ such that
  \[
    \redACplus{\config{\mt{M}}{\emptyset}}{\config{\return{\mte{V}{\eta}}}{\tau}}.
  \]
  Therefore, $\eval{\ac}{\mt{M}}$ terminates.

  Next, we prove the contrapositive of the ``if'' direction.
  There are two cases where
  $\eval{\reflang}{M}$ does not terminate: (1) $\eval{\del}{M}$
  diverges, or (2) $\eval{\del}{M}$ gets stuck.
  In each case, Lemmas~\ref{lem:reftoac:diverge} and
  \ref{lem:reftoac:stuck} imply that $\eval{\ac}{\mt{M}}$ also
  diverges or gets stuck, respectively.
  Thus, $\eval{\ac}{\mt{M}}$ does not terminate, which completes the
  proof.
  
\end{proof}

\end{document}